\documentclass[aps,prd,amsmath,amssymb,superscriptaddress,altaffillsymbol, preprintnumbers,preprint,nofootinbib,a4paper]{revtex4}
\pdfoutput=1
\usepackage{amsthm}
\usepackage{amsmath}
\usepackage{graphicx}
\usepackage{bbold}
\usepackage{color}

\usepackage{subfigure}
\usepackage{multirow}

\newcommand{\beq}{\begin{eqnarray}}
\newcommand{\eeq}{\end{eqnarray}}

\newcommand{\bmp}{\noindent\begin{minipage}{16cm}}
\newcommand{\emp}{\end{minipage}\vskip 7mm} 

\newcommand{\E}{\mathcal{E}}

\newcommand{\TeV}{\mbox{ ${\mathrm{TeV}}$}}

\usepackage{dcolumn}
\usepackage{bm}
\usepackage{slashed} 

\def\eq#1{{eq.~(\ref{#1})}}
\def\eqs#1#2{{eqs.~(\ref{#1})--(\ref{#2})}}
\def\fig#1{{fig.~(\ref{#1})}}
\def\sec#1{{sec.~(\ref{#1})}}
\def\tab#1{{tab.~(\ref{#1})}}

\theoremstyle{definition}

\theoremstyle{plain}

\usepackage{epsfig}

\usepackage[ margin=5pt, font=normalsize,labelfont=bf,justification=raggedright]{caption}

\usepackage{hyperref}
\definecolor{rossoCP3}{cmyk}{0,.88,.77,.40}
\definecolor{verdeCP3}{rgb}{0.09765625, 0.57421875, 0.1015625}
\definecolor{bluCP3}{rgb}{0, 0.23, 0.67}
\hypersetup{colorlinks, bookmarksopen, bookmarksnumbered,citecolor=verdeCP3, linkcolor=bluCP3, pdfstartview=FitH, urlcolor=rossoCP3}
\def\lsim{\mathrel{\rlap{\lower4pt\hbox{\hskip1pt$\sim$}}
    \raise1pt\hbox{$<$}}}                
\def\gsim{\mathrel{\rlap{\lower4pt\hbox{\hskip1pt$\sim$}}
    \raise1pt\hbox{$>$}}}                
\newcommand{\gt}{\widetilde{g}}

\preprint{LYCEN 2016-05}
\preprint{CP3-Origins-2016-022}

\newcommand{\bea}{\begin{eqnarray}}
\newcommand{\eea}{\end{eqnarray}}

\newcommand{\ba}{\begin{eqnarray}}
\newcommand{\ea}{\end{eqnarray}}
    
\newcommand{\Tr}{\mbox{Tr}\;}
        
\def\eq#1{{eq.~(\ref{#1})}}
\def\eqs#1#2{{eqs.~(\ref{#1})--(\ref{#2})}}
\def\fig#1{{fig.~(\ref{#1})}}
\def\sec#1{{sec.~(\ref{#1})}}
\def\fig#1{{fig.~(\ref{#1})}}
\def\tab#1{{tab.~(\ref{#1})}}

\newcommand{\be}{\begin{eqnarray}}
\newcommand{\ee}{\end{eqnarray}}

\begin{document}
\title{\Large  \color{rossoCP3} ~~\\ Vector and Axial-vector resonances in composite models of the Higgs boson}

\author{Diogo Buarque Franzosi}

\affiliation{II. Physikalisches Institut, Universit\"at G\"ottingen, Friedrich-Hund-Platz 1, 37077 G\"ottingen, Germany}

\author{Giacomo Cacciapaglia}
\author{Haiying Cai}
\author{Aldo Deandrea}

\affiliation{\mbox{Univ Lyon, Universit\'e Lyon 1, CNRS/IN2P3, IPNL, F-69622, Villeurbanne, France.}}

\author{Mads Frandsen}

\affiliation{{\color{rossoCP3} CP$^{\, 3}$-Origins} \& Danish Institute for Advanced Study {\color{rossoCP3} DIAS}, University of Southern Denmark, Campusvej 55, DK-5230 Odense M, Denmark}


\begin{abstract}
We provide a non-linear realisation of composite Higgs models in the context of the $SU(4)/ Sp(4)$ symmetry breaking pattern, where the effective Lagrangian of  the spin-0 and spin-1 resonances is constructed via the CCWZ prescription using the Hidden Symmetry formalism. We investigate the EWPT constraints by accounting the effects from reduced Higgs couplings and integrating out heavy spin-1 resonances.   
This theory emerges from an underlying theory of gauge interactions with fermions, thus
first principle lattice results predict  the massive spectrum in composite Higgs models. 
This model can be used as a template for the phenomenology of composite Higgs models at the LHC and at future 100 TeV colliders, as well as for other application. In this work, we focus on the formalism for spin-1 resonances and their bounds from di-lepton and di-boson searches at the LHC.
 \\[.1cm]

 \end{abstract}

\maketitle


\section{Introduction}

Effective Lagrangian approaches have played a major role in various physics applications to model
unknown sectors or situations that are simply difficult to treat. In more mature fields, they also provide a simple
and more easily calculable tool describing a detailed and complex underlying theory. This is, for example, the case for the 
strong interactions of Quantum ChromoDynamics (QCD) that are described, at low energy, by a chiral Lagrangian
incorporating the light composite degrees of freedom of the theory. The power of the chiral Lagrangian stands on a well defined expansion scheme
that allows for very accurate calculations.
A similar pattern can be followed in the electroweak sector, and attempts to describe the Standard Model (SM) in this way
have been numerous since the the beginning~\cite{Balachandran:1978pk,Eichten:1979ah}. 
In particular, shortly after the theoretical establishment of the SM, the idea that QCD itself may play the role of a template for a composite origin of the electroweak symmetry breaking has been gaining popularity, leading to rescaled-QCD Technicolour models~\cite{Weinberg:1975gm,Susskind:1978ms,Farhi:1980xs}.
In this set up, the longitudinal degrees of freedom of the massive $W$ and $Z$ are accounted for as Goldstone bosons, i.e. pions, of the strong sector. While nowadays it is accepted that rescaled-QCD does not describe the physical reality, especially due to its Higgs-less nature, issues with generating quark masses~\cite{Dimopoulos:1979es,Eichten:1979ah} and the correct flavour structure~\cite{Dimopoulos:134187} and precision tests~\cite{Peskin:1990zt}, new versions of such theories are gaining momentum.

The main break-through can be traced back to the idea that the Higgs too can be described as a pseudo Nambu-Goldstone boson (pNGB) of an enlarged flavour symmetry~\cite{Kaplan:1983fs,Georgi:1984af}, and the coset SU(5)/SO(5), containing a singlet and a 9-plet of the custodial SO(4)$\sim$SU(2)$_L \times$SU(2)$_R$ together with the Higgs doublet, has been one of the first candidates~\cite{Dugan:1984hq}.
For recent reviews on the developments occurred in the last decade, we refer the reader to Ref.s~\cite{Bellazzini:2014yua,Panico:2015jxa}.
The realisation that the minimal symmetry breaking, SO(5)/SO(4),  embedding custodial symmetry, contains only a Higgs boson at low energy~\cite{Agashe:2004rs} has inspired the construction of effective descriptions of the electroweak (EW) sector of the SM which do not contain more states~\cite{Giudice:2007fh,Contino:2013kra,Buchalla:2014eca}.
Other composite states which are not pNGBs, like spin-1 resonances, or the so-called top partners~\cite{Bellazzini:2014yua} (needed in the partial compositeness scenario~\cite{Kaplan:1991dc}), are typically heavier than a few TeV and thus their effect at low energy can be embedded in higher order operators. Yet, at the energy reached by the LHC, their direct production is a crucial test of the theories. A very rich literature is already available, and we list here a forcibly incomplete list of papers addressing various issues on the experimental tests of composite top partners~\cite{Contino:2008hi,Dissertori:2010ug,Li:2013xba,Flacke:2013fya,Matsedonskyi:2014mna, Cacciapaglia:2015dsa, Serra:2015xfa,Backovic:2015bca,Matsedonskyi:2015dns}, aka vector-like fermions~\cite{Buchkremer:2013bha,Barducci:2014ila},  spin-1 resonances~\cite{Contino:2011np,Low:2015uha,Niehoff:2015iaa}, or additional scalars~\cite{Cacciapaglia:2015eqa,Ferretti:2016upr}.  
In particular, the search for top partners at both Run--I and Run--II has produced bounds on their masses which are now approaching 1 TeV.

Models beyond the minimal case are interesting as they contain additional light scalars~\cite{Katz:2005au,Mrazek:2011iu,Bertuzzo:2012ya}, among which a Dark Matter candidate may arise~\cite{Frigerio:2012uc,Marzocca:2014msa}. One possible way to discriminate among them is the requirement that they arise from an underlying theory consisting on a confining fermionic gauge theory. In this sense, the coset SU(4)/Sp(4) (equivalent to SO(6)/SO(5)) can be considered the minimal one, also arising from a very simple underlying theory~\cite{Ryttov:2008xe,Galloway:2010bp} based on a confining SU(2) Yang-Mills theory with 2 Dirac fermions in the fundamental representation. 
Besides being a template for a Composite Higgs model~\cite{Cacciapaglia:2014uja}, this theory has also been used as a simple realisation of the SIMP Dark Matter candidate~\cite{Hochberg:2014kqa} (for a critical assessment, see~\cite{Hansen:2015yaa}).
In this paper we want to extend the effective field theory studies of this template by adding the lowest-lying spin-1 resonances, i.e. vector and axial-vector states~\cite{Appelquist:1999dq,Duan:2000dy}.
We follow the CCWZ~\cite{Coleman:1969sm,Callan:1969sn} prescription by employing the hidden symmetry technique~\cite{Bando:1987br}.
While we focus on the Composite Higgs scenario with the scope of studying the phenomenology of such states at the LHC and at future higher energy colliders, our construction can be also applied to other phenomenological uses of this simple theory~\cite{Batra:2007iz,Hochberg:2014kqa,Hietanen:2014xca,Drach:2015epq}.

After reviewing the basic properties of the SU(4)/Sp(4) coset, in Section~\ref{sec:lag} we construct an effective Lagrangian for the spin-1 states. In Section~\ref{sec:prop} we provide details of the properties of the physical states and connect them with the simplest underlying theory in Section~\ref{sec:UV}. Finally, we briefly study the collider phenomenology in Section~\ref{sec:pheno}, focusing both on di-lepton constraints at the LHC and on prospects for the future 100 TeV proton collider.

\subsection{Vacuum alignment structure and fermion mass generation}

The vacuum structure of the SU(4)/Sp(4) model has already been extensively studied~\cite{Katz:2005au,Gripaios:2009pe,Galloway:2010bp}, so here we will briefly recap the main features.
In this work, we will follow the prescription that the pNGB fields are defined around a true vacuum which includes the source of electroweak symmetry breaking, as in Ref.s~\cite{Galloway:2010bp,Cacciapaglia:2014uja}.
It can be thus shown that the vacuum alignment can be described in terms of a single parameter, $\theta$, and in the SU(4) space it looks like
\beq
\Sigma_0 = \left( \begin{array}{cc}
\cos \theta\ (i \sigma_2) & \sin \theta\ (\mathbb{1}) \\
- \sin \theta\ (\mathbb{1}) & - \cos \theta\ (i \sigma_2)
\end{array} \right) = u_H (\theta) \cdot \left( \begin{array}{cc}
(i \sigma_2) & 0 \\
0 & - (i \sigma_2)
\end{array} \right) \cdot u_H^T (\theta)\,,
\eeq 
where $u_H$ is an SU(4) rotation along a direction in the space defined by the generators transforming like a Brout-Englert-Higgs doublet.
For $\theta = 0$, i.e. $u_H (0) = \mathbb{1}$, the electroweak symmetry is unbroken once the SU(2)$_L$ and U(1)$_Y$ generators are embedded in SU(4) as
\beq
S^i = \frac{1}{2} \left( \begin{array}{cc}
\sigma^i & 0 \\ 0 & 0 \end{array} \right)\,, \quad Y = S^6 = \frac{1}{2} \left( \begin{array}{cc} 0 & 0 \\ 0 & - \sigma_3^T \end{array} \right)\,.
\eeq
In the phase where $\theta$ is non-zero, the gauged generators of SU(4) are no more aligned with the 10 unbroken generators $V^a$ defined as
\beq \label{eq:VY}
V^a \cdot \Sigma_0 + \Sigma_0 \cdot {V^a}^T = 0\,, \quad Y^a \cdot \Sigma_0 - \Sigma_0 \cdot {Y^a}^T = 0\,,
\eeq
where $Y^a$ are the 5 broken ones (explicit matrices can be found in~\cite{Cacciapaglia:2014uja}).
As mentioned above, the pNGBs are defined around the $\theta$-dependent vacuum $\Sigma_0$ as~\footnote{Other parameterisation have been used in the literature where the pNGBs are defined around the $\theta=0$ vacuum, and the ``Higgs'' one is then assigned a vacuum expectation value~\cite{Gripaios:2009pe}. The main differences lie in higher order interactions, see~\cite{Marzocca:2014msa}. We prefer this approach because it sequesters the explicit breaking of the Goldstone shift symmetry to the potential terms that generate the pNGB masses.}
\beq
U = \exp \left[ i \frac{\sqrt{2}}{f_\pi} \sum_a \pi_a Y^a \right],
\eeq
where the would-be Higgs boson is identified with $h = \pi_4$, the singlet $\eta = \pi_5$, and the remaining 3 are exact Goldstones eaten by the massive $W$ and $Z$.
Also, $f_\pi$ corresponds to the decay constant~\footnote{Here, we adopt the standard normalisation used in Technicolour literature, and other composite Higgs literature: the difference with the $f$ used in~\cite{Cacciapaglia:2014uja,Arbey:2015exa}  is $f_\pi = 2 \sqrt{2} f$.} of the pNGBs, and it is related to the electroweak scale via $\theta$:
\beq \label{eq:fpi}
f_\pi \sin \theta = v_{\rm SM} = 246~\mbox{GeV}.
\eeq

The alignment along $\theta$, together with the masses of the two physical pNGB, is then fixed by a potential generated by explicit breaking terms of the SU(4) flavour symmetry: the gauging of the electroweak symmetry, Yukawa couplings for the top (above all) and a mass for the underlying fermions (which is allowed by the symmetries as the underlying theory is vectorial).

The origin of the potential and the top mass is only relevant for the current study in so far as it modifies the spin-1 phenomenology. 
The top mass for this model may arise as in extended technicolor (ETC) descriptions via 4-fermion operators bilinear in the top-quark. This is discussed for the $SU(4)/Sp(4)$ coset in e.g.~\cite{Galloway:2010bp,Cacciapaglia:2014uja}. These bilinear 4-fermion interactions may arise from the exchange of heavy spin-1 bosons~\cite{Eichten:1979ah,Dimopoulos:1979es} or heavy scalars~\cite{'tHooft:1979bh} external to the strongly interacting sector considered here. A recent explicit example employing a chiral gauge theory is provided in \cite{Cacciapaglia:2015yra}.
These ETC interactions induce direct couplings of the spin-1 resonances with SM fermions and these couplings can provide a welcome negative contribution to the electroweak $S$-parameter~\cite{Chivukula:2005bn}. However these couplings are typically negligible relative to those induced by the mixing between the heavy spin-1 resonances, especially for the light SM fermion generations,  and the effects on $S$ consequently small.  We therefore ignore these small effects in the current study. 
%
%
%
Alternatively the top mass may also arise via fermion partial compositeness~\cite{Kaplan:1991dc}, a mechanism that requires the presence of fermionic bound states that mix linearly to the elementary fields. The realisation of this mechanism in terms of explicit 4d gauge theories with fermions, relevant for our coset, has been studied in~\cite{Ferretti:2013kya,Barnard:2013zea,Ferretti:2016upr}. 
Top partners may play a dual role of generating the top mass and stabilising the Higgs potential, in which case one would expect them to be parametrically lighter than the typical resonance scale~\cite{Matsedonskyi:2012ym}. Else, other spurions like a mass for the underlying fermions~\cite{Cacciapaglia:2014uja} can be used as a stabiliser, and the top partner can be heavy and irrelevant for the phenomenology of the Higgs~\footnote{For the potential, details can be found in~\cite{Cacciapaglia:2014uja} for the case where the fermion mass is used as a stabiliser, and in~\cite{Gripaios:2009pe} for the case where top partners are present and used to fine tune the top loops.} and vector resonances.
In both mechanisms sketched above, the model needs to face severe constraints from flavour observables, especially in the form of Flavour Changing Neutral Currents induced by four-fermion operators at the flavour scale.
The usual way out relies on the presence of a Conformal Theory in the UV, that generates large anomalous dimensions responsible for the separation of the scale of flavour violation from the scale of compositeness and the generation of hierarchies in the fermion masses. Obtaining large anomalous dimensions, however, has been proven to be very challenging for scalar operators~\cite{Rattazzi:2008pe,Rychkov:2009ij,Rattazzi:2010yc} (as needed in the case of bilinear 4-fermion operators), as well as for fermionic ones (as in partial compositeness, see for instance~\cite{Pica:2016rmv,Vecchi:2016whd}). A complete theory of flavour is thus still beyond the horizon.

In the following, we will assume the case of 4-fermion interaction, or of heavy top partners, and only focus on the dynamics responsible for the composite Higgs.
As realising partial compositeness requires extending the strongly interacting sector (need additional coloured techni-fermions), we leave this possibility and the study of the interplay between top partners and vectors for a future study.

\section{Effective Lagrangian} \label{sec:lag}

To describe the new strong sector and remain as general as possible, a chiral-type theory 
can be constructed on the basis of custodial symmetry and gauge invariance. 
The simplest construction one can imagine uses a local copy of the global SU(2) ``chiral'' symmetry and builds the relevant invariants \cite{Casalbuoni:1991cw}. 
The same results follow from the hidden gauge symmetry approach \cite{Bando:1987br}. Furthermore the global flavour symmetry can be enlarged in different ways,
depending on the required model-building features (see for example the early attempts in \cite{Casalbuoni:1992dd,Casalbuoni:1995qt}).
To this basic idea one can add the Higgs boson as a pseudo-Nambu-Goldstone boson \cite{Kaplan:1983fs,Georgi:1984af} or as a massive composite state, 
or as a superposition of both \cite{Cacciapaglia:2014uja}.
In the following we shall consider a model with vector and axial-vector particles: for a template description of these resonances based on the SU(2) group see  
\cite{Casalbuoni:1988xm}. In order to describe this kind of spectrum, we introduce a local copy of the global 
symmetry. When the new vector and axial-vector particles decouple, one obtains the non--linear sigma-model Lagrangian, describing the Goldstone bosons 
associated to the breaking of the starting symmetry to a smaller one. The approach we use is the standard one of the hidden gauge symmetry \cite{Bando:1987br} (for an alternative, equivalent, way, see~\cite{Marzocca:2012zn})~\footnote{In that approach only one set of pNGBs is used for the CCWZ prescription, and  the mass term for the axial-vectors $\frac{f_a^2}{2 \Delta^2}(g_a a_\mu - \Delta d_\mu)^2$ will give rise to a  bilinear mixing of  $a_\mu^4 \partial_\mu h$. }.

In our specific case, i.e. the minimal model with a fermionic gauge theory as underlying description, the global symmetry SU(4)/Sp(4) is extended in order to contain, initially, two SU(4)$_i$, $i=0,1$. The SU(4)$_0$ corresponds to the usual global symmetry leading to the Higgs as a composite pNGB, and the electroweak gauge bosons are introduced via its partial gauging. The new symmetry SU(4)$_1$ allows us to introduce a new set of massive ``gauge'' bosons, transforming as a complete adjoint of SU(4), which correspond to the spin-1 resonances in this model.

\subsection{Lagrangian}

Following the prescription of the hidden gauge symmetry formalism, we enhance the symmetry group 
${\mathrm{SU(4)}}$ to ${\mathrm{SU(4)}}_0\times {\mathrm{SU(4)}}_1$, and embed the SM gauge bosons in SU(4)$_0$ and the heavy resonances in SU(4)$_1$.
The low energy Lagrangian is then characterised in terms of the breaking of the extended symmetry down to a single Sp(4):
the ${\mathrm{SU(4)}}_i$ are spontaneously broken to 
${\mathrm{Sp(4)}}_i$ via the introduction of 2 matrices $U_i$ containing 5 pNGBs each. The remaining ${\mathrm{Sp(4)}}_0\times {\mathrm{Sp(4)}}_1$ is then spontaneously broken to ${\mathrm{Sp(4)}}$ 
by a sigma field $K$, containing 10 pNGBs corresponding to the generators of Sp(4). 

The 5+5 pNGBs associated to the generators in SU(4)/Sp(4) are parameterised by the following matrices:
\beq
	U_0 = \exp\left[ \frac{i \sqrt 2}{f_0} \sum_{a=1}^5 ( \pi_0^a  Y^a )  \right],   \qquad
        U_1 = \exp\left[ \frac{i \sqrt 2}{f_1} \sum_{a=1}^5 ( \pi_1^a  Y^a )  \right]\,,
\eeq
that transform nonlinearly as 
\beq
U_i\to U^\prime_i=g_i U_i h(g_i,\pi_i)^\dagger\,,
\eeq
where $g_i$ is an element of ${\mathrm{SU(4)}}_i$ and $h$ the corresponding transformation in the subgroup ${\mathrm{Sp(4)}}_i$.
It is convenient to define the gauged   Maurer-Cartan one-forms as
\begin{equation}
	\omega_{R\,i, \mu}  =  U_i^\dag D_{\mu} U_i\,,
\end{equation}
where $D_\mu$ are the appropriate covariant derivatives
\begin{eqnarray}
D_\mu U_0 &=& ( \partial_\mu -i g {\bf \widetilde{W}}_\mu - i g^\prime {\bf B_\mu} ) U_0\,,  \\
D_\mu U_1 &=& (\partial_\mu -i \widetilde{g} \bm{\mathcal{V}}_\mu - i \widetilde{g} \bm{\mathcal{A}}_\mu )U_1\,. 
\end{eqnarray}
The spin-1 fields are embedded in SU(4) matrices as
\begin{equation} \label{eq:vectors0}
{\bf B_\mu}=B_\mu\ S_6, \quad  {\bf\widetilde{ W}_\mu}=\sum_{a=1}^{3} \widetilde{W}_\mu^a\ S_a, \quad \bm{\mathcal{V}}_\mu = \sum_{a=1}^{10} {\cal V}_\mu^a\ V_a, \quad \bm{\mathcal{A}}_\mu = \sum_{a=1}^{5} {\cal A}_\mu^a\ Y_a\,,
\end{equation}
where $\widetilde{W}^k_\mu$ ($k=1,\,2,\,3$) and $B_\mu$ are the elementary electroweak gauge bosons associated with the $SU(2)_L$ and $U(1)$ hypercharge groups.  The vector $\mathcal{V}^j_\mu$ (j=1 to 10) and axial-vector $\mathcal{A}^l_\mu$ (l=1 to 5) are the composite resonances generated by the strong dynamics and associated to the unbroken $V_a$ and broken $Y_a$ generators as defined in \eq{eq:VY}.
The projections to the broken and unbroken generators are defined respectively by
\begin{eqnarray}
p_{\mu\,i}&=& 2\sum_a \Tr(Y_a \omega_{R\,i,\mu})\,Y_a\,, \\
v_{\mu\,i}&=&2\sum_a \Tr(V_a \omega_{R\,i,\mu})\,V_a \,,
\end{eqnarray}
so that $v_{\mu\,i}$ transforms inhomogeneously under $SU(4)_i$
\be
v_{\mu\,i}\to v_{\mu\,i}^\prime=h(g_i,\pi_i)\, (v_{\mu\,i}+i\partial_\mu)\, h^\dagger(g_i,\pi_i)\,,
\ee
while $p_{\mu\,i}$ transforms homogeneously
\be
p_{\mu\,i}\to p_{\mu\,i}^\prime=h(g_i,\pi_i)\, p_{\mu\,i}\, h^\dagger(g_i,\pi_i)\,
\ee
and can be used to construct invariants for the effective Lagrangian.

The $K$ field is introduced to break the two remaining copies of $Sp(4)$, $Sp(4)_0\times Sp(4)_1$ to the diagonal final $Sp(4)$:
\begin{equation}
	K = \exp\left[ i k^a V^a / f_K \right]\,,
\end{equation}
and it transforms like
\be
K\to K^\prime=h(g_0,\pi_0)\, K\, h^\dagger(g_1,\pi_1)\,,
\ee
thus its covariant derivative takes the form
\begin{equation}
D_\mu K = \partial_\mu K -i v_{0\mu} K + i K v_{1\mu} \,.
\end{equation}  
The 10 pions contained in $K$ are needed to provide the longitudinal degrees of freedom for the 10 vectors $\mathcal{V}^j_\mu$, while a combination of the other pions $\pi_i$ act as longitudinal degrees of freedom for the $\mathcal{A}^l_\mu$. It should be reminded that out of the 5 remaining scalars, 3 are exact Goldstones eaten by the massive $W$ and $Z$ bosons, while 2 remain as physical scalars in the spectrum: one Higgs-like state plus a singlet $\eta$.

To lowest order in momentum expansion, and including the scalar singlet $\sigma$, the effective Lagrangian is given by
\begin{eqnarray}
{\cal L} 
&=&-\frac{1}{2g^2}\ {\rm Tr}\ {\bf \widetilde  W}_{\mu\nu} {\bf \widetilde W}^{\mu\nu}
       -\frac{1}{2g^{\prime 2}}\ {\rm Tr}\ {\bf B}_{\mu\nu} {\bf B}^{\mu\nu} 
       -\frac{\kappa_F (\sigma)}{2\gt^2}\ {\rm Tr}\ \bm{\mathcal{F}}_{\mu\nu} \bm{\mathcal{F}}^{\mu\nu} \nonumber\\ 
&+&\frac{1}{2}\kappa_{G_0}(\sigma)f_0^2\ {\rm Tr}\ p_{0\mu} p_0^\mu
       + \frac{1}{2}\kappa_{G_1}(\sigma)f_1^2\ {\rm Tr}\ p_{1\mu} p_1^\mu
      + r(\sigma) f_{1}^2\ {\rm Tr}\ p_{0\mu} K p_1^\mu K^\dagger \nonumber \\
&+& \frac{1}{2} \kappa_{K} (\sigma) f_K^2\ {\rm Tr}\ {\cal D}^\mu K\ {\cal D}_\mu K^\dagger 
       + \frac{1}{2} \partial_\mu \sigma \partial^\mu \sigma - {\cal V}(\sigma) \nonumber\\
&+& {\cal L}_{fermions} \,.
\label{eq:DEWSB}
\end{eqnarray}
We have introduced the singlet field $\sigma$ for generality, as it may be light in some theories, via generic functions in front of the operators in the strong sector: in the following, however, we will be interested to the case where it's heavy and thus we will replace the functions by the first term in the expansion, i.e. $\kappa_X (\sigma) = 1$ and $r(\sigma) = r$. 
The field strength tensors are defined by
\beq
{\bf V}_{\mu\nu}=\partial_\mu {\bf V}_\nu -\partial_\nu {\bf V}_\mu -i[{\bf V}_\mu, {\bf V}_\nu]
\eeq
for ${\bf V}_\mu={\bf B}_\mu,\,{\bf \widetilde W}_\mu,$ and 
$\bm{\mathcal{F}}_\mu=\bm{\mathcal{V}}_\mu+\bm{\mathcal{A}}_\mu$.
The canonically normalised fields are $g^\prime\ B_\mu$, $g\ \widetilde{W}_\mu^k$, $\gt\  {\cal V}_\mu^k$ and $\gt\ {\cal A}_\mu^k$.

Due to the presence of the $r$ term in the Lagrangian, the pions $\pi_{0,1}$ do not have proper kinetic terms. Calling the normalised fields $\pi_A$ and $\pi_B$, they are given by
\beq
\pi_0^a&=&\frac{\pi_A^a}{\sqrt{2}\sqrt{1+r\,f_1/f_0}}-\frac{\pi_B^a}{\sqrt{2}\sqrt{1-r\,f_1/f_0}}, \\
\pi_1^a&=&\frac{\pi_A^a}{\sqrt{2}\sqrt{1+r\,f_1/f_0}}+\frac{\pi_B^a}{\sqrt{2}\sqrt{1-r\,f_1/f_0}}.
\label{eq:NGorthonorm}
\eeq
As already mentioned, a linear combination of the two sets of 5 pions is eaten by the vector states ${\cal A}_\mu$ once they pick up their mass.
The eaten Goldstones $\pi_U^a$, and the 5 physical ones $\pi_P^a$ before the EW gauging, are given by
\beq
\pi_A^a&=&\cos\alpha\,\pi_P^a-\sin\alpha\,\pi_U^a, \\
\pi_B^a&=&\sin\alpha\,\pi_P^a+\cos\alpha\,\pi_U^a, 
\eeq
where the mixing angle $\alpha$ is
\beq
\tan\alpha=-\sqrt{\frac{1+rf_1/f_0}{1-rf_1/f_0}}\,.
\eeq
Combining the above redefinitions, we get~\footnote{Note that for $r=0$, we would have $\pi_P = \pi_0$ and $\pi_U = \pi_1$, as expected seen that SU(4)$_1$ is associated with the massive vectors. Thus, $r$ parameterises the mixing between the two sectors.  The rotation defined  for $\pi_P$ and $\pi_U$  is divergent  for  $r=f_0/f_1$,  this point is not physical since  it will lead to $v =0$ thus no EWSB can be generated.}
\beq
\pi_0^a&=&\pi_P^a\frac{1}{\sqrt{1-r^2f_1^2/f_0^2}}, \\
\pi_1^a&=&\pi_U^a-\pi_P^a\frac{rf_1/f_0}{\sqrt{1-r^2f_1^2/f_0^2}}.
\eeq
Note that only the pions associated with $Y^4$ and $Y^5$ are physical, as the remaining 3 are exact Goldstones eaten by the $W$ and $Z$. In the following, we will associate  one with the Higgs boson, $\pi_P^4 = h$, and the other with the additional singlet $\pi_P^5 = \eta$ of the SU(4)/Sp(4) coset.

\subsection{Other terms}
The previous Lagrangian contains the low energy composite sector in terms of effective fields using the CCWZ formalism and the hidden symmetry one, allowing 
for a description of composite spin-0 pNGB and spin-1 vector and axial-vector resonances. The interactions among these states are, to a
large extent, described by this formalism, however some extra terms can potentially be added. 
While we leave a detailed study to a future work, it is worth 
mentioning how they can affect the phenomenology when added. 

A first set of contributions are those induced by the Wess-Zumino-Witten (WZW) anomaly~\cite{Wess:1971yu,Witten:1979vv}. In our case, these can be added in a 
similar way to what is done for chiral Lagrangians to describe, for example, the decay of a neutral pion into two photons. These terms are relevant for
di-boson final states, allowing a scalar pNGB to couple to the SM gauge bosons. Furthermore, anomalous couplings of the vectors will also be generated, thus potentially providing new decay channels.

Another set of possible terms are the ones allowing the spin-1 vector and axial-vector resonances to couple directly to fermions instead of getting their coupling
to the SM fermions only by mixing effects with the SM gauge bosons. Such couplings are allowed by the symmetries of the Lagrangian, as in a similar way to the SM,
the fermionic current couples to weak SU(2) gauge triplets. However the phenomenological constraints indicate that the new direct coupling should 
be small. Nevertheless this new source of direct decay to fermions for the new composite vectors can have a non-negligible impact on phenomenology.
Finally, in analogy with QCD, the new $\eta$ and $\sigma$ of the underlying strong dynamics require a detailed study. This part of the scalar sector is not the 
main focus here and a detailed description can be found in  \cite{Arbey:2015exa}. Other possible items in this list are symmetry breaking terms and kinetic 
mixing terms. All these points will not be discussed further here, and may deserve a separate study.

\section{Properties of vector states} \label{sec:prop}

The intrinsic properties of the 15 spin-1 states introduced in \eq{eq:vectors0} determine the structure of masses, mixing, couplings and their contributions to electroweak precision tests.
The vector fields can be organised as a matrix in SU(4) space, defined by
\begin{equation}
\bm{\mathcal{F}}_\mu = \bm{\mathcal{V}}_\mu+\bm{\mathcal{A}}_\mu = \sum_{a=1}^{10} {\cal V}_\mu^a V_a + \sum_{a=1}^{5} {\cal A}_\mu^a Y_a.
\label{eq:vectors}
\end{equation}
where the generators in the general vacuum, $V^a$ and $Y^a$, are defined in \eq{eq:VY}.
Under the unbroken Sp(4), the two multiplets transform as a ${\bf 10}$ and a ${\bf 5}$ respectively.
It is however more convenient to classify the states in terms of their transformation properties under a subgroup SO(4)$\subset$Sp(4), which corresponds to the custodial symmetry SU(2)$_L \times$ SU(2)$_R$ of the SM Higgs sector in the limit $\theta \to 0$:
\beq
\bm{\mathcal{V}} \to {\bf 10}_{\rm Sp(4)} = (3,1) \oplus (1,3) \oplus (2,2)\,, \qquad \bm{\mathcal{A}} \to {\bf 5}_{\rm Sp(4)} = (2,2) \oplus (1,1)\,.
\eeq
Physically, however, the SM custodial symmetry is broken to the diagonal SU(2)$_V$ in a generic vacuum alignment, under which symmetry the physical spectrum contains 4 triplets, plus additional singlets. A complete list of the states, and their classification, can be found in \tab{tab:class}.
We would like to remind the reader, here, of the generic properties of the vector resonances in a minimal case of composite EWSB: in fact, any model of compositeness necessarily contains the spontaneous breaking of SU(2)$_L \times$ SU(2)$_R \to$ SU(2)$_V$, under which one expects to have a vector triplet $\overrightarrow{\rho}_\mu$ and  an axial-vector triplet $\overrightarrow{a}_\mu$. In our case, more states are present (shown explicitly in their SU(4) embedding in \eq{eq:amu}), however one can always identify states corresponding to the minimal case.
In fact, the triplet $v^{0,\pm}_\mu$ can always be associated to the $\overrightarrow{\rho}_\mu$ of the ``vector'' SU(2); on the other hand the interpretation of the axial-vector depends on the specific realisation of the model. On one hand, in the Technicolor limit, $\theta = \pi/2$, we find that $a^{0,\pm}_\mu$, which has a component of axial-vector $SU(2)_A$ proportional to $\sin\theta$, can be associated to $\overrightarrow{a}_\mu$; on the other hand, in the pNGB Higgs limit, $\theta \to 0$, it is $s^{0,\pm}_\mu$, having a $SU(2)_A$ component proportional to $\cos\theta$, that transforms like the axial-vector states. All the above states mix with the elementary gauge bosons of the SM.
Simplified models describing vector triplets have been used in the composite Higgs literature: for instance, two triplets corresponding to our  $v^{0,\pm}_\mu$ and $s^{0,\pm}_\mu$ are usually considered in the minimal $SO(5)/SO(4)$ model \cite{Contino:2011np}, while in more simplified cases a single triplet is accounted for \cite{Belyaev:2008yj,Bellazzini:2012tv,Castillo-Felisola:2013jua,Low:2015uha}.
Although our complete Lagrangian contains such states, it is not possible to find limits where the other states decouple~\footnote{For instance, one may decouple the $\bm{\mathcal{A}}$ states by sending $f_1 \to \infty$ (and $r \to 0$), however the $\tilde{s}^{0,\pm}_\mu$ and $\tilde{v}^0_\mu$ will remain light.}. Thus, simplified models can only partially describe the phenomenology of the model under study.

\begin{table}[tb]
\begin{center}
\begin{tabular}{cc|c|c|c|c|}
&  &  SU(2)$_V$ & SU(2)$_L \times$ SU(2)$_R$ & TC & CH \\
\hline
\multirow{4}{*}{$\bm{\mathcal{V}}$} & $v^{0,\pm}_\mu$ & $3$ & \multirow{2}{*}{(3,1)$\oplus$(1,3)} & $\overrightarrow{\rho}_\mu$ & $\overrightarrow{\rho}_\mu$ \\
& $s^{0,\pm}_\mu$ & $3$ & & & $\overrightarrow{a}_\mu$ \\
& $\tilde{s}^{0,\pm}_\mu$ & $3$ & \multirow{2}{*}{(2,2)} & & \\
& $\tilde{v}^0_\mu$ & $1$ & & & \\
\hline \hline
\multirow{3}{*}{$\bm{\mathcal{A}}$} & $a^{0,\pm}_\mu$ & $3$ & \multirow{2}{*}{(2,2)} & $\overrightarrow{a}_\mu$ & \\
& $x^0_\mu$ & $1$ & & & \\
& $\tilde{x}^0_\mu$ & $1$ & (1,1) & & \\
\hline 
\end{tabular}
\caption{Classification of the spin-1 resonances in the model. } \label{tab:class}
\end{center}
\end{table}

The additional states $\widetilde{s}^{\pm,0}_\mu$, $\widetilde{v}^0_\mu$, $x^0_\mu$ and $\widetilde{x}^0_\mu$ do not mix with the elementary gauge bosons: their masses are given by
\begin{equation}
 M_{\widetilde{s}} = M_{\widetilde{v}^0} = M_V \quad\text{   and   }  \quad
 M_{x^0} = M_{\widetilde{x}^0} = M_A \,.
\end{equation}
where the mass parameters $M_A$ and $M_V$ are defined in terms of Lagrangian parameters as
\beq
M_A \equiv\frac{\widetilde{g}f_1}{\sqrt{2}} \quad \text{   and   }  \quad M_V\equiv\frac{\widetilde{g}f_K}{\sqrt{2}} \,.
\eeq

The other states mix among themselves and with the SM weak bosons ($\widetilde{W}^i_\mu$ and $B_\mu$).
The mass mixing Lagrangian is  
\begin{eqnarray}
{\cal L}_{\rm mass} =
\begin{pmatrix} \widetilde{W}^-_\mu & a^-_\mu & v^-_\mu & s^-_\mu \end{pmatrix} {\bf {\cal M}_{\rm C}^2}
\begin{pmatrix} \widetilde{W}^{+\mu} \\ a^{+\mu} \\ v^{+\mu} \\ s^{+\mu}  \end{pmatrix} +
\frac{1}{2}\begin{pmatrix} B_\mu & \widetilde{W}^3_\mu & a^0_\mu & v^0_\mu & s^0_\mu \end{pmatrix} {\bf {\cal M}_{\rm N}^2}
\begin{pmatrix} B^\mu \\ \widetilde{W}^{3\mu} \\ a^{0\mu} \\ v^{0\mu} \\ s^{0\mu} \end{pmatrix} .
\label{eq:mixmatrix}
\end{eqnarray}
The matrices ${\bf {\cal M}_{\rm C}^2}$ and ${\bf {\cal M}_{\rm N}^2}$ are given in \eqs{eq:massC}{eq:massN}.

Upon diagonalisation, the interaction eigenstates are rotated to the physical vector bosons 
\bea
\left(\begin{array}{c} \widetilde{W}^{+\mu} \\ a^{+\mu} \\ v^{+\mu} \\ s^{+\mu}  \end{array}\right)
= {\cal C}
\left(\begin{array}{c} W^{+\mu} \\ A^{+\mu} \\ V^{+\mu} \\ S^{+\mu} \end{array}\right)\ , \quad
\left(\begin{array}{c} B^\mu \\ \widetilde{W}^{3\mu} \\ a^{0\mu} \\ v^{0\mu} \\ s^{0\mu} \end{array}\right)
= {\cal N}
\left(\begin{array}{c} A^\mu \\ Z^{\mu} \\ A^{0\mu} \\ V^{0\mu} \\ S^{0\mu} \end{array}\right)\ .
\label{eq:mixN}
\eea
Approximate expressions for ${\cal C}$ and ${\cal N}$ are given in Appendix \ref{sec:pertdiag} in an expansion for large $\gt$.
The eigenstate in the neutral sector which is exactly massless is identified to be the photon, and it is related to the interactions eigenstates (exactly in $\gt$) as
\beq
{A_\mu } = \frac{e}{g}\ \widetilde{W}^3_\mu + \frac{e}{g'}\ B_\mu + \sqrt 2 \frac{e}{\widetilde g}\ v^0_\mu 
\eeq
with 
\beq
1/e^2 = 1/{g'}^2 + 1/g^2 + 2 /{\gt}^2\,. \label{eq:ee}
\eeq 
Besides the photon, all the massive states mix with each other with mixing angles typically of order $1/\gt$, with the exception of $v_\mu$ and $s_\mu$ whose mixing is controlled by the angle $\theta$. For instance, in the charged sector, see \eq{eq:massC}, it is clear that the combination $\cos \theta\ v^\pm_\mu - s^\pm_\mu$ decouples from the other states and has a mass equal to $M_V$. A similar situation is realised in the neutral sector where, however, residual mixings suppressed by $1/\gt$ are present.

Approximate expressions for the masses of the charged states are given below, including leading corrections in $1/\gt^2$: 
\begin{eqnarray}
 M_W^2&=& \frac{1}{4}g^2v^2 \left[1-\frac{1}{2} \left(\frac{g}{\gt}\right)^2 \left(\left(r^2-1\right) s_\theta^2+2\right)+{\cal O}\left(1/\gt^4\right) \right]\,,\\
M_{A^+}^2&=& M_A^2 \left[1+\frac{1}{2} \left(\frac{g}{\gt}\right)^2 r^2 s_\theta^2 +{\cal O}\left(1/\gt^4\right) \right]\,, \\
M_{S^+}^2&=& M_V^2\,, \\
M_{V^+}^2&=&M_V^2 \left[1+ \frac{1}{2} \left(\frac{g}{\gt}\right)^2  (2-s_\theta^2) +{\cal O}\left(1/\gt^4\right) \right]\,.
\label{eq:massesCharged}
\end{eqnarray}
Similarly, in the neutral sector, we find:
\begin{eqnarray}
M_Z^2&=&\frac{1}{4} (g^2+g'^2) v^2
      \left[1+ \frac{(g^2+g'^2)^2 (1-r^2) s_\theta^2 - 2(g^4+g'^4)}{2(g^2+g'^2)\gt^2}  +{\cal O}\left(1/\gt^4\right) \right]\,, \\
M_{A^0}^2&=& M_A^2 \left[1+\frac{r^2 (g^2+g'^2) s_{\theta }^2}{2 \gt^2}+{\cal O}\left(1/\gt^4\right)\right]\,, \\
M^2_{V^0/S^0}&=& M_V^2 \left[1+ \frac{g^2+g'^2}{4\gt^2} \left( 1+c_\theta^2\pm\sqrt{1+2 {\scriptstyle\frac{(g'^4-6g'^2g^2+g^4)}{(g^2+g'^2)^2}}c_\theta^2+c_\theta^4}  \right)
          +{\cal O}\left((1/\gt)^4\right) \right]\,.
\label{eq:massesNeutral}          
\end{eqnarray}
In all above expressions, $s_\theta = \sin \theta$ and $c_\theta = \cos \theta $. Furthermore, $v = 246$~GeV is the value of the effective EW scale, obtained from the definition of the Fermi decay constant as:
\be \label{eq:v1}
v^2\equiv \frac{1}{\sqrt{2}G_F}=\frac{-4}{g^2}\Pi_{W^+W^-}(0)= \frac{4}{g^2} \frac{1}{\left[ {{\cal M}_{\rm C}^2}^{-1} \right]^{11} } = 
   2 (M_{V}^2 \frac{f_0^2}{f_K^2} -M_{A}^2 r^2) s_\theta^2 /\tilde{g}^2\,.
\ee
Replacing the masses with the Lagrangian parameters, we also obtain the relation
\beq \label{eq:v2}
v^2 = \left( f_0^2 - r^2 f_1^2 \right) \sin^2 \theta = f_\pi^2 \sin^2 \theta
\eeq
where $f_\pi = \sqrt{f_0^2 - r^2 f_1^2}$ is the decay constant of the SU(4)/Sp(4) pions, as in \eq{eq:fpi}. As a consistency check, note that $f_\pi = f_0$ for $r=0$ and, as mentioned in the previous section, $f_\pi = 0$ for $r = f_0/f_1$.

\subsection{Couplings} 

We assume here that the SM fermions only couple to the SM weak bosons, $\widetilde{W}_\mu$ and $B_\mu$: this is a reasonable assumption, as direct couplings to the composite resonances can only be induced by interactions external to the dynamics. The interaction with the heavy vectors, therefore, are generated via mixing terms. 
For the charged currents, we have
\beq \label{eq:LCC}
{\cal L}_{\rm CC} = \frac{g}{\sqrt{2}}\ \sum_{i,f} \ {\cal C}_{1i}\ \bar{\psi}_f \gamma^\mu R_{i,\mu}^+ \psi_{f'} + h.c.\,,
\eeq
where $R^\pm_{i,\mu} = (W^\pm_\mu, A_\mu^\pm, V_\mu^\pm, S_\mu^\pm)$ and $f$ labels all the SM fermions.
For the neutral currents
\begin{eqnarray} \label{eq:LNC}
{\cal L}_{\rm NC}= \frac{1}{2}\
\sum_{i,f}\ R_{i,\mu}^0\ \bar{\psi}_f \gamma^{\mu} [ ( g_{L i}^f P_L +g_{R i}^f P_R)] \psi_f\,,
\end{eqnarray}
where $R^0_{i, \mu} = (A_\mu, Z_\mu, A_\mu^0, V_\mu^0, S^0_\mu)$, $f$ is a SM fermion, $P_{L,R}=(1\mp \gamma_5)/\sqrt{2}$,  and
\begin{equation}
g_{L j}=g T^3 {\cal N}_{2j} + g' Y_L {\cal N}_{1j}, \quad 
g_{R j}= g' Y_R {\cal N}_{1j}, \quad 
\end{equation}
with $T^3$ being the weak isospin and $Y_{L,R}$ the hypercharge of the left-handed doublet and the right handed singlet respectively. 
All the neutral vector couplings can be expressed like this, but for the photon gauge invariance requires that
\beq
g' {\cal N}_{11} = g {\cal N}_{21} = e\,.
\eeq
Note that \eq{eq:LCC} and \eq{eq:LNC} encode corrections to the couplings of SM fermions with respect to the SM predictions, that are strongly constrained by EW precision observables and can be encoded in the oblique parameters, as discussed in the following Section.

The Higgs couplings to weak bosons are phenomenologically important because they can constrain the model parameters, both from direct measurements and from its contribution to the electroweak parameters.
Schematically, the couplings can be written as
\beq \mathcal{L}_{h} &=& c_{hR_i^+R_j^-} \, h\ R_{\mu,i} ^+ R^{-,\mu}_j
 + \frac{1}{2}c_{hR_i^0 R_j^0} \, h \ R^0_{\mu,i}  R^{0,\mu}_j \,,
 \eeq
where $R^\pm_i$ and $R^0_i$ encode all the charged and neutral vectors.
In the gauge interaction basis, the couplings are provided in Appendix~\ref{app:formulas}, while in the mass eigenbasis we calculated expressions at leading order in $1/\gt$.
We find that  the Higgs couplings to at least one photon are automatically zero at the tree level, with the other couplings given by 
\beq \label{eq:chWW}
c_{hW^+W^-} \simeq   \frac{2M_W^2}{v}c_\theta = c_{hW^+W^-} ^{\rm SM} c_\theta\,, \quad
c_{hZZ} \simeq \frac{2M_Z^2}{v}c_\theta = c_{hZZ}^{\rm SM} c_\theta\,,
\eeq
in agreement with previous studies~\cite{Cacciapaglia:2014uja}, while the couplings to the resonances are (we list only the diagonal ones and the ones with one SM gauge boson)
\beq
 & c_{hA^+A^-} \simeq   \frac{ g^2  M_A^2 r^2 s^2_\theta }{\gt^2  v }\,, \quad c_{hV^+V^-} \simeq  -\frac{ g^2  M_V^2  s^2_\theta }{ \gt^2  v }\,, & \nonumber \\
 & c_{hW^+S^-}  \simeq \frac{ g  M_V^2 (r^2-1)  s^2_\theta  }{2 \gt v }\,, \quad   c_{hW^+V^-} \simeq \frac{ g  M_V^2 (r^2-1)  s^2_\theta  }{2 \gt v }\,; \label{eq:hVVcoup}
\eeq
\beq
& c_{hA^0A^0} \simeq  \frac{ ({g'}^2+ g^2)  M_A^2 r^2 s^2_\theta   }{\gt^2 v }\,, \quad c_{hV^0V^0} \simeq  -\frac{ g^2  M_V^2  s^2_\theta }{\gt^2  v }\,, \quad c_{hS^0S^0} \simeq  -\frac{ {g'}^2  M_V^2  s^2_\theta }{\gt^2  v } & \nonumber \\
& c_{hZS^0} \simeq \frac{ \sqrt{ {g'}^2 + g^2}  M_V^2 (r^2-1)  s^2_\theta  }{2 \gt v }\,, \quad  c_{hZV^0} \simeq \frac{ \sqrt{ {g'}^2 + g^2}  M_V^2 (r^2-1)  s^2_\theta  }{2 \gt v }\,. \label{eq:hVVcoup0}
\eeq

The charged heavy vector states contribute to the decay of the Higgs boson into two photons via loops. In general, computing loops of heavy resonances is not reliable, nevertheless  
we can approximate the contribution of the strong dynamics to $h \to \gamma \gamma$ by computing loops of the lightest spin-1 resonances.
While this is not a complete calculation, it can provide an estimate of the additional contributions and allow us to test their impact.
The partial width, including new physics effects, can be written as, 
\begin{eqnarray}
	\label{eq:gammahaa}
	\Gamma_{h\rightarrow\gamma\gamma}&=&
	\frac{\alpha^2 m_h^3}{256\pi^3v^2}|N_c Q_{top}^2 \kappa_t A_f(\tau_t)+\kappa_W A_V(\tau_W)  + \kappa_{\rm res} A_V (\infty)|^2,
    \end{eqnarray}
where we approximate the amplitude of the heavy states to the asymptotic value $A_V(\infty) = -7$ ($A_f(\tau_f)$ and $A_V (\tau_W)$ being the standard amplitudes~\cite{Spira:1995rr}), and $\kappa_{t,W}$ are the modification of the fermion and vector couplings of the Higgs normalised by the SM expectation (in our case, $\kappa_{W} \sim \cos \theta$, while $\kappa_t$ depends on the mechanism providing a mass for the top and is equal to $\kappa_t = \cos \theta$ in the simplest case):
\beq
\kappa_W = \frac{c_{hW^+W^-}}{2 M_W^2} v\,, \qquad \kappa_{\rm res} = \sum_{i=2,3,4} \frac{c_{h R_i^+ R_i^-}}{2 M_{R_i}^2} v\,,
\eeq
where both couplings and masses are defined in the mass eigenstate basis.
By analysing the mass and coupling matrices, we found that the following sum rule holds, at all orders in $1/\gt$:
\beq
\kappa_W + \kappa_{\rm res} = \frac{v}{2} \mbox{Tr} \left[ c_{hV^+V^-} \cdot \left({\cal M}_C^2\right)^{-1} \right] = \cos \theta\,,
\eeq
where $c_{hV^+V^-}$ are the couplings of the Higgs in the interaction basis (see \eq{eq:chVVpm}) and we used exact matrices.
At leading order in $1/\gt$, the sum rule is saturated by the $W$ coupling $\kappa_W = \cos \theta$, as shown in \eq{eq:chWW}.
However, corrections arise at order $1/\gt^2$: using \eq{eq:hVVcoup}, we find
\beq \label{eq:kappares}
\kappa_{\rm res} = \cos \theta - \kappa_W \simeq \frac{g^2}{\gt^2} \frac{1}{2} (r^2-1) \sin^2 \theta\,.
\eeq
We expect, therefore, the contribution of the mixing with the heavy resonances to be very small, as it is suppressed by $\sin^2 \theta$, $1/\gt^2$, and it also vanishes for $r = 1$: the latter is a reminder of the fact that the two SU(4)'s decouple in this limit.  
This analysis shows that the effect of the heavy resonances on the Higgs properties can be neglected, thus the bounds from the measured Higgs couplings are the same as in~\cite{Arbey:2015exa}, and they are typically less constraining than electroweak precision tests.

For completeness, a similar analysis can be done in the neutral sector, where we define
\beq
\kappa_Z = \frac{c_{hZZ}}{2 M_Z^2} v\,, \quad \kappa_{0,{\rm res}} = \sum_{i=3,4,5} \frac{c_{hR^0_i R^0_i}}{2 M_{R^0_i}^2} v\,,
\eeq
and the exact mass and coupling matrices entail the following sum rule:
\beq
\kappa_Z + \kappa_{0,{\rm res}} = \frac{v}{2} \mbox{Tr} \left[ \bar{c}_{hV^0V^0} \cdot \left(\bar{\cal M}_N^2\right)^{-1} \right] = \cos \theta\,,
\eeq
where the reduced coupling, $\bar{c}_{hV^0V^0}$, and mass matrices, $\bar{\cal M}_N^2$, are 4$\times$4 matrices obtained from the complete ones in \eq{eq:chVV0} and \eq{eq:massN} by removing the photon, i.e. the zero-mass eigenstate. Like for the charged case, $\kappa_Z$ is the deviation of the Higgs couplings to the $Z$ boson normalised by the SM value, while $\kappa_{0,{\rm res}}$ encodes the contribution of the heavy resonances:
\beq
\kappa_{0,{\rm res}} = \cos \theta - \kappa_Z = \sum_{i=3,4,5} c_{hR^0_iR^0_i} \frac{v}{2 M^2_{R^0_i}} \simeq \frac{g^2 + {g'}^2}{\gt^2} \frac{1}{2} (r^2-1) \sin^2 \theta\,.
\eeq
We can see that the custodial violation due to the gauging of the hypercharge emerges here.

The off-diagonal Higgs and gauge couplings  will contribute to the $H$-$Z$-$\gamma$ vertex~\cite{Cai:2013kpa}, but no bound  is available  at current  LHC precision~\cite{Aad:2015pla}. Other relevant interactions involve the $\eta$ state, which couples to the ``tilded'' vectors. The production of heavy vector states can go through a cascade of decays with rich phenomenology. The interaction Lagrangian involving $\eta$ and vector fields is given, at leading order in $g/\gt$ and $\sin \theta$, by
\begin{eqnarray}
{\cal L}_{\eta,C}&= & \frac{g  \left(r^2-1\right) s_\theta^2 M_V^2}{\sqrt{2}\gt v} \, \eta\, \widetilde{S}^+_\mu  W ^{-,\mu}  + \frac{ 
   \left(M_A^2-M_V^2\right) r s_\theta}{v} \, \eta\, \widetilde{S}^+_\mu  A ^{-,\mu}  \nonumber \\ &-& \frac{g^2 M_V^2 s_\theta^2}{\sqrt{2} \gt^2 v} \, \eta\, \widetilde{S}^+_\mu  V^{-,\mu}  + h.c  
\end{eqnarray}

\beq
{\cal L}_{\eta,N}&= &  \frac{ \sqrt{{g'}^2+g^2}  M_V^2 \left(r^2-1\right) s_\theta^2}{\sqrt{2}\gt v}  \, \eta\,\widetilde{S}^0_\mu  Z^{\mu} +  \frac{ 
   \left(M_A^2-M_V^2\right) r s_\theta}{v}  \eta \widetilde{S}^0_\mu A^{0, \mu}   \nonumber \\
&-& \frac{g^2  M_V^2 s_\theta^2 }{\sqrt{2} \gt^2 v} \, \eta\, \widetilde{S}^0_\mu  V ^{0,\mu}  -  \frac{{g'}^2   M_V^2  s_\theta^2 }{\sqrt{2} \gt^2 v}  \, \eta\, \widetilde{S}^0_\mu  S ^{0,\mu}  
+
    \frac{ 
   \left(M_A^2-M_V^2\right) r s_\theta}{\sqrt{2}
   v}  \eta \widetilde{V}^0_\mu X^{0, \mu}  
\eeq

An interesting collider signature would be the production of $A_\mu$ with subsequent decay into $\widetilde{S}+\eta$, then $\widetilde{S}\to\eta+Z$ and the two $\eta$ resonances decay for instance into top pairs.

\subsection{Electroweak Precision Tests}

The precise measurements near the $Z$-pole performed at several high energy experiments, especially at LEP~\cite{LEP-2}, are crucial tests for any kind of model of New Physics. These effects can be parameterised via the so--called oblique parameters~\cite{Peskin:1991sw,Barbieri:2004qk}, expressed explicitly in terms of the weak boson self energies in \eq{eq:ewposelfen}. 

At tree level, the vector contribution to the oblique parameters are given by 
\begin{eqnarray}
\hat{S} &=& -\frac{g^2 \left(r^2-1\right) s_{\theta }^2}{2\gt^2+g^2 \left[2+(r^2-1) s_{\theta }^2\right]}
      \ , \\ 
W &=& \frac{g^2 M_W^2 \left[s_{\theta }^2 \left(r^2 M_V^2-M_A^2\right)+2 M_A^2\right]}
{M_A^2 M_V^2 \left\{g^2 \left[\left(r^2-1\right) s_{\theta }^2+2\right]+2 \gt^2\right\}}\,, \\
Y &=& \frac{g'^2 M_W^2 \left[s_{\theta }^2 \left(r^2 M_V^2-M_A^2\right)+2 M_A^2\right]}
{M_A^2 M_V^2 \left\{2 \gt^2+g'^2 \left[\left(r^2-1\right) s_{\theta }^2+2\right]\right\}}\,, \\
X &=& \frac{g g' s_{\theta}^2 M_W^2 \left(M_A^2-r^2 M_V^2\right)}
{M_A^2 M_V^2 \sqrt{\left\{g^2 \left[\left(r^2-1\right) s_{\theta }^2+2\right]+2 \gt^2\right\} \left\{2 \gt^2+g'^2 \left[\left(r^2-1\right) s_{\theta }^2+2\right]\right\}}} \,,
\end{eqnarray}
where the other EW observables vanish, $\hat{T} =0 $, $\hat{U} =0$. For $\theta=\pi/2$ these expressions agree with~\cite{Foadi:2007ue} once one identifies $1-\chi=r$ and sets the hyper-charge $y=0$.  In our analysis, we are going to  use the notation adopted by the Particle Data Group (PDG) and rescale  $S = 4 s_W^2 \hat {S} / \alpha_{EW}$,  $T = \hat{T} / \alpha_{EW}$ and $U = - 4 s_W^2 \hat{U}/ \alpha_{EW}$. 
Note that 
the above contributions can replace the contribution of the strong dynamics, estimated in~\cite{Peskin:1990zt,Arbey:2015exa} as a loop of the underlying fermions. For $r\sim 1$ the $S$ parameter vanishes and higher order parameters, $W$,$Y$ and $X$ will play the dominant role. This situation is similar to the Custodial Vector Model described in~\cite{Casalbuoni:1992dd,Becciolini:2014eba}.

\begin{figure}
\includegraphics[width=0.45\textwidth]{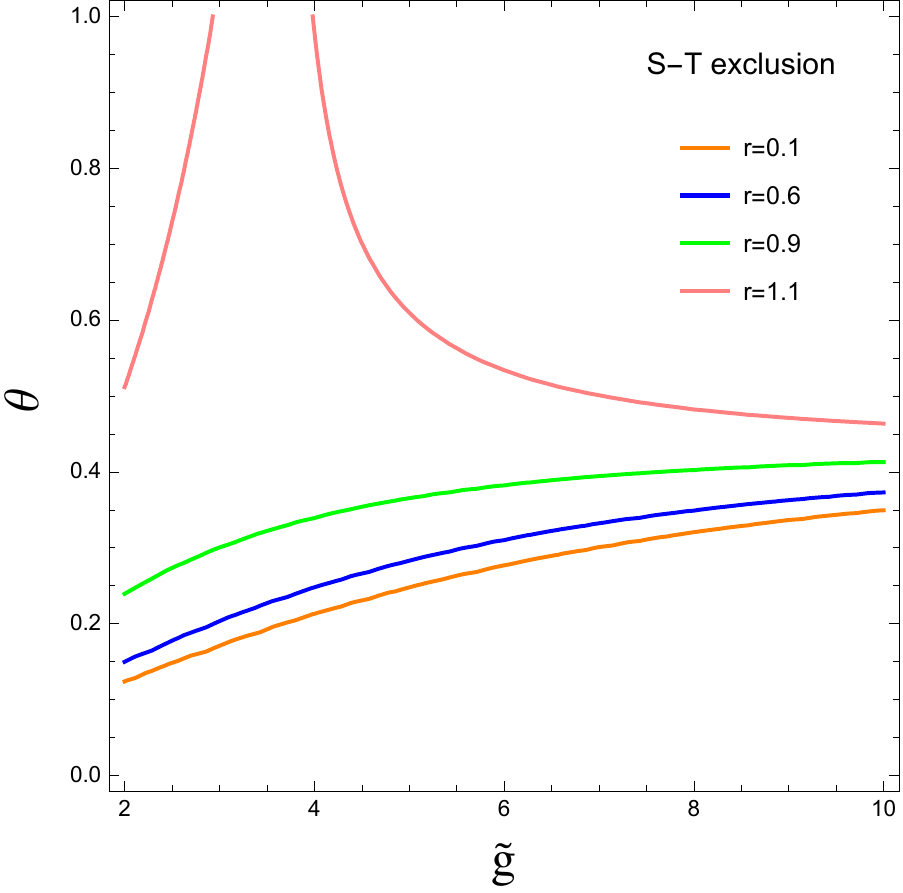}
\caption{The $S$-$T$ bound on the parameter space  at  $99 \%$ confidence level in the plane of $(\theta-\gt)$ for $r = 0.1, 0.6, 0.9, 1.1$,  with the cut off scale  $\Lambda = \sqrt{2} \pi v/\sin \theta$.  The region above the curves is excluded by EWPT. }
\label{fig:ewpt}
\end{figure}

Additionally, deviations in the Higgs coupling w.r.t. the SM ones also bring additional contributions to the $S$ and $T$ parameters. The modification in the Higgs coupling, \eq{eq:hVVcoup}, produce approximately the following deviations in the $S$ and $T$ parameters.
\begin{eqnarray}
\Delta S &=& \frac{1}{6 \pi} \left[ (1-\kappa_V^2) \log \left( \frac{\Lambda }{m_h} \right) + \log \left( \frac{m_h}{m_{h,ref}}\right) \right] \,, 
\\
\Delta T &=& - \frac{3}{8 \pi \cos^2 \theta_W} \left[ (1-\kappa_V^2)  \ln \frac{\Lambda}{m_h} + \log \left( \frac{m_h}{m_{h,ref}}\right)\right]\,.
\end{eqnarray} 
In the above formulas, the couplings of the SM gauge bosons to the Higgs, $\kappa_V$, include corrections up to order $1/\gt^2$ from \eq{eq:kappares} in order to be consistent with the tree-level effects (for simplicity, we neglect the term in ${g'}^2$ so that $\kappa_V \sim \kappa_W \sim \kappa_Z$).
In principle, the heavy resonances also contribute at one-loop level: naively, the loops with a Higgs boson have an additional suppression $1/\gt^2 \cdot m_W^2/M_V^2 \sim 1/\gt^4$. Pure loops of the heavy resonances may be unsuppressed, however their effect should be small as the dynamics is custodial invariant. Furthermore, due to the intrinsic strong interactions among resonances, such loop calculations are not reliable in general because perturbative expansions cannot be trusted. 
In this paper, therefore, we follow the philosophy of Vector Meson Dominance (which is experimentally tested in QCD) and assume that the tree level exchange is the dominant contribution, while effects due to loops or higher order operators can be neglected.
Alternative proposals have been put forward in the literature in order to render the theory more calculable, see for instance~\cite{Contino:2011np}. The main idea is to assume that the lowest lying resonances are weakly coupled to themselves and the rest of the dynamics, so that loops can be reliably calculated, together with the assumption of negligible higher order operators. As a result, potential cancellations have been observed that can relax the constraints from EW precision tests~\cite{Contino:2015mha,Ghosh:2015wiz}, including loop contributions from light top partners: these results, however, are not generic.
Loop corrections can be considered a modelling of the contribution of the strong dynamics (see also~\cite{Pich:2013fea,Contino:2015gdp}).
In principle, improved calculability in  Ref.~\cite{Contino:2011np} can be imposed on our model, however we prefer to stay with a more conservative  bound,  as an order of magnitude estimate of the resonance effects.


The experimental values from PDG for $S$ and $T$ (leaving $U$ to be free),  with a  strong  correlation coefficient  $0.90$, at $ 1 \sigma $  deviation are~\cite{Agashe:2014kda}:
\beq
S = -0.03 \pm 0.10  \, ,  \quad   T = 0.01 \pm 0.12\,.
\eeq
The corresponding limits on the model parameters are shown in \fig{fig:ewpt}.
For $r \gtrsim 1$ the vector partially cancels the Higgs contribution allowing a larger parameter space: in some areas of the parameter space, therefore, the most constraining bound comes form the measurements of the Higgs couplings, which give constraints on $\theta$ of the order of $\theta \lesssim 0.6$~\cite{Arbey:2015exa}.
Another effect that may significantly modify the EWPT is the presence of the $\sigma$ state, that will in general mix with $h$, with an un--calculable mixing angle $\alpha$, and can potentially alleviate the constraints from EWPT~\cite{Arbey:2015exa}.

\section{A minimal fundamental gauge theory} \label{sec:UV}

The effective model characterised in the previous sections, can originate from a very simple scalar--less underlying theory~\cite{Ryttov:2008xe,Galloway:2010bp}: it consists of a gauged and confining ${\cal G}_{\rm HC} = \mbox{SU(2)}$ with two light Dirac flavours transforming as the fundamental representation.
Following the notation of~\cite{Ryttov:2008xe,Cacciapaglia:2014uja}, the 2 Dirac fermions,  $U$ and $D$,  can be  arranged in a flavour SU(4) multiplet as
\begin{equation}
	Q_{\alpha}^{i , a} = \left(\begin{array}{c}
		U_L \\ D_L \\ \widetilde{U_L} \\ \widetilde{D_L}
	\end{array}\right),
	\label{eq:Q}
\end{equation}
where $\alpha$ is the spin Lorentz index, $i$ is a flavour index and $a$ is a hyper--colour index.  The tilded fields are left-handed spinors containing the right-handed components of the Dirac fields,  i.e. $\tilde{U}_L = - i \sigma^2 U_R^*$ and $\tilde{D}_L = - i \sigma^2 D_R^*$.

Following the embedding of the EW symmetry we chose in this work, the pair $(U_L, D_L)$ transforms as a doublet of the weak isospin SU(2)$_L$, while the other two $(\tilde{U}_L, \tilde{D}_L)$ as an anti-doublet of the custodial SU(2)$_R$.

\subsection{Scalar sector}

The scalar sector of the $SU(4)/Sp(4)$ models was studied in \cite{Cacciapaglia:2014uja}. 
In general we can write a scalar matrix
\begin{equation}
M_{ij}= Q^{\alpha A}_{i} Q_{\alpha A j} = Q_{\beta B i} Q_{\alpha A j} I^{AB} \epsilon^{\alpha \beta}
\end{equation}
where greek letters are Lorentz indices, capital letter are hyper--colour gauge indices and lower case latin are flavour indices. The gauge group invariant $I _{AB}$ depends on the gauge group and fermion representation.
If the gauge representation is pseudo-real, like in our case, $I _{AB}$ is antisymmetric (for fundamentals of SU(2)$_{\rm HC}$, $I^{AB}=\epsilon^{AB}$). 
Accordingly, $M$ is flavour anti-symmetric, and it transforms as a {\bf 6}$_{\rm SU(4)}$. In general, this matrix contains both the light pNGBs and heavier scalar resonances.

\subsection{Vector sector}

The composite spin-1 states can be defined in terms of the underlying fermions via the flavour adjoint left-current:
\bea
({\cal F}^\mu)_{i}^j &\sim& 
 \left( Q_{i  }^{ \alpha} \sigma^\mu_{\alpha \dot{\beta}} Q^{\dagger j \dot{\beta}} - \frac{1}{4} \delta_{i}^j \, 
			Q_k^{\alpha} \sigma^\mu_{\alpha \dot{\beta}} Q^{ \dagger k \dot{\beta}} \right) \\
       &=& \left( Q_{i  }^{ \alpha a} \sigma^\mu_{\alpha \dot{\beta}} Q_j^{b \dagger \dot{\beta}} - \frac{1}{4} \delta_{ij} \, 
		 Q_k^{a\alpha} \sigma^\mu_{\alpha \dot{\beta}} Q_k^{b \dagger \dot{\beta}} \right) \E_{ab} \\
       &=& - \left( Q_{j  \dot{\alpha}}^{\dagger } \overline{\sigma}^\mu  Q_{\beta i} - \frac{1}{4} \delta_{ij} \, 
			Q_{k \dot{\alpha}}^{\dagger}  \overline{\sigma}^\mu Q_{k \beta} \right) \E_{ab}	
\eea
where $\E_{ab}$ is the antisymmetric tensor making a hyper-colour singlet. 
Note the first line is non-standard notation for the left bilinears, but the one that directly implements the the flavour transformation structure ${\cal F}\to g\cdot {\cal F}\cdot g^\dagger$. The last line is the standard bilinear notation.
After some current algebra, the components in the vector matrix from \tab{tab:class} can be associated to currents in terms of the underlying quarks, as detailed in \tab{tab:currents}, where the  notation is used: $\Re (J^\mu)= \frac{1}{2}(J^\mu + J^{\mu \dag})$ and $\Im (J^\mu)= - \frac{i}{2}(J^\mu - J^{\mu \dag})$.
\begin{table}[tb]
\begin{center}
\begin{tabular}{|c|c||c|c|c||c|}
	\hline
	Field & Fermion currents & \texttt{P} & \texttt{C} & \texttt{G} & \texttt{GP}\\
         \hline \hline
	\multicolumn{6}{|l|}{Massive spin-1 ${\cal V}_\mu$ (unbroken generators)} \\
	\hline
	$v^+$ & {\small $\overline{D} \gamma^\mu  U$} &&&&\\
	\cline{1-2}\cline{4-4}
	$v^0$ & {\small $\frac{1}{\sqrt{2}} \left( \overline{U} \gamma^\mu U - \overline{D} \gamma^\mu D \right)$} & $-$ & $-$ & $-$ & $+$  \\
	\cline{1-2}\cline{4-4}
	$v^-$ & {\small $\overline{U} \gamma^\mu  D$}
		&&&& \\
	\hline
	$\tilde{v}^0$ & {\small $\sqrt{2} \cos\theta \, \Im \left( U^T C \gamma^\mu D \right)
		+ \frac{1}{\sqrt{2}} \sin\theta \, \left( \overline{U} \gamma^\mu  U
		+ \overline{D} \gamma^\mu D \right)$}
		& $-$ & $-$ & $+$ & $-$ \\
	\hline
	$s^+$ & {\small $\cos\theta \, \left( \overline{D} \gamma^\mu \gamma^5 U \right)
		+ \frac{i}{2} \sin\theta \, \left( U^T C \gamma^\mu \gamma^5 U
		- \overline{D} \gamma^\mu C \gamma^5 \overline{D}^T \right)$} &&&& \\
	\cline{1-2}\cline{4-4}
	$s^0$ & {\small $-\frac{1}{\sqrt{2}} \cos\theta \, \left( \overline{U} \gamma^\mu \gamma^5 U
		- \overline{D} \gamma^\mu \gamma^5 D \right)
		- \sqrt{2} \sin\theta \, \Im \left( U^T C \gamma^\mu \gamma^5 D  \right)$}
		& $+$ & $+$ & $+$ & $+$ \\
	\cline{1-2}\cline{4-4}
	$s^-$ & {\small $\cos\theta \, \left( \overline{U}  \gamma^\mu \gamma^5 D \right)
		+ \frac{i}{2} \sin\theta \, \left( \overline{U} \gamma^\mu C \gamma^5 \overline{U}^T
		- D^T C \gamma^\mu \gamma^5 D  \right)$} &&&& \\
	\hline
	$\tilde{s}^+$ & {\small $\frac{i}{2} \left( U^T C \gamma^\mu\gamma^5 U
		+ \overline{D} \gamma^\mu C  \gamma^5 \overline{D}^T \right)$}
		&&&& \\
	\cline{1-2}\cline{4-4}
	$\tilde{s}^0$ & {\small $\sqrt{2} \, \Re \left( U^T C \gamma^\mu \gamma^5 D  \right)$}
		& $+$ & $-$ & $-$ & $-$ \\
	\cline{1-2}\cline{4-4}
	$\tilde{s}^-$ & {\small $\frac{i}{2} \left( \overline{U} \gamma^\mu  C \gamma^5 \overline{U}^T
		+ D^T C \gamma^\mu \gamma^5 D \right)$} &&&&\\
	\hline
	\hline
	\multicolumn{6}{|l|}{Massive spin-1 ${\cal A}_\mu$ (broken generators)} \\
	\hline
	$a^+$ & {\small $\frac{i}{2} \cos\theta \, \left( U^T C \gamma^\mu  \gamma^5U
		- \overline{D} \gamma^\mu C \gamma^5 \overline{D}^T \right)
		-\sin\theta \, \left( \overline{D} \gamma^\mu \gamma^5  U \right)$} &&&& \\
	\cline{1-2}\cline{4-4}
	$a^0$ & {\small $-\sqrt{2} \cos\theta \, \Im \left( U^T C \gamma^\mu  \gamma^5 D \right)
		+ \frac{1}{\sqrt{2}} \sin\theta \, \left( \overline{U} \gamma^\mu \gamma^5 U - \overline{D} \gamma^\mu  \gamma^5D \right)$}
		& $+$ & $+$ & $+$ & $+$ \\
	\cline{1-2}\cline{4-4}
	$a^-$ & {\small $\frac{i}{2} \cos\theta \, \left( \overline{U} \gamma^\mu C  \gamma^5 \overline{U}^T
         	- D^T C \gamma^\mu \gamma^5  D \right)
		-\sin\theta \, \left( \overline{U}  \gamma^\mu  \gamma^5 D \right)$} &&&& \\
	\hline
	$x^0$ & {\small $\sqrt{2} \, \Re \left( U^T C \gamma^\mu  D \right)$}
		& $-$ & $+$ & $-$ & $+$ \\
	\hline
	$\tilde{x}^0$ & {\small $\frac{1}{\sqrt{2}} \cos\theta  \, \left( \overline{U} \gamma^\mu  U
		+ \overline{D} \gamma^\mu  D \right)
		- \sqrt{2} \sin\theta \, \Im \left( U^T C \gamma^\mu  D \right)$}
		& $-$ & $-$ & $+$ & $-$  \\
	\hline
\end{tabular}
\end{center}
\caption{Classification of composite vectors in terms of the underlying fermionic currents,  using  a notation  $C= \gamma^0 \gamma^2$.   We quote transformation properties in terms  of   spacial parity \texttt{P},  charge conjugation \texttt{C},  and   pion parity \texttt{G} defined from listed currents. The combination \texttt{GP} is a good symmetry from the  strong dynamics. In our notation,   P-parity is defined in spacial direction, i.e. $(-1)^\mu$ is ``$-$" parity. } \label{tab:currents}
\end{table}

\begin{table}[tb]
\begin{center}
\begin{tabular}{|c|c||c|c|c||c|}
	\hline
	Field & Fermion currents & $P$ & $C$ & $G$ & $GP$\\
         \hline \hline
         \multicolumn{6}{|l|}{Scalar pNGBs} \\
         \hline
         $h$ & $\frac{1}{2} \cos\theta \, \left(\overline{U} U + \overline{D} D \right)
		+ \sin\theta \, \Im\left(U^T C D \right)$
		& $+$ & $+$ & $+$ & $+$ \\
	\hline
	$\eta$ & $\Re\left(U^T C D\right)$
		& $+$ & $-$ & $-$ & $-$ \\
	\hline
\end{tabular}
\end{center}
\caption{Classification of the pNGB states in terms of the underlying fermionic currents, and their parities. } \label{tab:currents0}
\end{table}

We first notice that this decomposition matches with the interpretation we provided at the beginning of the previous section: the triplet $\overrightarrow{v}_\mu$ corresponds to the ``vector'' current $\bar{Q} \gamma_\mu Q$, typically associated to the $\rho_\mu$ meson in QCD, while $\overrightarrow{s}_\mu$ and $\overrightarrow{a}_\mu$ contain an ``axial'' current component, associated with $a_\mu$ meson in QCD, proportional to the $\cos \theta$ and $\sin \theta$ respectively.  More precisely, due to the symmetry relating the Technicolor limit to the composite Higgs limit,  the $\cos \theta$ component of $s_\mu^{\pm, 0}$ and $\tilde{x}_\mu^0$ can be exactly  mapped from  the $\sin \theta $ component of $a_\mu^{\pm,0}$ and $\tilde{v}_\mu^0$. 

\subsection{Discrete symmetries}

The action of space-time discrete symmetries on the composite states can be derived from the transformation properties of the underlying quarks.
However, the gauging of the EW interactions break $P$ and $C$ individually, but preserves $CP$ in the strong confining sector. Under $CP$, the bound state fields transform as
\beq
M  \xrightarrow{CP} M^\dagger\,, \qquad {\cal F}_\mu \xrightarrow{CP}  - (-1)^\mu ({\cal F}_\mu)^T\,,
\eeq
where $(-1)^\mu = 1$ for $\mu = 0$, and $-1$ on spacial directions. The parities associated with the spin--1 resonances are summarised in \tab{tab:currents}: in our notation, a vector has $CP = +$, while $CP = -$ for a pseudo-vector. In the scalar sector, as expected, the Higgs $h$ is defined as a scalar, while $\eta$ transforms as a pseudo-scalar, see \tab{tab:currents0}.

However, $CP$ is not a convenient symmetry to label states as it maps charged states in their complex conjugate (particles into anti-particles). In terms of composite states, it is thus convenient to define a new parity $G$,  defined as $C$ plus an internal rotation in the flavour symmetry, which corresponds to an SU(2) rotation in our case:
\beq 
U \xrightarrow{G} - \gamma^2 D^\ast\,, \qquad D  \xrightarrow{G}  \gamma^2 U^\ast\,.
\label{G-parity}
\eeq
with  its action on fermion current  illustrated in appendix \ref{sec:parity}. Once combined with $P$, the new symmetry defines a parity acting as:
\beq
M  \xrightarrow{GP} \Omega_{GP} \cdot M^\dagger \cdot \Omega_{GP}^T\,, \quad {\cal F}_\mu \xrightarrow{GP}  (-1)^\mu \Omega_{GP} \cdot ({\cal F}_\mu)^T \cdot \Omega_{GP}\,, \quad \Omega_{GP} = \left( \begin{array}{cc} \sigma^2 & 0 \\ 0 & \sigma^2 \end{array} \right)\,.
\eeq
From \tab{tab:currents} and \tab{tab:currents0} we see that all the tilded fields are odd under $GP$, as well as $\eta$, thus this is the symmetry preventing decays of such field directly into SM ones. Note, however, that $GP$ is violated by the anomalous WZW term which generates decays for $\eta$. Additional decay channels will also be generated for the vectors.

\section{Phenomenology at the LHC and future 100 TeV colliders}
\label{sec:pheno}

\subsection{Model implementation}
\label{sec:tools}

To study the phenomenology of this model, we implemented the Lagrangian in {\sc MadGraph}~\cite{Alwall:2014hca}, using the {\sc Mathematica} package {\sc FeynRules}~\cite{Alloul:2013bka}. 
The implementation of the {\sc FeynRules} model file is sketched in this section, while the model files are publicly available on the {\sc HEPMDB} website~\footnote{\href{http://hepmdb.soton.ac.uk/hepmdb:0416.0200}{http://hepmdb.soton.ac.uk/hepmdb:0416.0200}}.

The neutral resonances $A$, $Z$, $A^0$, $V^0$, $S^0$, $\tilde{S}^0$ and $\tilde{V}^0$  are introduced  as one  particle class of \texttt{VN}, with two additional neutral states $X^0$, $\tilde{X}^0$ into another particle class of  \texttt{VX}.  The charged resonances $W^+$, $A^+$, $V^+$, $S^+$ and $\tilde{S}^+$  are put into  one particle class of \texttt{VC}. The effective Lagrangian for the strong sector is written in terms of  physical pions, thus in the Unitary gauge, and  vector bosons in gauge basis, with the latter rotated to their mass eigenstates via mixing matrices. The quarks and leptons only couple to $\widetilde{W}$ and $B$, thus the Yukawa structure is exactly the same as in the  Standard Model.  In the model implementation, the rotation matrices  \texttt{CM}$_{4\times4}$ and  \texttt{NM}$_{5\times5}$  are provided  in  two independent Les Houches blocks of \texttt{VCMix} and \texttt{VNMix} as external parameters,  whose numerical values are calculated by a specific Fortran routine.  Note that  for the rotation,  the eigenstates are ordered  such that  the diagonal element in \texttt{CM}$_{4\times4}$ or  \texttt{NM}$_{5\times5}$ are maximal in each corresponding column, to make sure  each one carries the largest component of the original gauge state as described in \eq{eq:mixN}.

Five model parameters $M_V$, $M_A$, $r$, $\theta$ and $\gt$ are introduced into the Les Houches block  \texttt{DEWSB}, with all associated  decay constants $f_1= \sqrt{2} M_A /\gt$, $f_K = \sqrt{2} M_V/\gt$ and $f_0 = \sqrt{v^2/\sin^2 \theta + f_1^2 r^2}$  defined as internal parameters in the model file. Note that the latter relation derives directly from fixing the value of $G_F$ (i.e. the EW scale $v$), as shown in \eq{eq:v1}. Furthermore, we have imposed  the SM values of   $e$ and $M_Z$  into the following analytic expressions  to  calculate  $g'$ and $g$ in terms of the independent model parameters.
\beq
\frac{1}{e^2} = \frac{1}{g'^2 } + \frac{1}{g^2} + \frac{2}{\gt^2}  \,, \quad  \det \left(M_N^2 -M_Z^2 I_{5} \right) =0  
\eeq

The model file is loaded using {\sc FeynRules} package which exports the Lagrangian into {\sc UFO} format~\cite{Degrande:2011ua}. We  implement one python code as the parameter card calculator,  to conduct the numerical rotation and write  all  block information into a \texttt{param\_card.dat}.

\subsection{LHC Run--II}

\begin{figure}
\begin{tabular}{cc}
\includegraphics[width=0.45\textwidth]{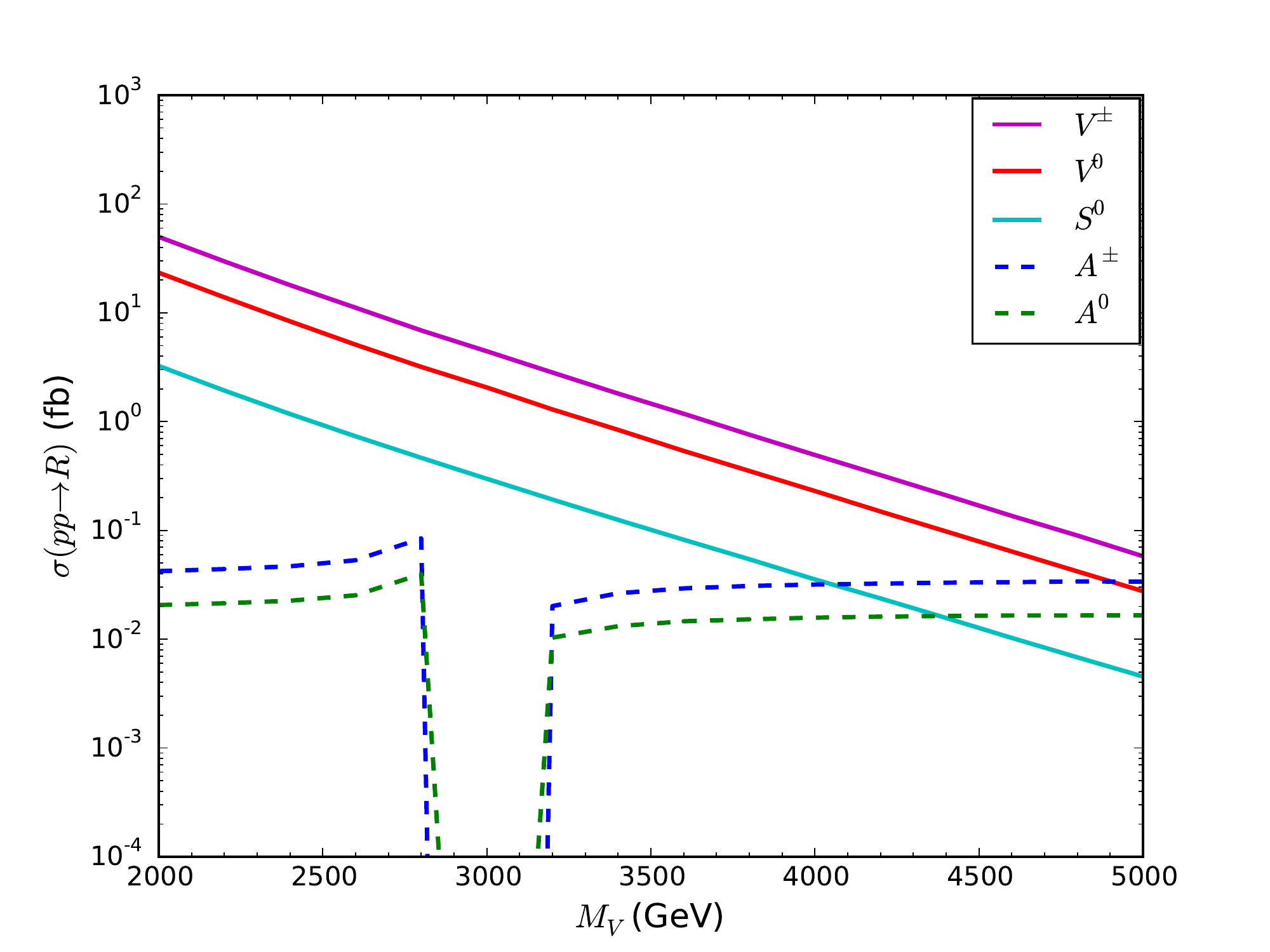} &
\includegraphics[width=0.45\textwidth]{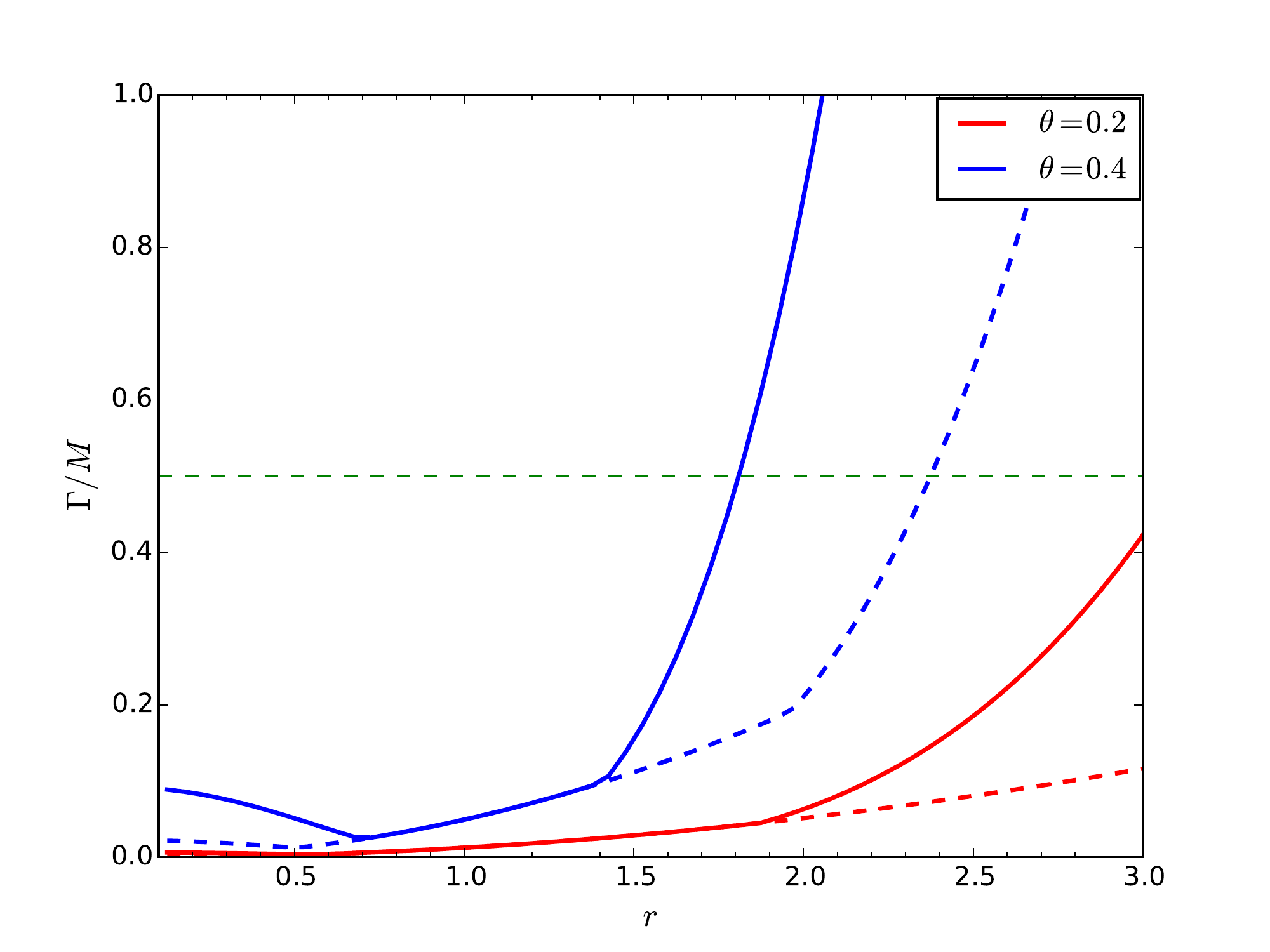} \\
a) & b)
\end{tabular}
\caption{Left: Drell--Yan cross section of composite states for $M_A=3\TeV$, $\gt =3.0$, $r =0.6$  and $\theta=0.2$  at LHC Run--II with  $\sqrt{s}=13\TeV$. Right: The largest $\Gamma/M$ of composite resonances as a function of $r$, with $M_V = 2.5$ TeV, $M_A = 3$ TeV, $\gt =3.0$ ~ (solid line) and $\gt =6.0$ (dashed line).} 
\label{fig:ratiolhc}
\end{figure}

At the LHC Run--II, several resonances may be produced via the Drell-Yan production mechanism, with $q q^\prime$ as initial states, therefore unfolding a delighting and rich phenomenology just like hadron spectroscopy in QCD but with completely new challenges and opportunities. Here we  briefly discuss what would be the first probable observations in the vector sector of our model, by investigating cross sections and  experimental bounds from a $\sqrt s =$13 TeV LHC. 
The calculation is conducted in {\sc MadGraph} 5~\cite{Alwall:2014hca}, using the PDF set NN23LO~\cite{Ball:2012cx}. 

We present  the cross section for each resonance at the LHC Run II  in \fig{fig:ratiolhc}a 
  by varying the parameter of  $M_V$, with fixed $M_{A}=3\TeV$, $ r = 0.6$ and $\theta =0.2$. The leading production channel is for the resonance $V^{\pm}$, followed by the neutral resonances $V^0$ and $S^0$. The vector  resonance $S^{\pm}$ is defined as the one  with largest  portion of $s^+$ state, with exact mass of $M_V$. This state is rotated out from the matrix ${\cal {C}}^a$ in \eq{eq:Ca},  as  a linear combination of $v^+$ and $s^+$, thus it can not be directly produced due to current model set up.  Increasing $\gt$ will result in a smaller cross section as the couplings to quarks, generated by the mixing, are suppressed. Furthermore, only $\sigma_{A^{\pm, 0}}$  shows  clear dependence on the  other parameters, $\theta$ and $r$, and in the case of  $M_V < M_A$,  $\sigma_{A^{\pm, 0}}$ will always  be subleading to  $\sigma_{V^{\pm,0},S^0}$  by several orders of magnitude.  Note that for the ``axial'' resonances $A^{\pm}$ and $A^{0}$, the cross sections turn out to be zero at the point of $M_V =M_A$, since the mixing does not contain any component of  $\widetilde{W}$ and $B$ and they decouple from SM quarks.   We also check the parameter space where the narrow width approximation (NWA) can be used, as shown in \fig{fig:ratiolhc}b where we find that the relevant parameter is $r$. We set the benchmark point to be $M_V =2.5$ TeV and $M_A =  3$ TeV,  and  vary the other parameters  $(\gt, \theta, r)$ to inspect the region where the largest $\Gamma/M$ among all  resonances is less than $50\%$. Generally, in order to use  NWA as an approximate analysis for the event line shapes (e.g.  di-lepton invariant mass distribution) we require $\Gamma/M < 10 \%$ so that interference effects with $Z, \gamma$ can be safely neglected. Due to the small mass split between many resonances, off-diagonal width effects may also be important~\cite{Cacciapaglia:2009ic,deBlas:2012qp}. Furthermore, the small width region will be favoured in order to resolve the compressed multi-peaking structure in the spectrum.  According to this criterion, for a small $\theta =0.2$, the  NWA  applies  very well  for $  0< r < 2.0$, but with a larger value $\theta =0.4$, the resonance will become broad and we need at least to tune $\gt > 6.0$ for the NWA to be effective.

\begin{figure}
\includegraphics[width=0.45\textwidth]{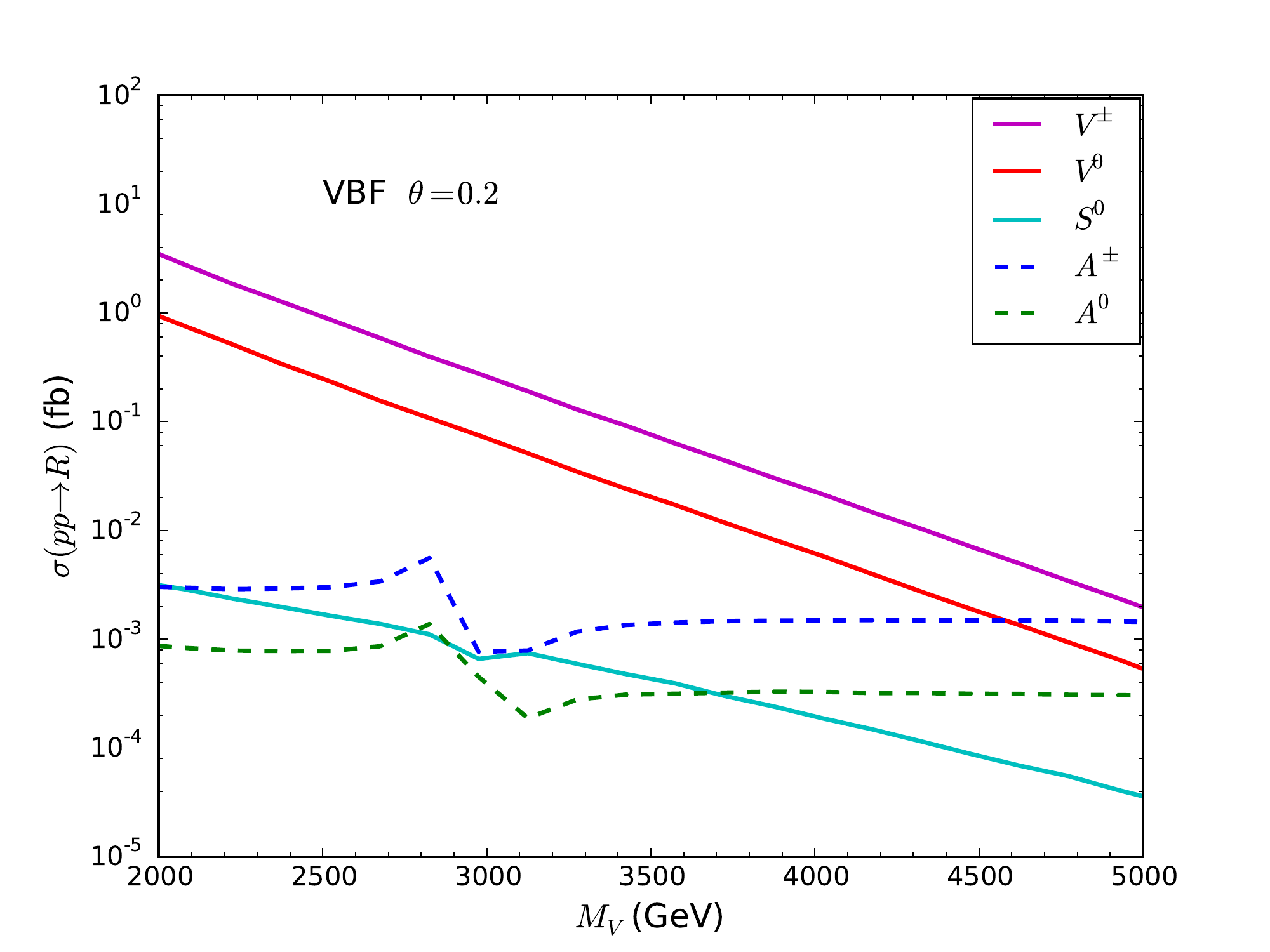}
\includegraphics[width=0.45\textwidth]{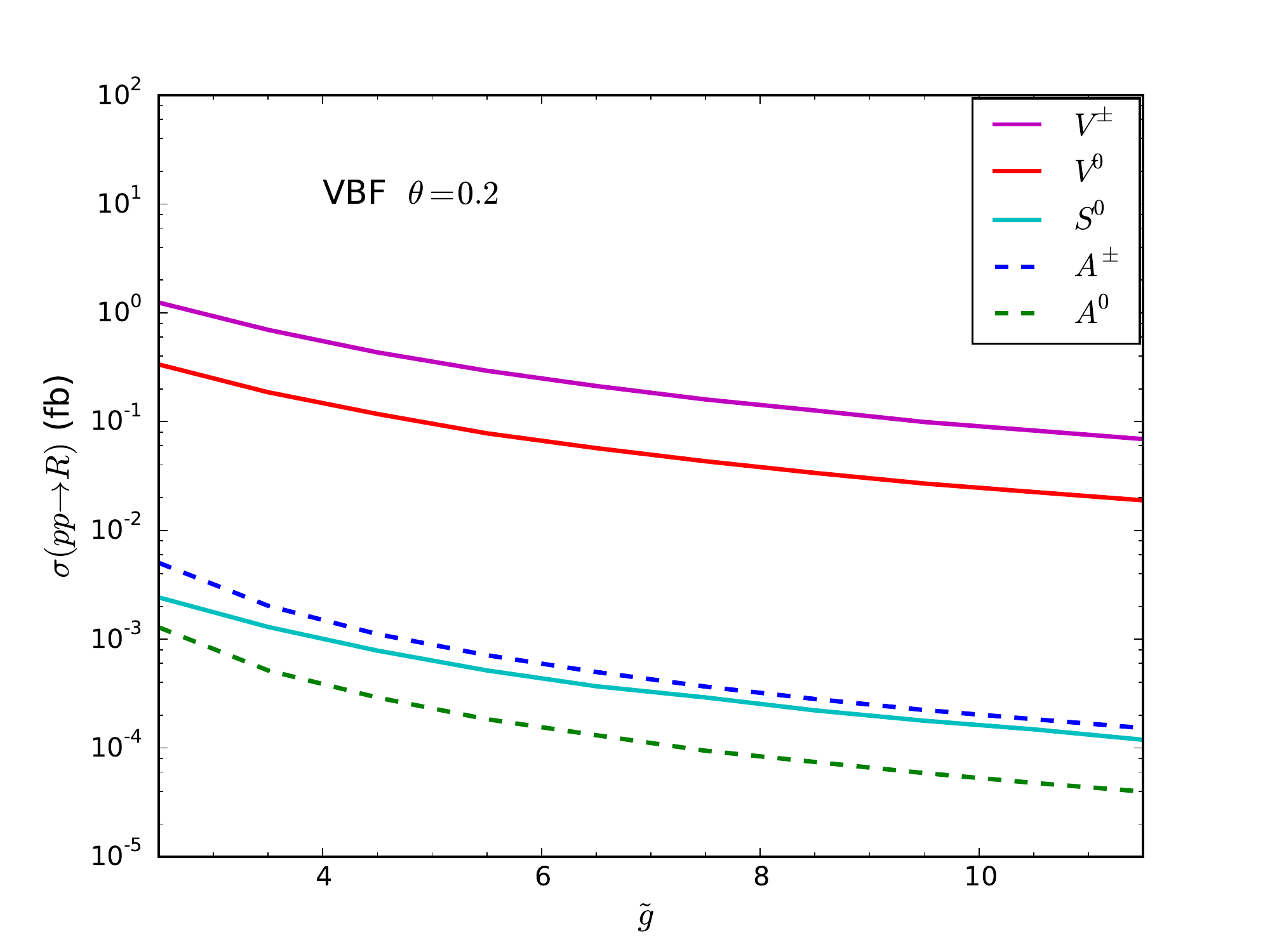}
\caption{Vector Boson Fusion (VBF) cross sections  of composite states for $M_A=3\TeV$, $r =0.6$  and $\theta=0.2$ at LHC Run--II with  $\sqrt{s}=13\TeV$. On the \emph{left}:  $\gt =3.0$ is fixed and $M_V$ varies; On the \emph{right}: $M_V=2.5\TeV$ and  $\gt$ varies. }  
\label{fig:VBFxs}
\end{figure}

Alternatively the composite resonances can  be produced via vector boson fusion (VBF)\cite{Belyaev:2008yj,Franzosi:2012ih, Mohan:2015doa}, with the production cross sections  shown in  \fig{fig:VBFxs} for the same benchmark scenario.
 In the calculation for $pp\to R$+2 jets ($R=A^{0,\pm},V^{0,\pm},S^{0}$),  we consider all pure EW diagrams which form a gauge invariant set  with the VBF topology, including diagrams with one $t$-channel weak boson exchange following a composite resonance  emitted from a quark line. Although the signal definition is ambiguous we expect that the VBF topology dominates.
  It was required $p_T(j)>20$ GeV in order to avoid the $t$-channel singularity of a photon exchange.  
The longitudinal weak bosons, $W_L$, coupled to composite vectors through partial compositeness, play a less important role due to small mixing angles. This is  noticed that in the right hand panel of  \fig{fig:VBFxs},  the cross section decreases with $\gt$.
Therefore, in general VBF is a subdominant production mechanism.
As previously remarked in \cite{Belyaev:2008yj}, the exception to this trend occurs in the special parameter space region $M_A\simeq M_V$, where the Drell--Yan production of $A^{0, \pm}$ is highly suppressed. 

\begin{figure}
\begin{tabular}{cc}
\includegraphics[width=0.49\textwidth]{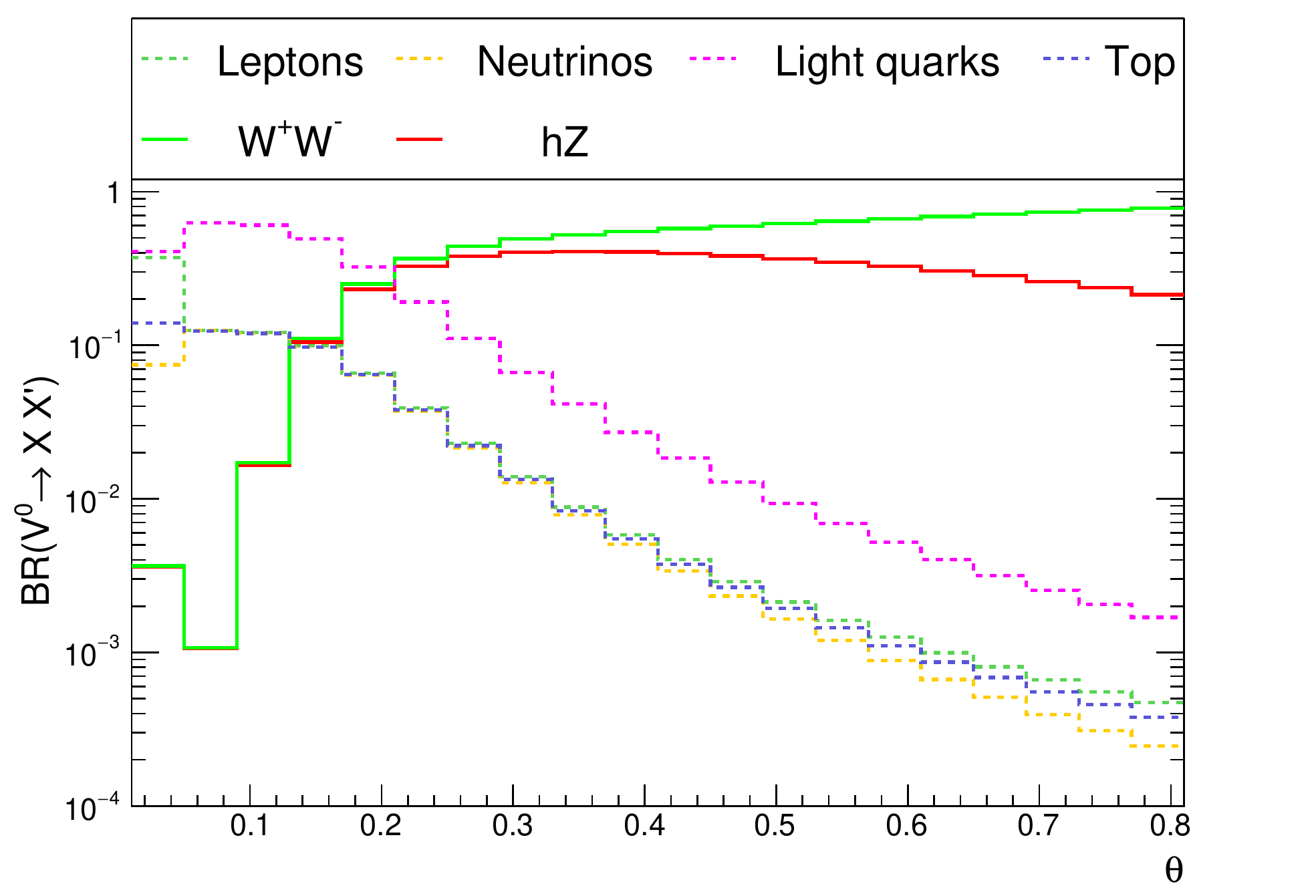} &
\includegraphics[width=0.49\textwidth]{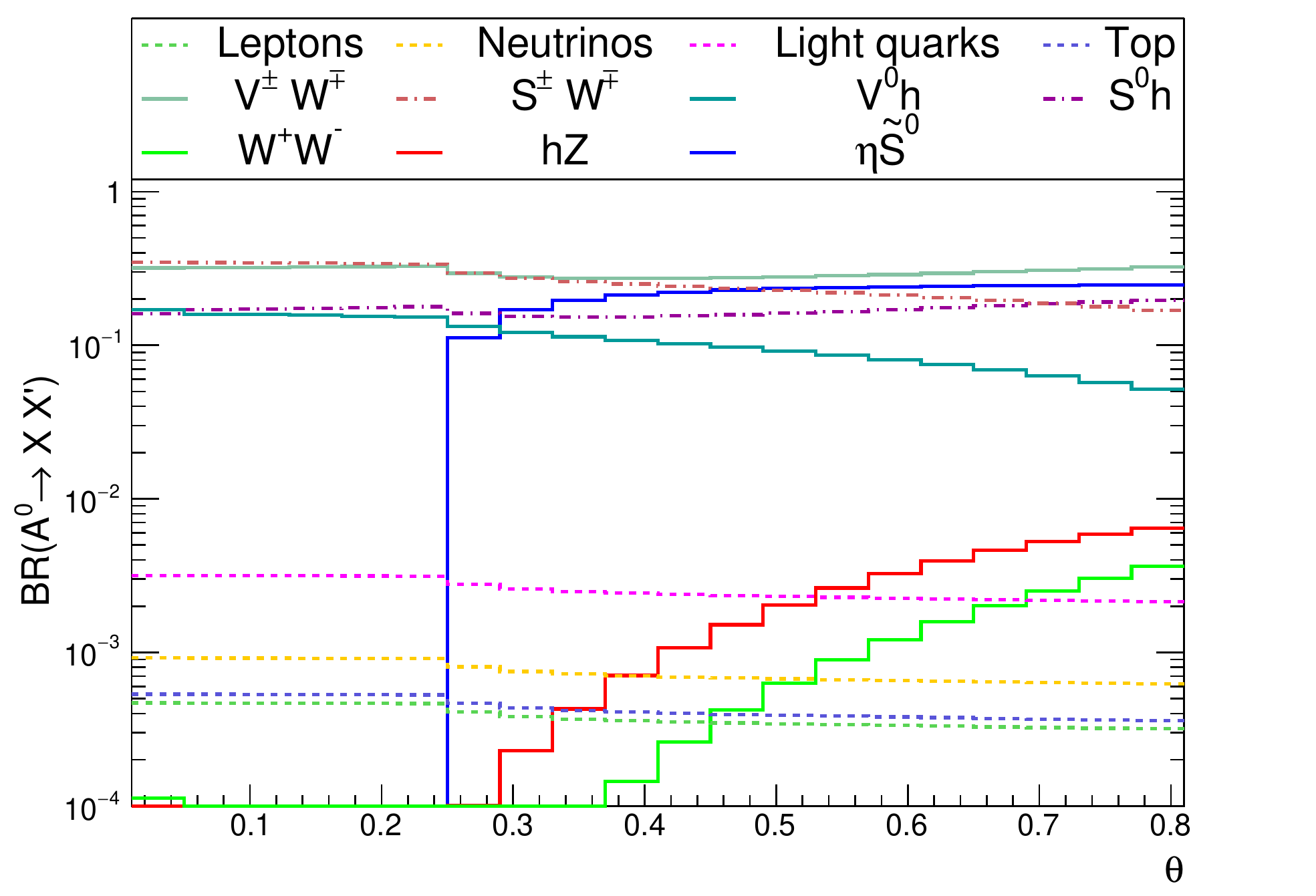} \\
(a) & (b) \\  \\
\includegraphics[width=0.49\textwidth]{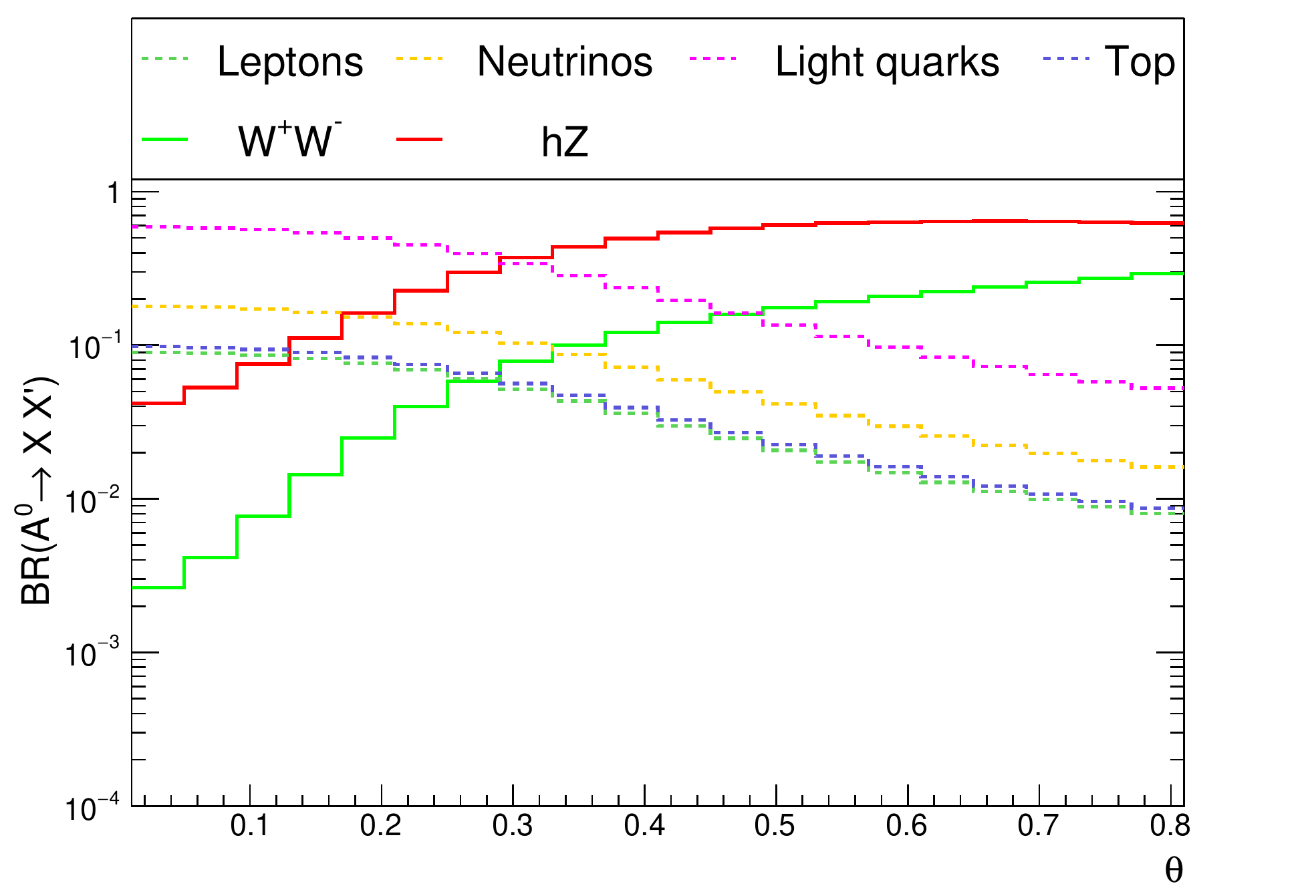}&
\includegraphics[width=0.49\textwidth]{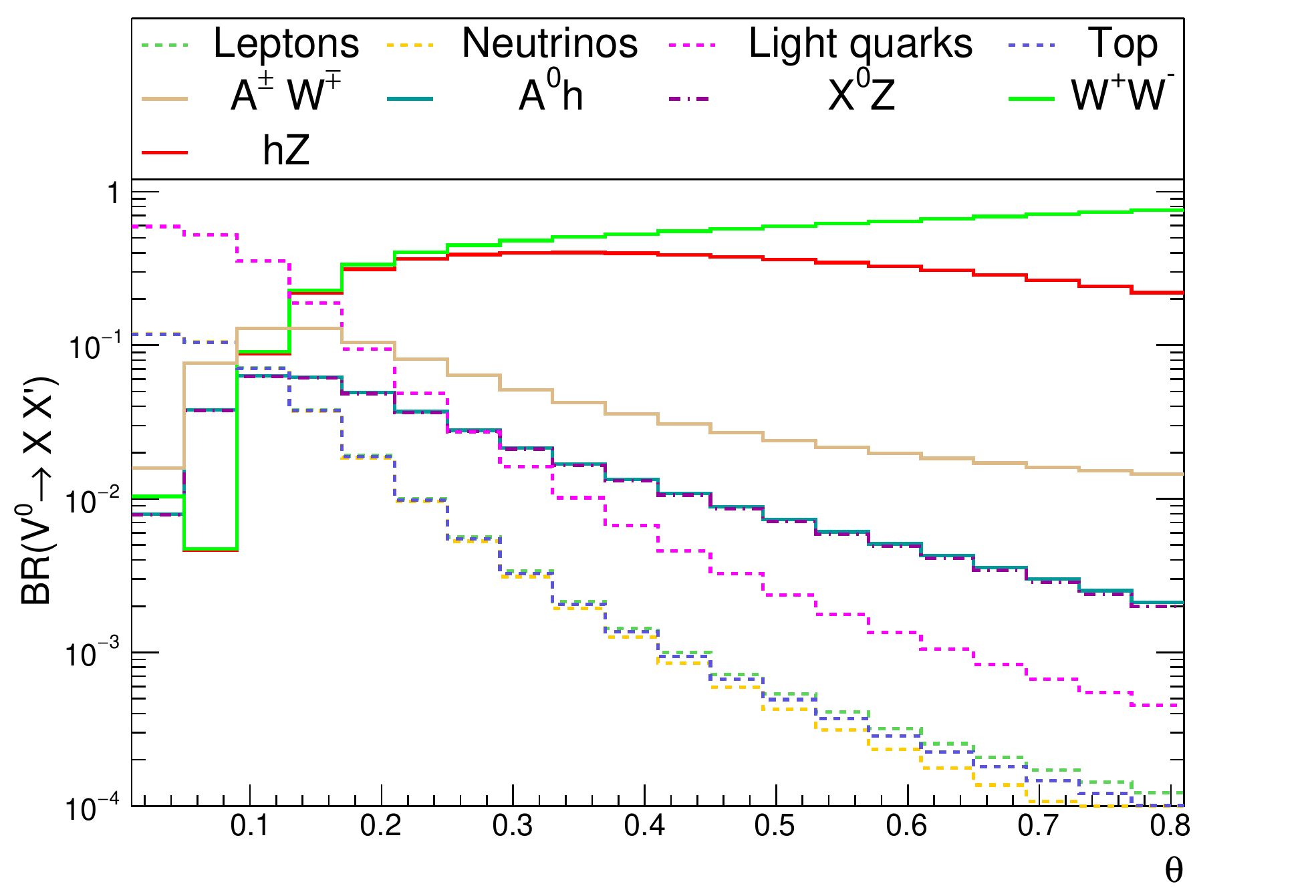} \\
(c) & (d) \\  
\end{tabular}
\caption{Branching ratios (BR) of composite states $V^0$  and $A^0$, with the dependence on $\theta$ for $\gt =3.0$, $r =0.6$, $M_V=2.5$ TeV (top row) and  $M_V=3.5$ TeV (bottom row),  with fixed $M_A=3$ TeV.}
\label{fig:BRVtheta}
\end{figure}

In \fig{fig:BRVtheta},  we show the typical  branching ratios for $V_0$ and $A_0$,  with all the decays into SM fermions drawn in dotted line ($V^+,\, A^+$ show similar decay pattern).  The  entry in the legend is well  patterned, each mode arranged in the same colour and line-style in order to easily compare the differences in each scenario,  with  top  standing for  $t \bar t$,  light quarks for  $u_{1,2}\bar u_{1,2}$+$d_{1,2} \bar d_{1,2}$ (Cabibbo CKM mixing used), leptons for $l^+ l^-$ and neutrino for $\nu \bar \nu$.  For the decay of $A^0$ in the case of $M_V <  M_A$, we   draw  the mode with $V^{\pm,0}$ in the solid line while the mode with $S^{\pm, 0}$ is in the dash-dotted line, since there is certain overlap between the decay modes of $V^0 h$ and $S^0 h$,  similar for  $V^\pm W^\mp$ and $S^\pm W^\mp$, in the low $\theta$ region, but start to split from $\theta \gtrsim 0.3$.  An analogous situation happens to the decay of $V^0$ in the case of $M_A <  M_V$,  where branching ratios into  $A^0 Z$  and $X^0 h$,   mostly overlap with each other in the range of $0 <\theta < 0.8$ due to the  global symmetry. We also explore the branching ratios as a function of $r$: the fermions spectrum goes to a maximum at $r =1$,  while the  $WW$ or $hZ$ spectrum,  instead, goes to a minimum since the coupling is  $\propto (r^2-1)$ at $1/\gt$ order.  In either  vector or axial resonance dominant case,   the lower mass state displays  a larger  branching ratio into  $l^+ l^-$ and $W^+W^-$  rather than into final states containing a composite vector, therefore we can exploit the most recent LHC Run--II  results to constrain the model parameters.

\begin{figure}
\includegraphics[width=0.44\textwidth]{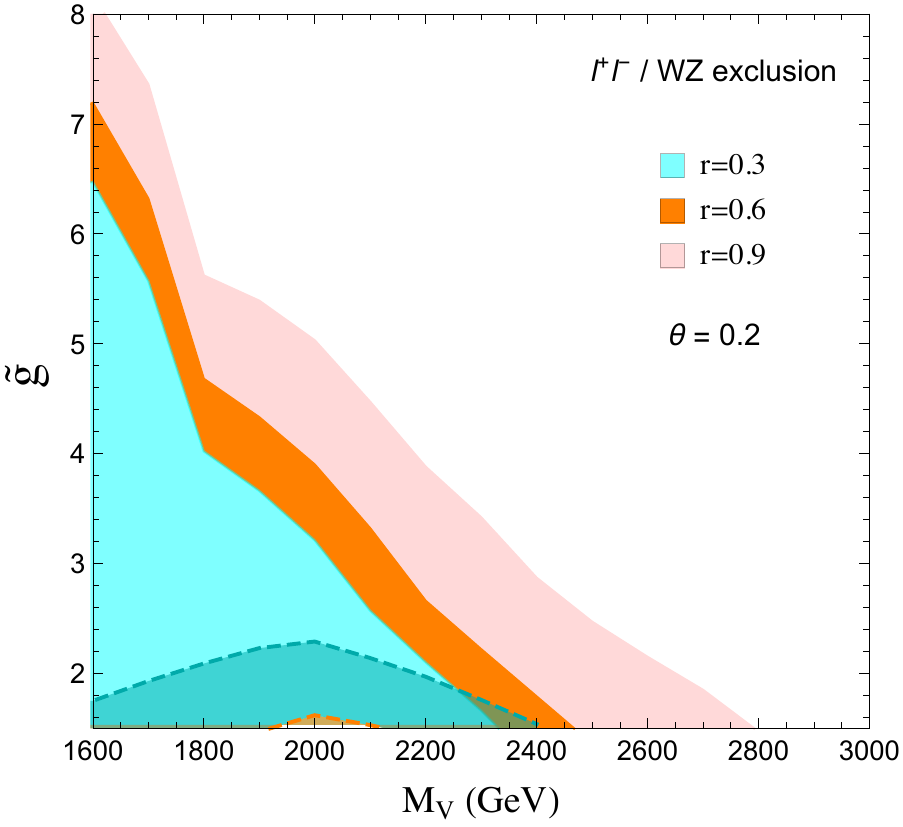}
\includegraphics[width=0.45\textwidth]{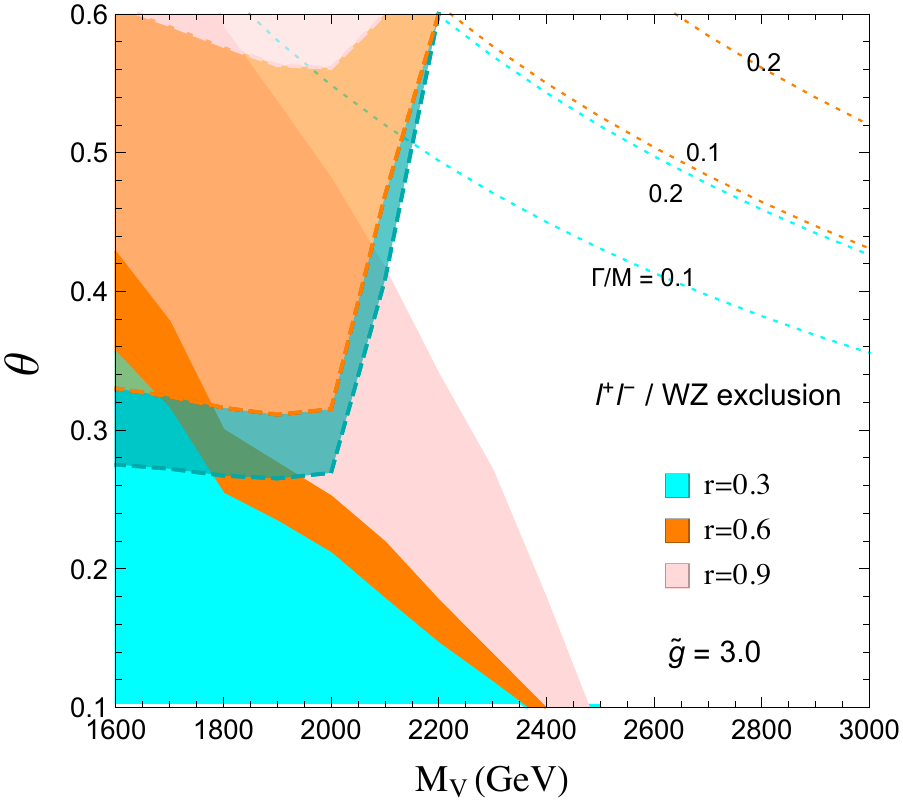}
\caption{ Excluded region with $M_V=M_A$, recast from the $95 \%$ observed limit for the di-lepton $l^+l^-$ and di-boson $WZ$  channel  measured by ATLAS at LHC Run--II. The $l^+l^-$ exclusion is in solid lines, and  the $WZ$ exclusion  in dashed ones. The parameter $r$ is varied in the range of  $[0, 1]$, with contours in the $(\gt-M_V)$ plane  on the \emph{left} and in  the $(\theta-M_V)$ plane on the \emph{right}. The region with $\Gamma/M > 0.1$, where NWA is not applicable, is identified by dotted contour lines.  }
\label{fig:ExclusionRegion}
\end{figure}

Since our model provides several candidates as a heavy $Z^\prime$ or $W^\prime$,  the LHC measurement for the Drell-Yan process  and di-boson process  would impose a stringent constraint on the parameter space. We calculate the theoretical cross section for $pp \to R^0 \to l^+ l^- $  and $p p \to R^\pm \to WZ $ in this $SU(4)/Sp(4)$ model and compare them with the $95 \% $ upper bound observed from the latest ATLAS measurement \cite{dilepton, diboson}. Similar results can be obtained by using the corresponding CMS searches~\cite{CMS:2015nhc,CMS:2015nmz}. The single lepton plus MET process is expected to require similar constraints to the di-lepton ones, thus we do not consider the $l \nu $ channel in detail for simplicity.  We derive the exclusion limits in the  parameter space specified by ($\gt$, $\theta$, $r$ ) after assuming $M_V = M_A$.  Since we have not included the acceptance factor into this  analysis,   our result   would be  stronger than the exact $95 \%$ exclusion from the LHC Run--II searches.  We show  the  exclusion contours from $l^+ l^-$ and $W Z$ in \fig{fig:ExclusionRegion},  with the di-lepton bound drawn in  solid line, and the di-boson bound in  dashed line. The plot shows  that the two channels are  complementary to each other.
Notice that for an increasing $r$ (in range of $[0, 1]$), the di-lepton channel imposes a stronger exclusion limit than the di-boson. The left panel of~\fig{fig:ExclusionRegion} shows that the lower limit for the mass of the resonances approaches $M_V \gtrsim 2.5$ TeV for a coupling constant $\gt \simeq 3.0$ and small angle $\theta = 0.2$.  
The right panel of~\fig{fig:ExclusionRegion} also shows that the di-lepton  limit is more sensitive to the small $\theta$ area, while the di-boson channel mainly probes the large $\theta$  area.
To summarise, for small $\theta < 0.2$, as expected in composite pNGB Higgs limit of the model~\cite{Arbey:2015exa}, the di-lepton searches impose a lower bound on $M_V$ between $2$ and $2.5$ TeV, depending on the value of $r$.

\subsection{Future 100 TeV proton colliders}

\begin{figure}
\begin{tabular}{cc}
\includegraphics[width=0.405\textwidth]{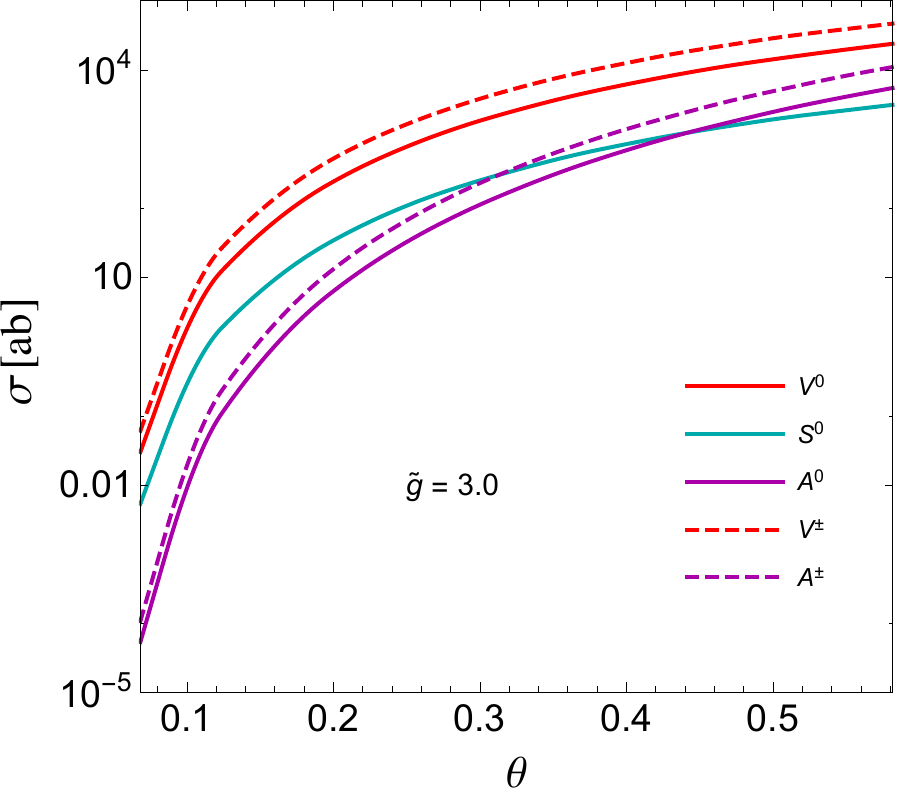} &
\includegraphics[width=0.4\textwidth]{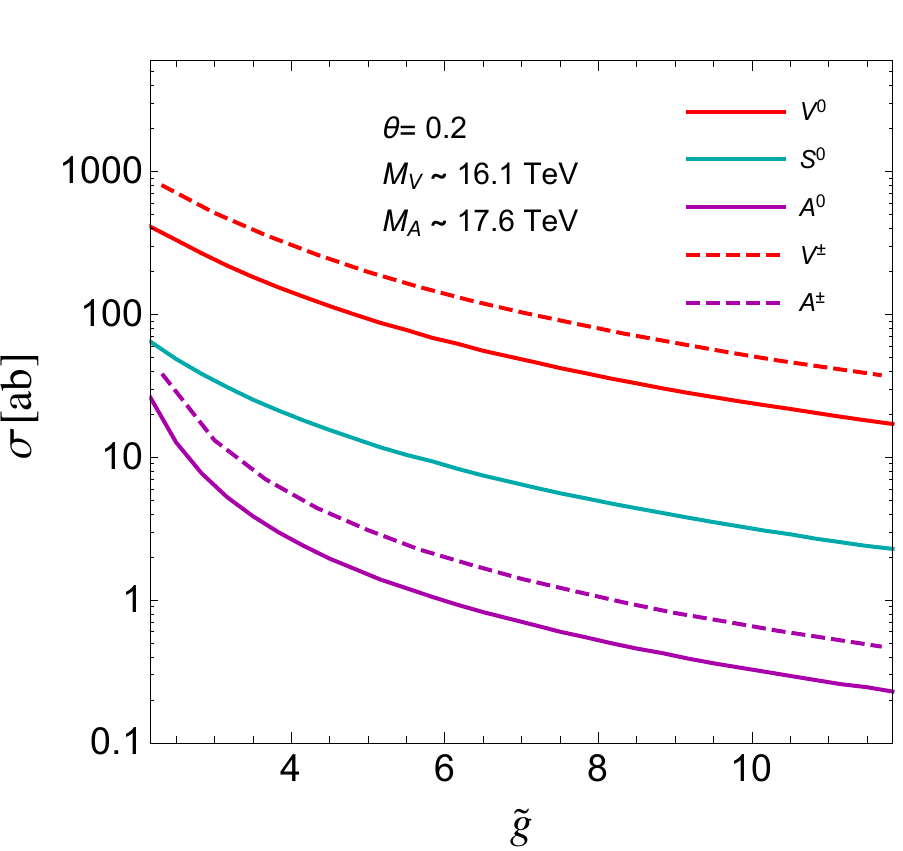} \\
(a) & (b) \\
\includegraphics[width=0.405\textwidth]{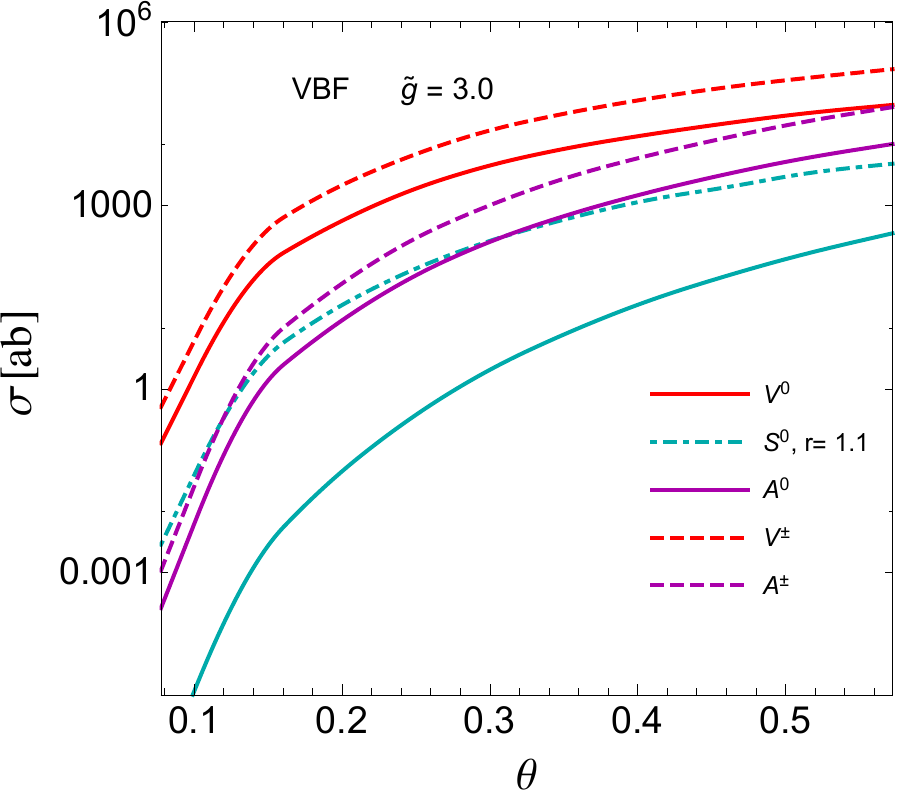} &
\includegraphics[width=0.4\textwidth]{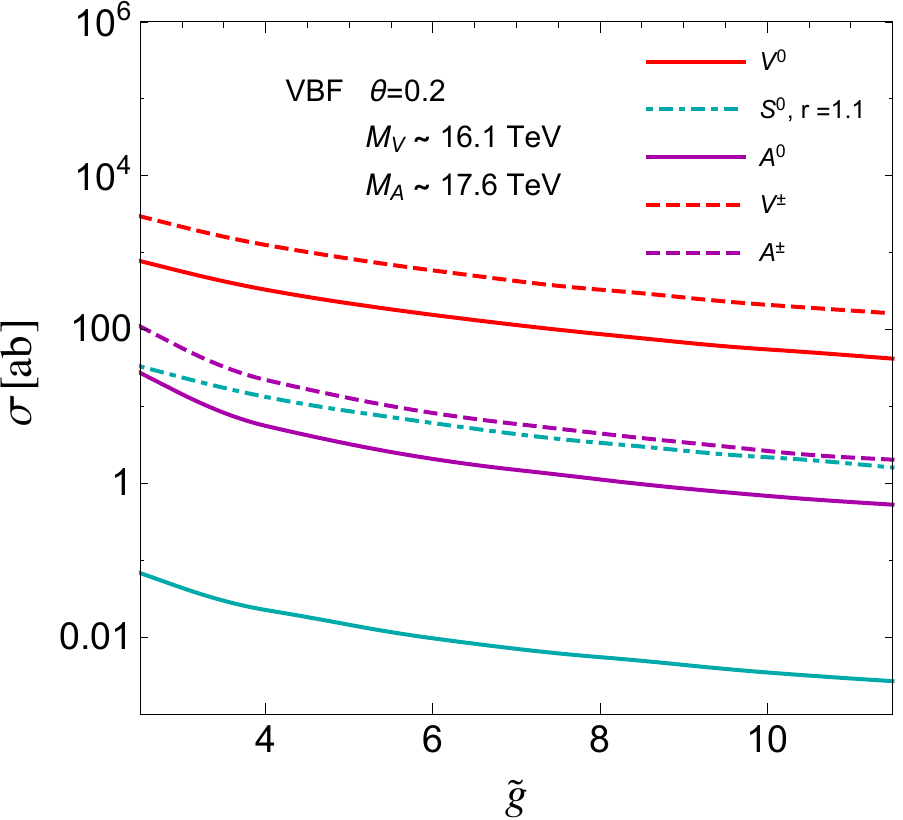}  \\
(c) & (d)
\end{tabular}
\caption{Drell-Yan  ( plots (a) and  (b) ) and  Vector Boson Fusion ( plots (c) and (d) )  cross sections of composite states with $M_V=3.2/\sin(\theta) $ TeV, $M_A=3.5/\sin(\theta) $ TeV at $\sqrt{s}=100\TeV$. The \emph{left} panel shows  the dependence on $\theta$  for $\gt =3.0$, $r =1.0 $.  The \emph{right} panel shows the dependence on $\gt$ for $r=1.0$ and $\theta=0.2$.  For the $S^0$ VBF production, the dot-dashed cyan line is using $r = 1.1$, compared with the solid cyan line of $r =1.0$.
}
\label{fig:xs100tev}
\end{figure}

As shown in the previous section, current LHC bounds on the resonances range in the $2$ TeV ballpark.
However, the naive expectation is that the resonances populate this mass range only in the Technicolour limit, where $f_\pi \sim v$, in the composite pNGB limit, all the resonances' masses would be enhanced by a factor $1/\sin \theta$ due to the increase in the compositeness scale. Thus, the most natural mass range seem to lie above tens of TeV, thus more relevant for a future 100 TeV collider than for the LHC. 
For the simplest underlying gauge theory realising $SU(4)/Sp(4)$ global symmetry, namely $SU(2)$ with 
2 Dirac fermions, lattice results have recently been published~\cite{Arthur:2016dir}, providing a first numerical prediction for the masses of the spin-1 resonances, found to be $M_A=3.5\TeV/\sin(\theta)$ and $M_V=3.2\TeV/\sin(\theta)$, far from LHC reach in the small $\theta$ limit. Thus, a machine  colliding protons at $\sqrt{s}=100 \TeV$ would be a perfect stage to probe its vast spectrum.  It should be noted, however, that the masses can be lighter in different underlying gauge theories.
In such case, even though the mass scales as $1/\sin \theta$, the resonances might be at the reach of LHC.

The Drell-Yan production of the states $V^{0,\pm}$, $S^{0}$ and $A^{0,\pm}$  are shown in the top row of  \fig{fig:xs100tev} as functions of $\theta$ and $\gt$. When we use  {\sc Madgraph} for simulation, only the PDF of the first two generations of quarks are taken into account. However, at the high energy collider, the top and bottom quark PDFs can be important and need to be included  to conduct a reliable prediction at 100 TeV~\cite{Han:2014nja}. Nonetheless, the cross sections present here can serve as a guideline. Similarly to the scenario described in the last section, the production rate for these states with $r\sim 1$  is not large, around ${\cal{O}}(1)$ fb for $\theta \sim 0.2 $, since the resonance coupling to SM quarks are generated via  mixing.

At 100 TeV, Vector Boson Fusion  plays a more important role due to the enhancement of collinear radiated weak bosons from the spectator quarks, which translates into a large effective luminosity of weak bosons inside the proton in the language of the Effective W approximation \cite{Chanowitz:1985hj}.
Indeed, the importance of VBF can be appreciated in the bottom row of  \fig{fig:xs100tev}, which shows  that   for   $V^{0, \pm}$ and $A^{0, \pm}$ resonances the VBF cross section is dominant over the Drell--Yan production, with very mild dependence on $r$.  However the $S^0$ production is much more sensitive to $r$.  Since the $W^+ W^- S^0$ coupling almost vanishes at the point of $r=1.0$,  with the main contribution to VBF  from the  $V^\pm W^\mp S^0$ fusion,  this makes the $S^0$ production particularly small.  But once departing from $r=1.0$,  the $W^+ W^-S^0$ fusion turns back to be important,  therefore  the  VBF cross section for $S^0$ is actually  two orders of magnitude larger in the case of  $r=1.1$.



It is also important to note that SM physics, jets, top production and other important background for the process will present quite peculiar aspects at a 100 TeV collider (see \emph{e.g.} \cite{Bothmann:2016loj}) and must be taken into consideration for a more precise phenomenological analysis.

\begin{figure}
\includegraphics[width=0.50\textwidth]{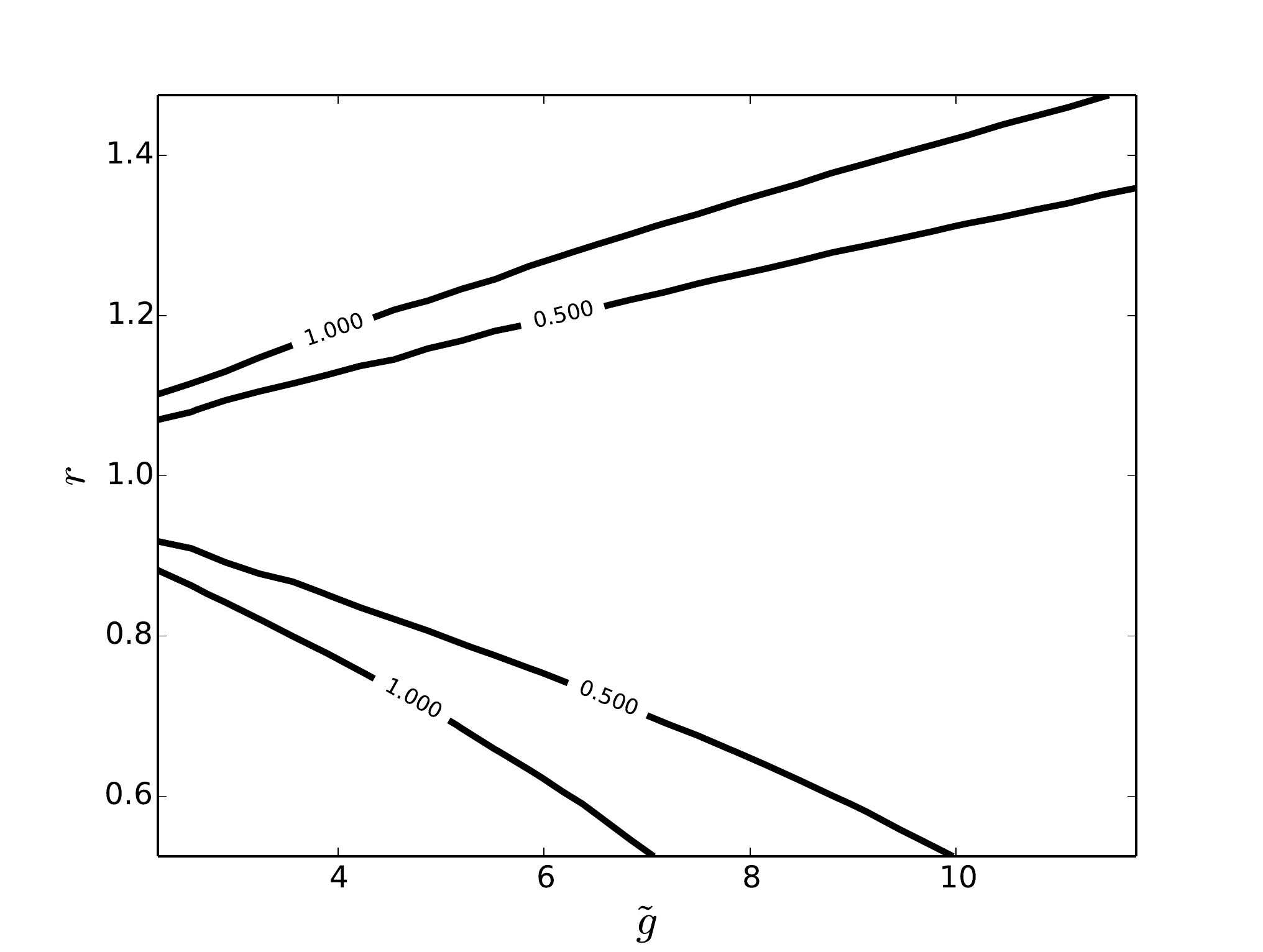}
\caption{The largest ratio of width over mass  $max(\Gamma_{R}/M_{R})$  among all  the resonances. For large $r$,  some resonances become too broad. The value of $\theta$ is set to 0.2, and  the dependence on $\theta$ is very small in the perturbative region.}
\label{fig:wom}
\end{figure} 

The value of $r$ is constrained by perturbativity of the effective description.  The consistent region  is illustrated in \fig{fig:wom}, with the largest ratio of width over mass  extracted in the plane of $(\gt-r)$. We find that the region of $r$ close to one  is where  all the resonances are narrow, thus it is valid to apply the NWA  for event analysis.  For $r\neq 1$ the coupling of heavy vector to longitudinal bosons rapidly grows  as the width of the resonance approaches its mass, jeopardising perturbativity and the validity of the description~\cite{Bando:1987br}.

\begin{figure}
\begin{tabular}{cc}
\includegraphics[width=0.49\textwidth]{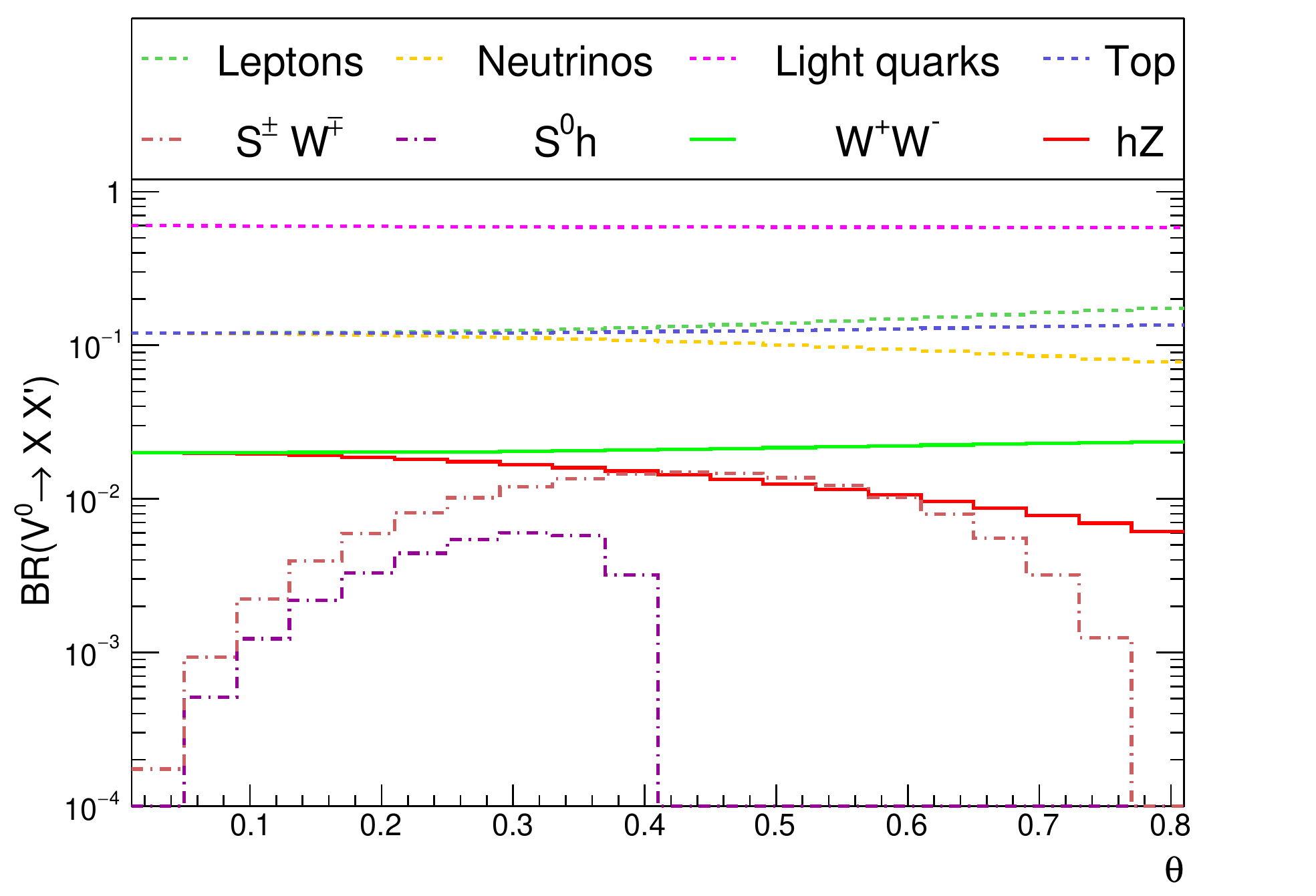} &
\includegraphics[width=0.49\textwidth]{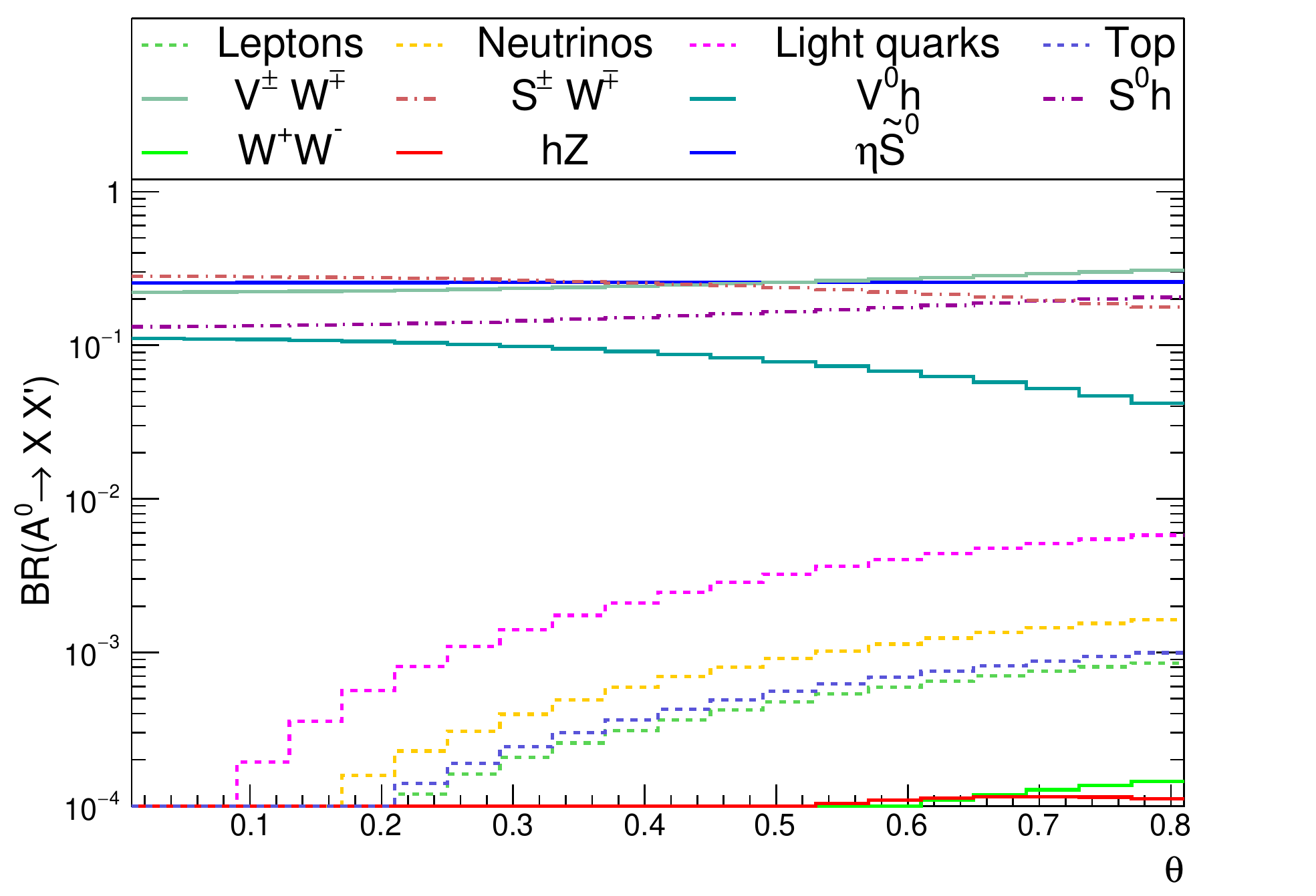} \\
(a) & (b) \\  \\
\includegraphics[width=0.49\textwidth]{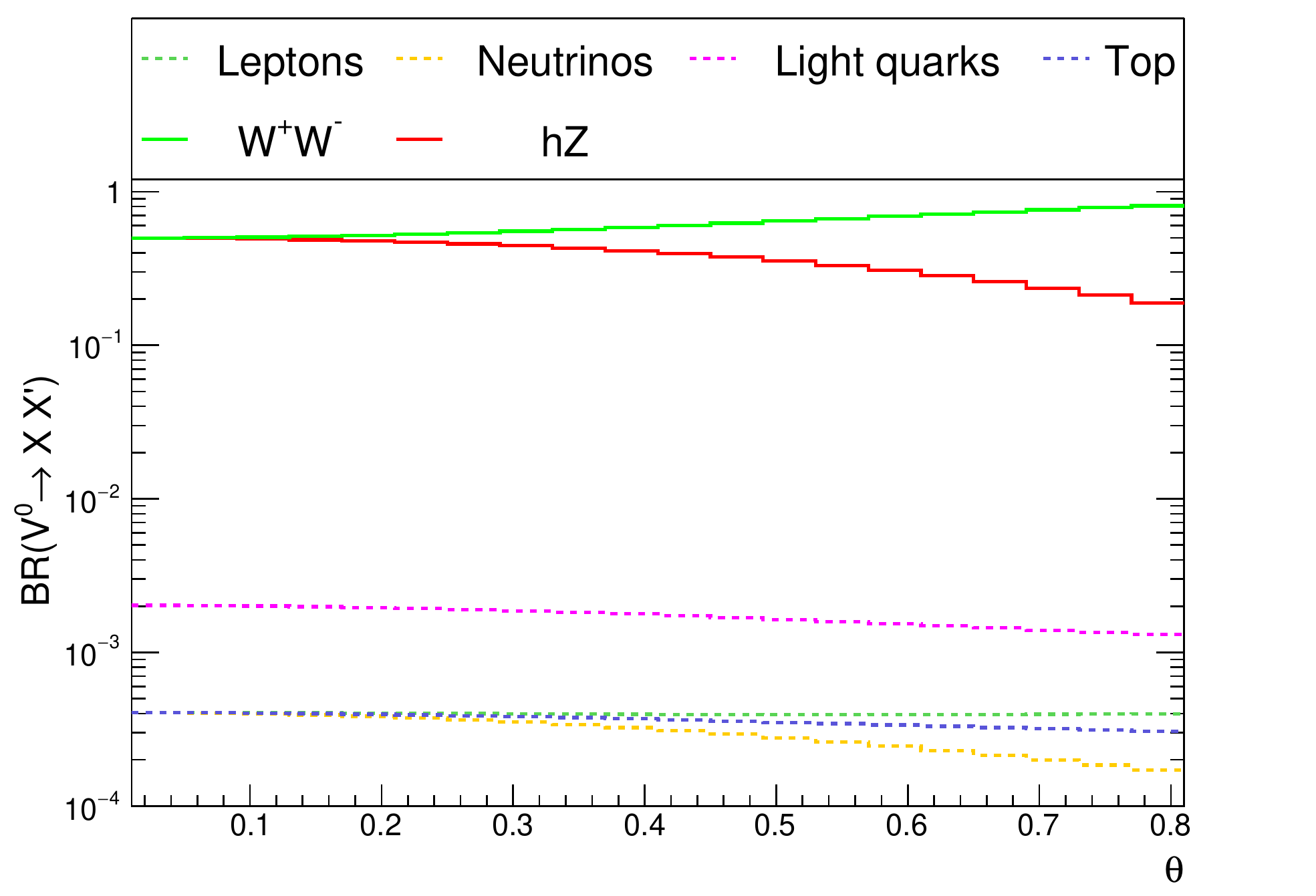} &
\includegraphics[width=0.49\textwidth]{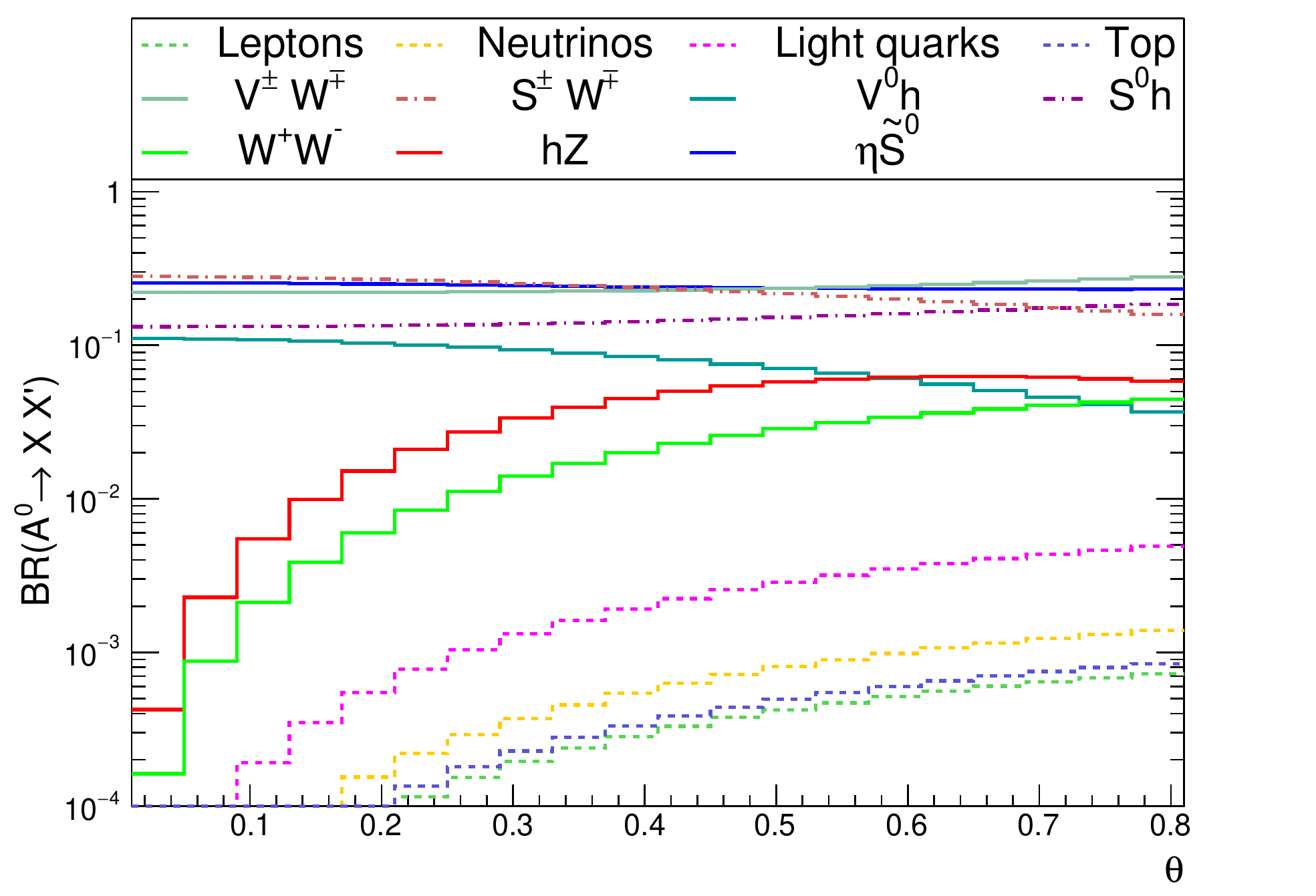} \\
(c) & (d) \\  
\end{tabular}
\caption{Branching ratios (BR) of composite states  $A_0$  and $V_0$, with the dependence on $\theta$  for $M_V=3.2/\sin(\theta) $ TeV, $M_A=3.5/\sin(\theta)$ TeV,   $\gt =3.0$ and $r=1.0$ (top row) and $r=1.1$  (bottom row).}
\label{fig:BR100TeV}
\end{figure}

We show the BR of $V^0$ state as a function of $\theta$ in the left panel of  \fig{fig:BR100TeV}. $S^0$  has similar decay structure as $V^0$, while charged states present similar pattern, thus we do not show them here.  At $r=1$ they mainly decay into SM fermions, in particular into di-jets. There will be small differences in  the BR spectrum between $V^0$ and $S^0$.   We find that, in the channel of  di-leptons, $V_0$ decays at $\sim 10\%$ and $S_0$ at $\sim 40\%$. Moreover, for $V_0$,  the decay into $W^+W^-$ is larger than  $hZ$, while for $S_0$ the decay of  $hZ$ turns to  dominate over  $W^+W^-$.  Varying $r$ to be slightly larger than $1$,  notable changes happen as the di-boson and $hZ$ channels rapidly overcome the fermion ones. For $r=1.1$ the branching ratios are close to $45\%$, equally split between $W^+W^-$ and $hZ$ at small $\theta$. Only  small variation can be observed in  $\theta\gtrsim 0$,  but the two channels will start to split exactly till  $\theta\lesssim 0.8$. 

The decay pattern of the $A_0$ resonance is shown in the right panel of  \fig{fig:BR100TeV},  with more channels opened. At $r=1$,  the fermion channel is subdominant,  while the $WW$, $hZ$ channels almost disappear. The decay into  $V^\pm/S^\pm W^\mp$ and $V^0/S^0 h$  become competitive, and  we can  observe  dominant decays into  $\eta \tilde{S}$, with $\eta$  further decaying into a pair of tops, or gauge bosons via the WZW anomaly term (as discussed in \cite{Arbey:2015exa}).  Since the $\tilde{S}$ decay is nearly $100 \%$ to $Z\eta$ for the lattice benchmark point, this  will give rise to novel  collider signature of  $4 t + Z$ final states.  For  $r =1.1$ the di-boson and Higgs-strahlung channels enter into play,  but  this does not alter the picture dramatically. 

It has been argued that the luminosity of this future machine should be at least a factor 50 larger than the LHC luminosity in order to profit from its full potential to find new physics~\cite{Richter:2014pga,Rizzo:2015yha}. An integrated luminosity of $3-30~ab^{-1}$ per year is therefore expected, leading to several heavy vector bosons produced and a promising phenomenology.
We also stress that probing masses up to $\sim 50$ TeV indirectly tests the models at small values of $\theta \sim 0.05$, where the high level of fine tuning renders the models unnatural and unappealing. While an ultimate exclusion is not possible due to a decouplings limit $\theta \to 0$ (like in supersymmetry), in our opinion a 100 TeV collider can ultimately probe the ``motivated'' region of the parameter space in this class of composite Higgs models.

\section{Conclusions} \label{sec:concl}

In the present work we construct an effective Lagrangian that allows to describe vector spin-1 resonances in composite 
models of the Higgs boson. The framework adopted is the one of the hidden gauge symmetry approach, and we focus 
on a case with global 
symmetry structure based on the minimal case of an SU(4) symmetry broken to Sp(4). The chosen coset both satisfies the requirement of a custodial 
Higgs sector and allows for a fundamental composite  description of the new resonances in terms of fermionic bound states. The 
SU(4) structure is promoted to SU(4)$_0 \times$SU(4)$_1$ in order to apply the hidden gauge symmetry idea and to obtain the vector and 
axial-vector states in the adjoint of the second SU(4). The paper discusses in detail the effective Lagrangian for these states and their properties including mass matrices, mixing and couplings. The underlying fundamental realisation of the theory in terms of fermionic bound states is also discussed, together 
with the associated discrete symmetries, such as parity. 

Schematically, the model contains 3 triplets that mix with the standard model gauge bosons, plus additional states that do not mix. Therefore, the phenomenology is much richer than in the minimal case containing just a single isospin triplet.
We outline the main properties of the spin-1 states and their role in the phenomenology of the basic model.
At the LHC, the most sensitive channel for searching for the new resonances is di-lepton, which already imposes a bound on their mass around 2 TeV.
The unmixed states, on the other hand, tend to decay into the singlet pion, $\eta$, thus providing new signatures compared to the minimal cases studied in the literature.
Furthermore, in the case of a pseudo-Goldstone Higgs, where the compositeness scale is raised, the masses are expected to be higher, in the 10 TeV range. We show that a future 100 TeV collider may be able to probe the most interesting parameter space for naturalness. We focus on a minimal underlying description, where the masses have been computed on the Lattice, and detail the cross sections and branching ratios. This scenario can thus be one of the benchmark models for the 100 TeV collider.

This overview of the model, 
and its phenomenology, that we present is a template for the study of fundamental strong dynamics in the electroweak sector. Besides the specific case under study, which corresponds to the minimal fundamental model, it can be applied to other scenarios like, for instance, the case of composite strongly interacting Dark Matter candidates.

\section*{Acknowledgements}

We thank Marc Gillioz for collaboration at an early stage of this work. GC, HC and AD acknowledge partial support from the Labex-LIO (Lyon Institute of Origins) under grant ANR-10-LABX-66 and FRAMA (FR3127, F\'ed\'eration de Recherche ``Andr\'e Marie Amp\`ere''). MTF is partially funded by the Danish National Research Foundation, grant number DNRF90.

\appendix

\section{Explicit formulas} \label{app:formulas}

The explicit embedding of the vectors in SU(4) matrix form, in terms of charge eigenstates (see \tab{tab:class}), is
\beq
\bm{\mathcal{F}}_{\mu} \equiv \bar{\bm{\mathcal{F}}}_{\mu}+\widetilde{\bm{\mathcal{F}}}_{\mu} \,,
\label{eq:amu}
\eeq
with
{\begin{footnotesize}
\begin{eqnarray}
\bar{\bm{\mathcal{F}}}_{\mu} =
\begin{pmatrix}
{\displaystyle \frac{s_\theta a^0-c_\theta s^0 +v^0}{2 \sqrt{2}} } & 
{\displaystyle \frac{c_\theta s^+ - s_\theta a^+ + v^+}{2} }& 
{\displaystyle \frac{c_\theta a^+ +s_\theta s^+}{2} } & 
{\displaystyle \frac{c_\theta a^0+ s_\theta s^0- i x^0}{2 \sqrt{2}} } \\ \smallskip 
\\
{\displaystyle  \frac{-s_\theta a^-+c_\theta s^-+v^-}{2}  } & 
{\displaystyle \frac{c_\theta s^0 - s_\theta a^0 - v^0}{2 \sqrt{2}} }&
{\displaystyle \frac{c_\theta a^0 + s_\theta s^0+ i x^0}{2 \sqrt{2}} } & 
{\displaystyle \frac{-c_\theta a^- -s_\theta s^-}{2} }  \\ \smallskip 
\\
{\displaystyle \frac{c_\theta a^-+s_\theta s^-}{2}} & 
{\displaystyle   \frac{ c_\theta a^0+s_\theta s^0 - i x^0}{2 \sqrt{2}} } & 
{\displaystyle -\frac{ c_\theta s^0 - s_\theta a^0 +v^0}{2 \sqrt{2}}  } & 
{\displaystyle \frac{-s_\theta a^- +c_\theta s^- -v^-}{2} } \\ \smallskip 
\\
{\displaystyle   \frac{ c_\theta a^0+ s_\theta s^0 + i x^0}{2 \sqrt{2}} } & 
{\displaystyle  \frac{-c_\theta a^+ -s_\theta s^+}{2} }&
{\displaystyle    \frac{-s_\theta a^+ +c_\theta s^+ -v^+}{2} } & 
{\displaystyle \frac{c_\theta s^0 -s_\theta a^0+v^0 }{2 \sqrt{2}} }
\end{pmatrix} \ ,
\end{eqnarray}    

\begin{eqnarray}
\widetilde{\bm{\mathcal{F}}}_{\mu} =
\begin{pmatrix}
{\displaystyle \frac{s_\theta \widetilde{v}^0+c_\theta \widetilde{x}^0}{2 \sqrt{2}} } & 
0 & 
{\displaystyle \frac{\tilde{s}^+}{2} } & 
{\displaystyle \frac{c_\theta \widetilde{v}^0 - s_\theta\widetilde{x}^0 + i \widetilde{s}^0}{2 \sqrt{2}} } \\ \smallskip 
\\
0 & 
{\displaystyle \frac{c_\theta \widetilde{x}^0+ s_\theta \widetilde{v}^0}{2 \sqrt{2}} }&
{\displaystyle \frac{-c_\theta \widetilde{v}^0 + s_\theta\widetilde{x}^0 +i \widetilde{s}^0}{2 \sqrt{2}} } & 
{\displaystyle \frac{\widetilde{s}^-}{2} }  \\ \smallskip 
\\
{\displaystyle \frac{\widetilde{s}^-}{2}} & 
{\displaystyle   \frac{ -c_\theta\tilde{v}^0+s_\theta\widetilde{x}^0-i \widetilde{s}^0 }{2 \sqrt{2}} } & 
{\displaystyle -\frac{ c_\theta\widetilde{x}^0 +s_\theta \widetilde{v}^0}{2 \sqrt{2}}  } & 
0 \\ \smallskip 
\\
{\displaystyle   \frac{ c_\theta\widetilde{v}^0- s_\theta\widetilde{x}^0-i\widetilde{s}^0}{2 \sqrt{2}} } & 
{\displaystyle  \frac{\widetilde{s}^+}{2} }&
0 & 
{\displaystyle \frac{-c_\theta\widetilde{x}^0-s_\theta\widetilde{v}^0 }{2 \sqrt{2}} }
\end{pmatrix} \ .
\end{eqnarray}  
\end{footnotesize}}

In the gauge eigenbasis, the vector mass matrices in the charged ${\cal M}_{\rm C}$ and neutral ${\cal M}_{\rm N}$ sectors are 
\begin{eqnarray}
{\bf {\cal M}_{\rm C}^2} =
\begin{pmatrix}
{\displaystyle \frac{g^2 M_V^2 (1+\omega s_\theta^2)}{\widetilde{g}^2} } & 
{\displaystyle \frac{- g r M_A^2 s_{\theta }}{\sqrt{2} \widetilde{g}} }&
{\displaystyle \frac{- g M_V^2}{\sqrt{2} \widetilde{g}} }& 
{\displaystyle \frac{- g M_V^2 c_{\theta }}{\sqrt{2} \widetilde{g}}} 
  \\
{\displaystyle \frac{- g r M_A^2 s_{\theta }}{\sqrt{2} \widetilde{g}} }& 
{\displaystyle M_A^2 }& 
0 & 
0 
 \\
 {\displaystyle  \frac{- g M_V^2}{\sqrt{2} \widetilde{g}} }& 
 0 & 
{\displaystyle  M_V^2 }& 
 0 \\
{\displaystyle  \frac{- g M_V^2 c_{\theta }}{\sqrt{2} \widetilde{g}} }& 
 0 & 0 & 
 {\displaystyle M_V^2 }\\
\end{pmatrix} \ ,
\label{eq:massC}
\end{eqnarray}

\begin{eqnarray}
{\bf {\cal M}_{\rm N}^2} =
\begin{pmatrix}
{\displaystyle \frac{g'^2 M_V^2 (1+\omega s_\theta^2)}{\tilde{g}^2}} &
{\displaystyle -\frac{g' g M_V^2 \omega s_{\theta }^2}{\tilde{g}^2} }& 
{\displaystyle \frac{-g' M_A^2 r s_{\theta }}{\sqrt{2} \tilde{g}} }& 
{\displaystyle \frac{-g' M_V^2}{\sqrt{2} \tilde{g}} }& 
{\displaystyle \frac{-g' c_{\theta } M_V^2}{\sqrt{2} \tilde{g}} }\\
\smallskip \\
{\displaystyle  -\frac{g' g M_V^2 \omega  s_{\theta }^2}{\tilde{g}^2} }& 
{\displaystyle  \frac{g^2 (1+\omega s_\theta^2) M_V^2}{\tilde{g}^2} }& 
{\displaystyle  \frac{ g r M_A^2 s_{\theta }}{\sqrt{2} \tilde{g}} }& 
{\displaystyle  \frac{- g M_V^2}{\sqrt{2} \tilde{g}} }& 
{\displaystyle  \frac{g c_{\theta } M_V^2}{\sqrt{2} \tilde{g}} }\\
\smallskip \\
{\displaystyle \frac{-g' M_A^2 r s_{\theta }}{\sqrt{2} \tilde{g}} }& 
{\displaystyle \frac{ g M_A^2 r s_{\theta }}{\sqrt{2} \tilde{g}} }& 
{\displaystyle M_A^2 }& 
       0 & 
       0 \\
\smallskip \\
{\displaystyle \frac{- g' M_V^2}{\sqrt{2} \tilde{g}} }& 
{\displaystyle \frac{- g M_V^2}{\sqrt{2} \tilde{g}} }& 
	       0 & 
{\displaystyle M_V^2 }& 
	       0 \\
\smallskip \\
{\displaystyle \frac{- g' c_{\theta } M_V^2}{\sqrt{2} \tilde{g}} }& 
{\displaystyle \frac{g c_{\theta } M_V^2}{\sqrt{2} \tilde{g}} }& 
 	       0 & 
 	       0 & 
{\displaystyle M_V^2} \\
\end{pmatrix}  \ , \nonumber \\
\label{eq:massN}
\end{eqnarray}
where $2\omega=f_0^2/f_K^2-1$.

In the same basis, we provide the couplings of one Higgs with charged vector bosons
\beq
&& c_{hV^+V^-}= \nonumber \\
&& \left(
\begin{array}{cccc}
 \frac{\sqrt{2} g^2 M_V^2 \omega  \cos \theta
   \sin \theta }{\tilde{g} \sqrt{M_V^2 (2 \omega
   +1)-M_A^2 r^2}} & \frac{g
   \left(M_V^2-M_A^2\right) r \cos \theta }{2
   \sqrt{M_V^2 (2 \omega +1)-M_A^2 r^2}} & 0 &
   \frac{g (M_V^2-M_A^2 r^2) \sin \theta }{2
   \sqrt{M_V^2 (2 \omega +1)-M_A^2 r^2}} \\
 \frac{g \left(M_V^2-M_A^2\right) r \cos
   \theta }{2 \sqrt{M_V^2 (2 \omega +1)-M_A^2
   r^2}} & 0 & 0 & \frac{\tilde{g} (M_A^2-M_V^2) r}{\sqrt{2} \sqrt{M_V^2 (2
   \omega +1)-M_A^2 r^2}} \\
 0 & 0 & 0 & 0 \\
 \frac{g (M_V^2-M_A^2 r^2) \sin \theta }{2
   \sqrt{M_V^2 (2 \omega +1)- M_A^2 r^2}} &
   \frac{\tilde{g} (M_A^2-M_V^2) r}{\sqrt{2} \sqrt{M_V^2 (2
   \omega +1)-M_A^2 r^2}} & 0 & 0 \\
\end{array}
\right) \label{eq:chVVpm}
\eeq
and for the neutral ones
{\begin{footnotesize}
\beq
&& c_{hV^0V^0}= \nonumber  \\
&& \left(
\begin{array}{ccccc}
 \frac{\sqrt{2} g'^2 M_V^2 \omega  \cos \theta
    \sin \theta }{\tilde{g} \sqrt{M_V^2 (2 \omega
   +1)-M_A^2 r^2}} & -\frac{\sqrt{2} g'
   g M_V^2 \omega  \cos\theta  \sin \theta
   }{\tilde{g} \sqrt{M_V^2 (2 \omega +1)-M_A^2
   r^2}} & \frac{g'
   \left(M_V^2-M_A^2\right) r \cos \theta }{2
   \sqrt{M_V^2 (2 \omega +1)-M_A^2 r^2}} & 0 &
   \frac{g' (M_V^2-M_A^2 r^2) \sin\theta}{2
   \sqrt{M_V^2 (2 \omega +1)-M_A^2 r^2}} \\
 -\frac{\sqrt{2} g' g M_V \omega  \cos
   \theta  \sin \theta }{\tilde{g} \sqrt{M_V^2 (2
   \omega +1)- M_A^2 r^2}} & \frac{\sqrt{2} g^2
   M_V \omega  \cos \theta  \sin \theta
   }{\tilde{g} \sqrt{M_V^2 (2 \omega +1)-M_A^2
   r^2}} & \frac{g (M_A^2-M_V^2) r \cos \theta }{2
   \sqrt{M_V^2 (2 \omega +1)-M_A^2 r^2}} & 0 &
   -\frac{g (M_V^2-M_A^2 r^2)  \sin\theta }{2
   \sqrt{M_V^2 (2 \omega +1)-M_A^2 r^2}} \\
 \frac{g' \left(M_V^2-M_A^2\right) r \cos
   \theta }{2 \sqrt{M_V^2 (2 \omega +1)-M_A^2
   r^2}} & \frac{g (M_A^2-M_V^2) r \cos \theta }{2
   \sqrt{M_V^2 (2 \omega +1)-M_A^2 r^2}} & 0 & 0
   & \frac{\tilde{g} (M_A^2-M_V^2) r}{\sqrt{2} \sqrt{M_V^2 (2
   \omega +1)-M_A^2 r^2}} \\
 0 & 0 & 0 & 0 & 0 \\
 \frac{g' (M_V^2-M_A^2 r^2) \sin \theta}{2
   \sqrt{M_V^2 (2 \omega +1)-M_A^2 r^2}} &
   -\frac{g (M_V^2-M_A^2 r^2) \sin \theta }{2
   \sqrt{M_V^2 (2 \omega +1)-M_A^2 r^2}} &
   \frac{\tilde{g} (M_A^2-M_V^2) r}{\sqrt{2} \sqrt{M_V^2 (2
   \omega +1)-M_A^2 r^2}} & 0 & 0 \\
\end{array}
\right)\,. \label{eq:chVV0}
\eeq
\end{footnotesize}}

Similarly, the $\eta$-$V$-$V$ interaction in gauge eigenstate are provided below:

 \begin{eqnarray}
{\cal L}_{\eta,C}^G&= &  -\frac{g \sin^2 \theta 
   \left(M_V^2-M_A^2 r^2\right)}{\sqrt{2}\gt v } \, \eta\, \widetilde{s}^+_\mu  \tilde{W} ^{-,\mu}  +  \frac{\sin\theta 
   \left(M_A^2-M_V^2\right) r}{v} \, \eta\, \widetilde{s}^+_\mu  a ^{-,\mu}  + h.c   \nonumber
\end{eqnarray} 

\begin{eqnarray}
{\cal L}_{\eta,N}^G&= & \frac{g' \sin^2 \theta 
   \left(M_V^2-M_A^2 r^2\right)}{\sqrt{2}
   \gt v} \, \eta\,\widetilde{s}^0_\mu B^{\mu}  - \frac{g \sin^2 \theta 
   \left(M_V^2-M_A^2 r^2\right)}{\sqrt{2}
   \gt v} \,  \eta \widetilde{s}^0_\mu \widetilde{W}^{3, \mu}   \nonumber \\
&+& \frac{\sin\theta r
   \left(M_A^2-M_V^2\right)}{v}  \eta \widetilde{s}^0_\mu a^{0, \mu}  +  \frac{\sin\theta r
   \left(M_A^2-M_V^2\right)}{v}  \eta \widetilde{v}^0_\mu x^{0, \mu}  \,.
\end{eqnarray}

The above couplings are provided in the gauge eigenbasis, so one need to include the mixing matrices in 
order to extract couplings in the mass eigenstate basis. Approximate expressions for the mixing matrices are provided in the following section.

\subsection{Perturbative diagonalisation of the mass matrices}
\label{sec:pertdiag}

The label of the physical states, $W^{+\mu}$, $A^{+\mu}$, $V^{+\mu}$ and $S^{+\mu}$ in the charged sector, and  $A^{\mu}$, $Z^{\mu}$, $A^{0\mu}$,  $V^{0\mu}$ and $S^{0\mu}$ in the neutral sector  (left hand side of \eq{eq:mixmatrix}), are defined as the ones with predominant component of the corresponding interaction eigenstates,    $\widetilde{W}^{+\mu}$, $a^{+\mu}$, $v^{+\mu}$ and $s^{+\mu}$ in the charged sector, and  $B^{\mu}$, $\widetilde{W}^{3\mu}$, $a^{0\mu}$,  $v^{0\mu}$ and $s^{0\mu}$ in the neutral sector respectively. 
Therefore, in theory, the columns in ${\cal C}$ and ${\cal N}$ do not assume fixed expressions which can swap depending on the largest entry, \emph{i.e}, the matrix is reorganised in such a way that the diagonal entry is the largest in each column. 
In practice, however, for the parameter values we consider, the columns 1 and 2 in ${\cal C}$ and 1,2 and 3 in ${\cal N}$ have fixed expressions, even though there are significant mixing between the photon, $A^\mu$ and $Z^\mu$. On the other hand, the states $V^{0,\pm}_\mu$ and $S^{0,\pm}_\mu$ are highly mixed, and columns 3 and 4 in ${\cal C}$ and 4 and 5 in ${\cal N}$ can be swapped, depending on the parameters, to fulfil our definition of these states.

In the following we provide expressions for these mixing matrices, ${\cal C}$ and ${\cal N}$, defined in \eq{eq:mixmatrix}, keeping in mind that the last two columns may be swapped depending on the values of their entries.

The charged rotation matrix can be split like
\be
{\cal C}= {\cal C}^a{\cal C}^b
\ee
where
\begin{eqnarray}
{\bf {\cal C}^a} =
\left(
\begin{array}{cccc}
 1 & 0 & 0 & 0 \\
 0 & 1 & 0 & 0 \\
 0 & 0 & \frac{\cos \theta}{\sqrt{\text{cos}^2(\theta )+1}} & \frac{1}{\sqrt{\text{cos}^2(\theta )+1}} \\
 0 & 0 & -\frac{1}{\sqrt{\text{cos} ^2(\theta )+1}} & \frac{\cos \theta}{\sqrt{\text{cos} ^2(\theta )+1}} \\
\end{array}
\right)
\label{eq:Ca}
\end{eqnarray}
rotates away a state with mass exactly $M_V$. The other part ${\cal C}^b$ at leading order in $g/\gt$ is given by:
\begin{eqnarray}
 {\cal C}^b_{11}&=& 1-\frac{1}{4} \left(\frac{g}{\gt}\right)^2 \left(\cos ^2(\theta )+r^2 \sin ^2(\theta )+1\right) \\
 {\cal C}^b_{12}&=& \frac{g r \sin (\theta)}{\gt\sqrt{2}} \\
 {\cal C}^b_{14}&=&  \frac{g \sqrt{1+ \cos^2(\theta)}}{ \gt\sqrt{2}} \\
 {\cal C}^b_{21}&=& \frac{g r  \sin (\theta )}{\gt\sqrt{2}} \\
 {\cal C}^b_{22}&=&  -1 +\frac{1}{4} \left(\frac{g}{\gt}\right)^2 \left(r^2 \sin^2(\theta) \right) \\
 {\cal C}^b_{24}&=&  \left(\frac{g}{\gt}\right)^2\frac{ M_A^2 r  \sin (\theta) \sqrt{1+ \cos^2(\theta)}}{2 \left(M_A^2- M_V^2\right)} \\
 {\cal C}^b_{41}&=& \frac{g \sqrt{1+ \cos^2(\theta)}}{ \gt\sqrt{2}} \\
 {\cal C}^b_{42}&=&  -\left(\frac{g}{\gt}\right)^2\frac{ M_V^2 r  \sin (\theta) \sqrt{1+ \cos^2(\theta)}}{2 \left(M_A^2- M_V^2\right)}\\
 {\cal C}^b_{44}&=&  -1 +   \left(\frac{g}{\gt}\right)^2\frac{(1+ \cos^2 (\theta))}{4}  
\end{eqnarray} 
and ${\cal C}^b_{3i}={\cal C}^b_{i3}=0,\,i\neq3$, ${\cal C}^b_{33}=1$.

For the neutral  gauge bosons,  we  define:
 \begin{equation}
 \mathcal{N} =\mathcal{N}^a\cdot \mathcal{N}^b\cdot \mathcal{N}^c 
 \end{equation}
At leading order in $1/ \tilde{g} $, each  matrix  has  the following explicit  expression:

\begin{eqnarray}
\mathcal{N}^a_{11} &=& 1- \frac{1}{4} \left(\frac{g'}{\gt}\right) \left( 1+ \cos ^2(\theta)+ r^2 \sin ^2(\theta ) \right), \\
\mathcal{N}^a_{21} &=&0, \\
\mathcal{N}^a_{31} &=& \frac{g' r   \sin (\theta )}{\gt\sqrt{2}}, \\ 
\mathcal{N}^a_{41} &=&\frac{g' }{\gt\sqrt{2}}, \\
\mathcal{N}^a_{51} &=&\frac{g'   \cos (\theta )}{\gt\sqrt{2}} 
\end{eqnarray}

\begin{eqnarray}
\mathcal{N}^a_{12} &=& - \frac{1}{2}\left(\frac{g' g }{\gt^2}\right) \left(1- r^2\right)  \sin ^2(\theta ), \\
\mathcal{N}^a_{22} &=&  1- \frac{1}{4} \left(\frac{ g }{\gt}\right)^2 \left( 1+ \cos ^2(\theta)+ r^2 \sin ^2(\theta ) \right), \\
\mathcal{N}^a_{32} &=&-\frac{ g r   \sin (\theta )}{\gt\sqrt{2}}, \\
\mathcal{N}^a_{42} &=&\frac{ g }{\gt\sqrt{2}}, \\
\mathcal{N}^a_{52} &=& -\frac{ g   \cos (\theta )}{\gt\sqrt{2}}
\end{eqnarray}

\begin{eqnarray}
\mathcal{N}^a_{13} &=& -\frac{ g' r  \sin (\theta )}{\gt\sqrt{2}}, \\
\mathcal{N}^a_{23} &=&\frac{g r  \sin (\theta )}{\gt\sqrt{2}},  \\
\mathcal{N}^a_{33} &=& 1- \frac{1}{4} \frac{ \left( g'^2 + g^2 \right) }{\gt^2} r^2  \sin^2(\theta ), \\
\mathcal{N}^a_{43} &=& \frac{1}{2}  r \frac{ \left( g^2 - g'^2 \right) }{\gt^2} \sin (\theta) \frac{ M_V^2 }{ \left( M_V^2- M_A^2\right)}, \\
\mathcal{N}^a_{53} &=&-\frac{1}{2} r \frac{ \left( g'^2 + g^2 \right) }{\gt^2} \sin (\theta ) \cos (\theta ) \frac{ M_V^2}{\left(M_V^2- M_A^2\right)} 
\end{eqnarray}

\begin{eqnarray}
\mathcal{N}^a_{14} &=& -\frac{ g'  }{\gt\sqrt{2}}, \\
\mathcal{N}^a_{24} &=&- \frac{ g }{\gt\sqrt{2}},  \\
\mathcal{N}^a_{34} &=& -\frac{1}{2}  r \frac{ \left( g^2 - g'^2 \right) }{\gt^2} \sin (\theta) \frac{  M_A^2 }{ \left( M_V^2- M_A^2\right)},\\
\mathcal{N}^a_{44} &=& 1 - \frac{1}{4} \frac{ \left( g'^2 + g^2 \right) }{\gt^2}, \\
\mathcal{N}^a_{54} &=&0
\end{eqnarray}

\begin{eqnarray}
\mathcal{N}^a_{15} &=&- \frac{g'  \cos (\theta )}{\gt\sqrt{2}}, \\
\mathcal{N}^a_{25} &=& \frac{ g \cos (\theta )}{\gt\sqrt{2}}, \\
\mathcal{N}^a_{35} &=& \frac{1}{2} r \frac{ \left( g'^2 + g^2 \right) }{\gt^2} \sin (\theta ) \cos (\theta ) \frac{  M_A^2}{\left(M_V^2- M_A^2\right)} , \\
\mathcal{N}^a_{45} &=&  -\frac{1}{2} \frac{ \left( g'^2 - g^2 \right) }{\gt^2}  \cos (\theta), \\
\mathcal{N}^a_{55} &=& 1-\frac{1}{4} \frac{\left(  g'^2 + g^2 \right)}{\gt^2} \cos ^2(\theta)
\end{eqnarray}

\begin{eqnarray}
\mathcal{N}^b= \left(
\begin{array}{ccccc}
 \frac{g}{\sqrt{g'^2+ g^2}} &
   \frac{g'}{\sqrt{g'^2+ g^2}} & 0 & 0 & 0 \\
 \frac{ g'}{\sqrt{ g'^2+ g^2}} &
   -\frac{ g}{\sqrt{g'^2+g^2}} & 0 & 0 & 0 \\
 0 & 0 & 1 & 0 & 0 \\
 0 & 0 & 0 & 1 & 0 \\
 0 & 0 & 0 & 0 & 1 \\ 
\end{array}
\right) 
\end{eqnarray}

$\mathcal{N}^a\cdot \mathcal{N}^b$  will take the mass matrix  into the following form:
\begin{eqnarray}
\left(
\begin{array}{ccccc}
 0 & 0 & 0 & 0 & 0 \\
 0 & M_Z^2 & 0 & 0 & 0 \\
 0 & 0 & M_{A^0}^2 & 0 & 0 \\
 0 & 0 & 0 & \frac{1}{2\gt^2}  M_{V}^2 \left( \left(g'^2+g^2\right) 
   +2\right) & \frac{1}{2\gt^2} \left({g'}^2-g^2\right)
   M_{V}^2  \cos (\theta ) \\
 0 & 0 & 0 & \frac{1}{2\gt^2} \left(g'^2-g^2\right) M_{V}^2  \cos (\theta ) & \frac{1}{2\gt^2} M_{V}^2
   \left(\left(g'^2+g^2\right) \cos ^2(\theta )+2\right)
   \\
\end{array}
\right)
\end{eqnarray}

For  the  vector bosons $V^0$ and $S^0$,  we take  a  further  approximation  $ \sin^2 \theta \sim 0$ , and we define:
\begin{eqnarray}
\mathcal{N}^c= \left(
\begin{array}{ccccc}
 1 & 0 & 0 & 0 & 0 \\
 0 & 1 & 0 & 0 & 0 \\
 0 & 0 & 1 & 0 & 0 \\
 0 & 0 & 0 & \frac{\left( g'^2+ g^2\right) \sin^2 \theta}{4\sqrt{2}
   \left( g'^2- g^2\right)}+\frac{1}{\sqrt{2}} &
   -\frac{1}{\sqrt{2}}-\frac{\left( g'^2+ g^2\right) \sin^2 \theta}{4\sqrt{2}
   \left(g'^2- g^2\right)} \\
 0 & 0 & 0 & \frac{1}{\sqrt{2}}-\frac{\left(g'^2+ g^2\right) \sin^2 \theta}{4\sqrt{2}
   \left( g'^2- g^2\right)} &
   \frac{\left(g'^2+ g^2\right)\sin^2 \theta}{4\sqrt{2}
   \left(g'^2- g^2\right)}+\frac{1}{\sqrt{2}} \\
\end{array}
\right)
\end{eqnarray}

The  rotation  $\mathcal{N} =\mathcal{N}_1\cdot \mathcal{N}_2 \cdot \mathcal{N}_3$ will fully diagonalize 
the  mass matrix to be:
\begin{eqnarray}
\left(
\begin{array}{ccccc}
 0 & 0 & 0 & 0 & 0 \\
 0 & M_Z^2 & 0 & 0 & 0 \\
 0 & 0 &M_{A^0}^2 & 0 & 0 \\
 0 & 0 & 0 & \frac{1}{2} M_{V}^2 \left(2+  \frac{g'^2}{\tilde{g}^2} \left(2 - \sin
   ^2 \theta \right)\right) & 0 \\
 0 & 0 & 0 & 0 & \frac{1}{2} M_{V}^2 \left( 2+ \frac{g^2}{\tilde{g}^2} \left(2-\sin
   ^2 \theta  \right)\right) \\
\end{array}
\right)
\end{eqnarray}

\subsection{EW Precision parameters}

The oblique parameters are related to the polarisation functions of the EW gauge bosons:
\begin{eqnarray}
\hat{S} &\equiv & \frac{\Pi_{W^3B}^\prime (0)}{\Pi_{W^+W^-}^\prime (0)} \ , \\
\hat{T} &\equiv & \frac{1}{M_W^2}\frac{\Pi_{W^3W^3}(0) -
\Pi_{W^+W^-}(0) }{\Pi_{W^+W^-}^\prime (0)}\ , \\
\hat{U} &\equiv & -\frac{\Pi^\prime_{W^3W^3}(0)-
\Pi^\prime_{W^+W^-}(0)} {\Pi_{W^+W^-}^\prime (0)}\ , \\
W &\equiv & \frac{M_W^2}{2} \frac{\Pi^{\prime\prime}_{W^3W^3}(0)}{\Pi_{W^+W^-}^\prime (0)} \ , \\
Y &\equiv & \frac{M_W^2}{2} \frac{\Pi^{\prime\prime}_{BB}(0)}{\Pi_{BB}^\prime (0)} \ , \\
X &\equiv & \frac{M_W^2}{2} \frac{\Pi_{W^3B}^{\prime\prime}(0)}{\sqrt{\Pi_{W^+W^-}^\prime (0)\Pi_{BB}^\prime (0)}}  
\label{eq:ewposelfen}
\end{eqnarray}

\section{G- parity transformation}
\label{sec:parity}
The convention we are using here are:
\beq
& &\gamma^\mu = \left(\begin{array}{cc}
		& \sigma^\mu \\ \overline{\sigma}^\mu & 
	\end{array}\right), 
	\hspace{1cm}
	C = \gamma^0 \gamma^2 = \left(\begin{array}{cc}
		-\sigma^2 & \\ & \sigma^2
	\end{array}\right),  \\
& & \sigma^\mu = \left( 1, \sigma^i \right),
	\hspace{2cm} 
	\overline{\sigma}^\mu = \left( 1, -\sigma^i \right). 
\eeq

with  the conjugate of  fermion currents  derived to be:
\beq 
& \left( \overline{U} \gamma^\mu \gamma^5 D \right)^\dag =  \overline{D} \gamma^\mu \gamma^5 U  & \\ 
&  \left( U^T C \gamma^\mu D \right)^\dag =  - \overline{D} \gamma^\mu C \overline{U}^T  &\\
& \left( U^T C  D \right)^\dag =  - \overline{D} C \overline{U}^T , ~~ \left( U^T C \gamma^\mu \gamma^5 D \right)^\dag =  - \overline{D} \gamma^\mu C \gamma^5 \overline{U}^T &
\eeq

Using the definition for the  G-parity in eq. (\ref{G-parity}), we can derive its action on fermionic currents to be:

\beq
&  \overline{U} U   \xrightarrow{G}   \overline{D} D,   \quad  \overline{D} D   \xrightarrow{G}    \overline{U} U& \\
& \overline{D} \gamma^\mu U   \xrightarrow{G}  - \overline{D} \gamma^\mu U, \quad \overline{D} \gamma^\mu \gamma^5 U   \xrightarrow{G}    \overline{D} \gamma^\mu \gamma^5 U & \\
& \overline{U} \gamma^\mu U   \xrightarrow{G}   \overline{D} \gamma^\mu D, \quad   \overline{D} \gamma^\mu D  \xrightarrow{G}    \overline{U} \gamma^\mu U & \\
& \overline{D} \gamma^\mu \gamma^5 D   \xrightarrow{G}  -  \overline{U} \gamma^\mu \gamma^5 U, \quad \overline{U} \gamma^\mu \gamma^5 U   \xrightarrow{G}  -  \overline{D} \gamma^\mu \gamma^5 D & \\
& U^T C \gamma^\mu \gamma^5 U   \xrightarrow{G} - \overline{D}  \gamma^\mu C \gamma^5 \overline{D}^T  , \quad
D^T C \gamma^\mu \gamma^5 D   \xrightarrow{G} -  \overline{U}  \gamma^\mu C \gamma^5 \overline{U}^T  &
\eeq

\beq
& U^T C  D   \xrightarrow{G}  - (U^T C D)^\dag  & \\
& U^T C \gamma^\mu D   \xrightarrow{G} - ( U^T C \gamma^\mu D)^\dag   & \\
& U^T C \gamma^\mu \gamma^5 D   \xrightarrow{G}  -(U^T C \gamma^\mu \gamma^5 D)^\dag  &
\eeq

\beq
& \Re (U^T C  D )  \xrightarrow{G} - \Re (U^T C  D), ~~  \Im (U^T C  D  ) \xrightarrow{G}  \Im (U^T C  D)&
\\
& \Re (U^T C \gamma^\mu D )  \xrightarrow{G}  - \Re (U^T C \gamma^\mu D), ~~  \Im (U^T C \gamma^\mu D  ) \xrightarrow{G}  \Im (U^T C \gamma^\mu D) &
\\
& \Re (U^T C \gamma^\mu \gamma^5 D )  \xrightarrow{G}  - \Re (U^T C \gamma^\mu \gamma^5 D), ~~ \Im (U^T C \gamma^\mu \gamma^5 D  ) \xrightarrow{G}   \Im (U^T C \gamma^\mu \gamma^5 D) &
\eeq

\section{Branching ratios}
\label{sec:br}

\begin{figure}
\includegraphics[width=0.49\textwidth]{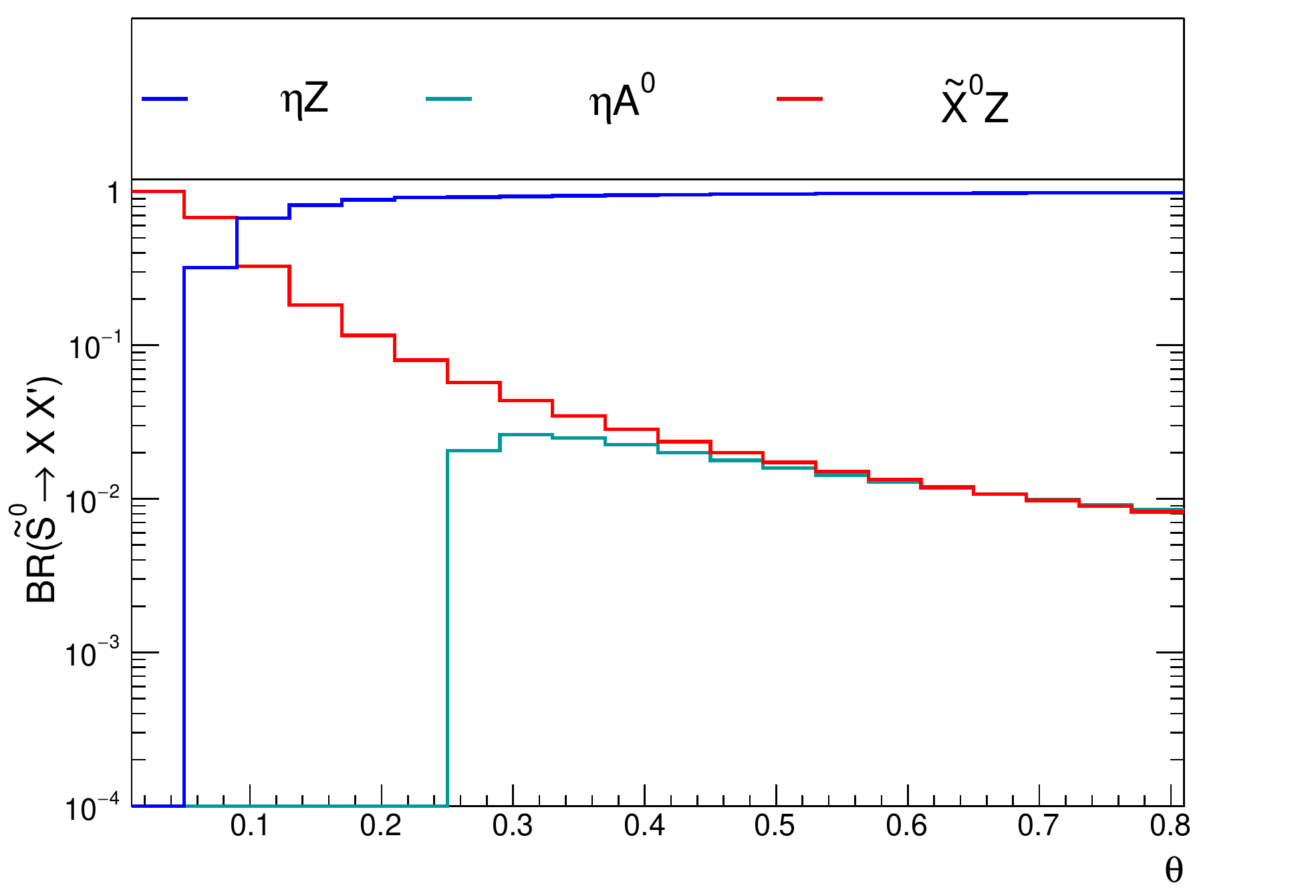}
\includegraphics[width=0.49\textwidth]{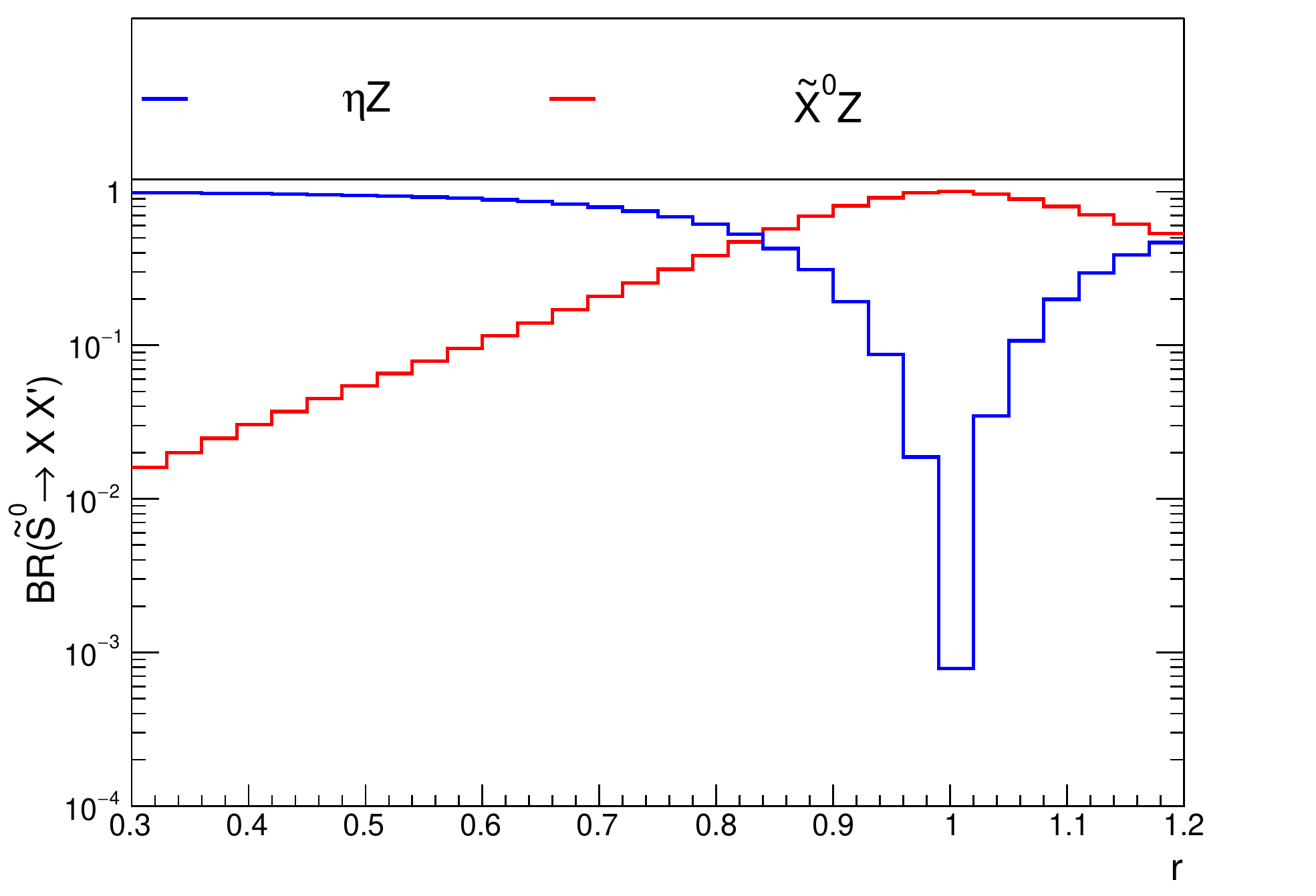}
\caption{Branching ratios (BR) of $\tilde{S}^0$ for $M_A=3$ TeV, $M_V=3.5$ TeV and $\gt=3$. On the \emph{left} as a function of $\theta$ with $r=0.6$ and on the \emph{right} as a function of $r$ with $\theta=0.2$. The behaviour of the charged $\tilde{S}^\pm$ is analogous, replacing $Z$ by $W^\pm$ and $A^0$ by $A^\pm$.}
\label{fig:BRSt0}
\end{figure}

\begin{figure}
\includegraphics[width=0.49\textwidth]{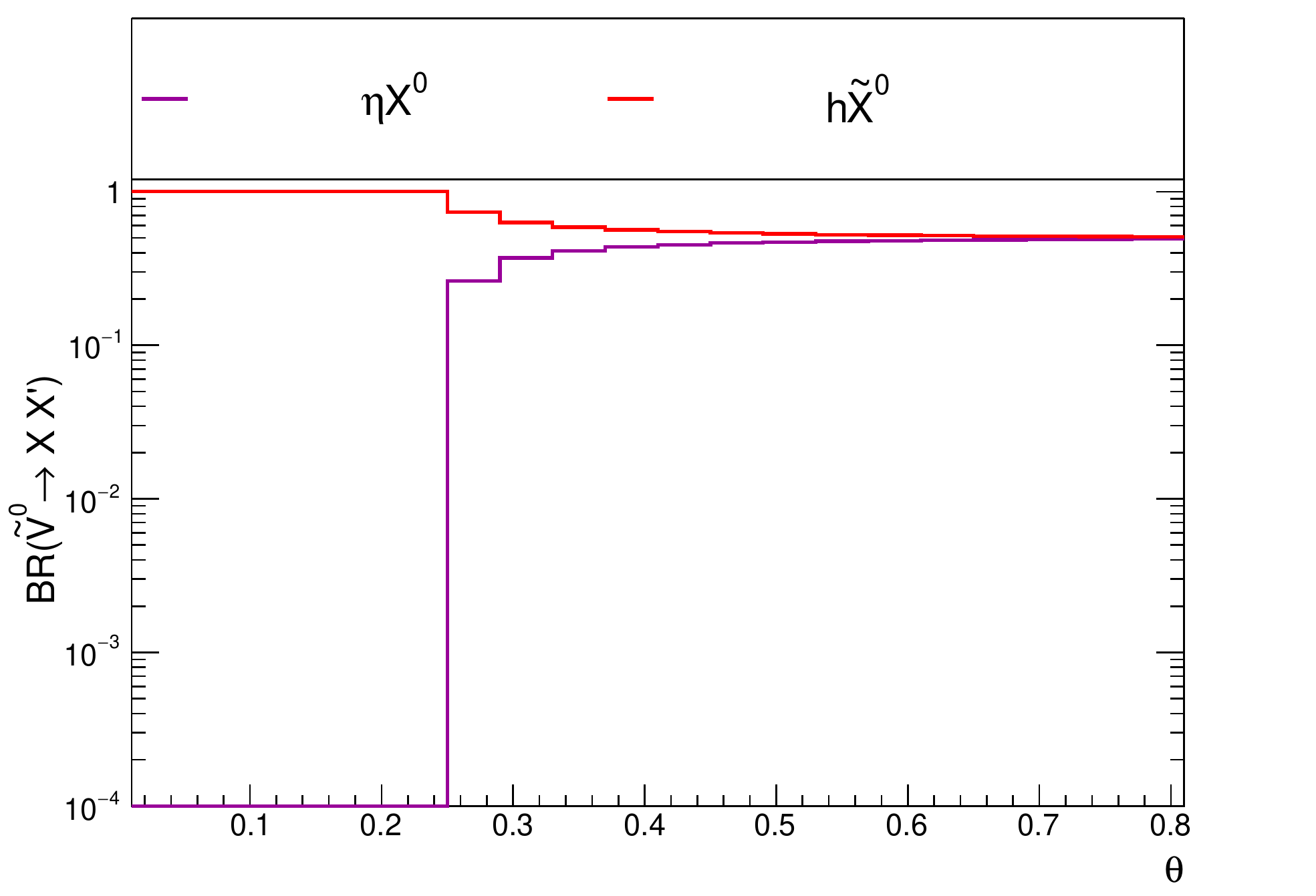}
\includegraphics[width=0.49\textwidth]{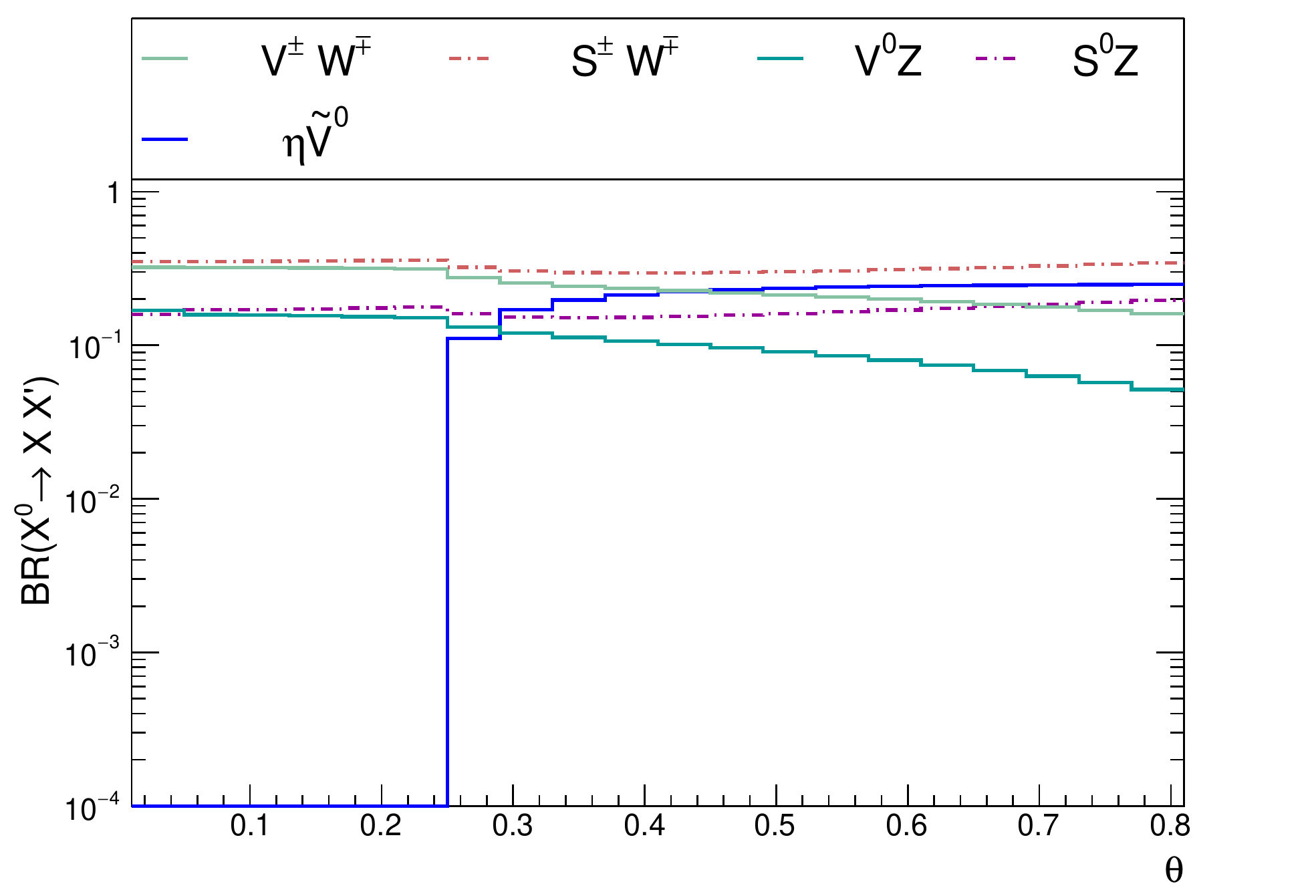}
\caption{\emph{Left:} ratios (BR) of $\tilde{V}^0$ for $M_A=3$ TeV, $M_V=3.5$ TeV. It is independent of  $\gt$ and $r$. This state decays only to $h \tilde{X}^0$ and $\eta X^0$. 
\emph{Right:}Branching ratios (BR) of $X^0$ for $M_A=3$ TeV and $M_V=2.5$ TeV. It is independent of  $\gt$ and $r$.}
\label{fig:BRV0}
\end{figure}

Since in \sec{sec:pheno} we discussed the production and decay of $V$, $S$ and $A$ triplets,  here in Fig.( \ref{fig:BRSt0}) and Fig. (\ref{fig:BRV0})  we show some representative branching ratio distributions of the more exotic vector states.

The branching ratio of $\tilde{X}^0$  is  independent on $\tilde{g}$, $r$ and $\theta$, and this state will decay  into $h\tilde{V}^0$, $Z\tilde{S}^0$, $W^\pm\tilde{S}^\mp$ in the ratio of $(1:1:2)$. 

\newpage

\bibliographystyle{JHEP}

\bibliography{vectors}

\end{document}